\documentclass[aps,prb,twocolumn]{revtex4-1}
\usepackage{graphicx}
\usepackage{float}
\usepackage{times}
\usepackage{color}
\usepackage{gensymb}
\input pdfcolor.tex
\usepackage{graphicx}
\usepackage{epsfig,color}

\begin{document}

\author{Sauri Bhattacharyya and Pinaki Majumdar}

\affiliation{Harish-Chandra Research Institute, HBNI,\\
  Chhatnag Road, Jhunsi, Allahabad 211 019, India}

\title{Dynamics of magnetic collective modes in 
the square and triangular lattice \\
Mott insulators at finite temperature}

\date{\today}

\begin{abstract}
We study the equilibrium dynamics of magnetic moments in the Mott insulating 
phase of the Hubbard model on the square and triangular lattice. We rewrite 
the Hubbard interaction in terms of an auxiliary vector field and use a 
recently developed Langevin scheme to study its dynamics. A thermal noise, 
derivable approximately from the Keldysh formalism, allows us to study 
the effect of finite temperature. At strong coupling, $U \gg t$, 
where $U$ is the local repulsion and $t$ the nearest neighbour hopping, 
our results reproduce the well known dynamics of the nearest neighbour 
Heisenberg model with exchange $J \sim {\cal O}(t^2/U)$. 
These include crossover from weakly damped dispersive 
modes at temperature $T \ll J$ to strong damping at $T \sim {\cal O}(J)$, 
and diffusive dynamics at $T \gg J$. The crossover temperatures are 
naturally proportional to $J$. To highlight the progressive deviation from 
Heisenberg physics as $U/t$ reduces we compute an effective exchange scale 
$J_{eff}(U)$ from the low temperature spin wave velocity. We discover two 
features in the dynamical behaviour with decreasing $U/t$: (i)~the low 
temperature dispersion deviates from the Heisenberg result, as expected, due 
to longer range and multispin interactions, and (ii)~the crossovers between 
weak damping, strong damping, and diffusion take place at noticeably lower 
values of $T/J_{eff}$. We relate this to enhanced mode coupling, in particular 
to thermal amplitude fluctuations, at weaker $U/t$. A comparison of the square 
and triangular lattice reveals the additional effect of geometric frustration 
on damping.
\end{abstract}

%
%
%
%
\keywords{Hubbard model, Mott insulator, spin waves, Heisenberg model}
\maketitle

\section{Introduction}

The Hubbard model at half-filling 
provides a minimal description
of an interaction driven Mott metal-insulator transition 
\cite{zhang,imada,rozenberg,
capone,park,ohashi1,ohashi2,furukawa,sahebsara,yamada1} 
(MIT).
The Mott phase generally has some kind of antiferromagnetic order 
\cite{sahebsara,yamada1}, except in fully frustrated lattices like 
the Kagome or pyrochlore where it has only short range correlations
\cite{bulut,ohashi1,furukawa,yamada2,kita,fujimoto,normand,swain}. 
The static charge and magnetic correlations are reasonably well understood 
in the various lattices \cite{hirsch,white,hrk,zimmermann,fazekas,tasaki,
hochkeppel,watanabe,yoshioka,tocchio1,tocchio2,kokalj,goto,shirakawa,li}. 

Theoretical results on dynamics are more limited. 
Approaches like dynamical mean field theory (DMFT) or its
extensions, which provide a detailed description of the MIT,
focus on the single particle spectral function 
\cite{DMFT,bulla,zitko,eckstein,peters,kitatani}.
The collective mode dynamics associated with 
the magnetic degrees of freedom is much less explored
\cite{kotliar1,schulz,ho,singh1,singh2,singh3,gunnarsson,chern,leblanc1}, 
although in the Mott phase, where single particle excitations 
are gapped, these are in fact the relevant degrees of freedom.

Deep in the insulating phase, where the Hubbard model maps on to 
the nearest-neighbour Heisenberg model \cite{fazekas,cleveland}, 
the spin dynamics is well documented  
\cite{blume1,blume2,takahashi,gouvea,volkel,peczak,costa,moessner,
taillefumier,sherman}. 
However, on decreasing the electron-electron 
interaction two effects occur simultaneously:
(i)~the coupling among magnetic moments become progressively 
longer ranged, multi-spin, and begin to 
involve ring-exchange terms 
\cite{capriotti,yang}, and (ii)~the moments begin to ``soften'', {\it i.e},
become more prone to amplitude fluctuations.
The first effect affects mainly 
the low temperature spin-wave dispersion.
The second effect is important for the thermal physics since 
amplitude fluctuations generate additional scattering of the
magnetic modes. In a Mott insulator where the charge gap is
$\sim 10^3-10^4$K, say, and the effective exchange
is $\sim 10-100$K these effects would be visible
over an accessible temperature window.

Experiments on dynamics in Mott insulating materials have mostly 
concentrated on quasi-2d systems like layered cuprates 
\cite{aeppli,coldea,stock}, organics \cite{kurosaki,powell}, 
ruthenates \cite{friedt,steffens} and fully 3d systems like
iridates \cite{shapiro,choi,tomiyasu,bahr}, 
doped V$_{2}$O$_{3}$ \cite{bao}, NiO \cite{kim} and 
Sr$_{2}$Mn$_{3}$As$_{2}$O$_{2}$ \cite{chen}. 
In the Mott phase, 
inelastic neutron scattering (INS) studies on La$_{2}$CuO$_{4}$ 
find substantial non-Heisenberg features in the dispersion. 
In the iridate experiments,  
one infers no long-range magnetic order \cite{shapiro} in some cases, 
while in certain others \cite{choi,tomiyasu,bahr}, sharp 
low-energy spin waves originating from complex magnetic order are observed. 
Near the transition, NMR measurements on organics have found 
strong suppression of spin fluctuations in the Mott phase.
By contrast, in Ca$_{2-x}$Sr$_{x}$RuO$_{4}$, 
one finds enhanced magnetic fluctuations in the metallic phase
at an incommensurate wave vector.

A reliable estimate of the magnetic excitation spectrum requires
several ingredients: (i)~one should be able to handle correlation
effects away from the Heisenberg limit, in particular as the
system heads towards an insulator-metal transition, (ii)~the
dimensionality and lattice geometry needs to be respected since
the magnetic order and excitations depend crucially on them,
(iii)~the approach should access thermal effects well beyond the
reach of linear spin wave theory, and (iv)~the theory should
yield real time (or real frequency) information - a rarity in 
finite temperature schemes. Most approaches unfortunately fall short.

The tools currently available to study equilibrium dynamics of the
Hubbard model include exact methods like quantum Monte Carlo 
\cite{QMC} (QMC), 
approximate numerical strategies like DMFT \cite{DMFT,kotliar2}
and its cluster extensions \cite{park,hochkeppel},
slave boson techniques \cite{fresard}, 
and semi-analytic schemes like the 
random phase approximation (RPA) or $1/S$ expansion.
More recently, dual fermion method \cite{leblanc1,li}, 
and semiclassical Langevin dynamics \cite{chern} 
have entered the scenario. 
A recent review covers most of the existing
approaches used for the 2d model \cite{leblanc2}.
Both QMC and DMFT are usually formulated in 
imaginary time, and hence the results need analytic 
continuation. QMC also has 
size limitations and often the "fermion sign problem". 
DMFT neglects spatial correlations at the 
single site level, but its cluster variants alleviate
the problem in some cases.
The RPA approach yields reasonable low-temperature spin wave dispersion 
($\Omega_{\bf q}$) on magnetically ordered states 
\cite{singh1,singh2,singh3} and also captures 
high-energy features like the two-particle continuum.
However, as order is suppressed with increasing temperature,
and large angular fluctuations become relevant, 
the RPA results lose validity.

An approximate strategy well suited for this problem is the
Langevin dynamics approach, first introduced by
Chern et. al. \cite{chern}. This method does make some
simplifying assumptions but meets all the requirements 
that we had defined earlier. Using this we address the following
questions: (i)~how are the crossover scales in magnetic 
dynamics affected as we move to lower values of $U/t$ from
the Heisenberg limit? 
(ii)~what is the role of amplitude fluctuations on the 
lineshape of excitations, and (iii)~what is the effect
of increasing geometric frustration on the spectrum?

There are two "reference calculations" that define what
is known in this problem.
(a)~For $U/t \gg  1$ and for nearest neighbour hopping 
the Hubbard model maps on to the nearest neighbour
Heisenberg model. The ground state on the square lattice 
is N\'{e}el ordered with ${\bf Q}=(\pi,\pi)$, while on
the triangular lattice ${\bf Q}=(2\pi/3,2\pi/3)$.
The relevant exchange scale is $J = 4t^2/U$,
for moments with $S=1/2$. The thermal dynamics of the 
Heisenberg model is well known
\cite{blume1,blume2,takahashi,gouvea,volkel,peczak,costa,moessner,
taillefumier,sherman},
albeit numerically. 
(b)~On the mean field ground state, RPA provides a 
reasonable excitation spectrum at any $U/t$. 

We have confirmed that the Langevin scheme captures the 
dynamics of the 2d \textit{classical} Heisenberg model on 
both lattices, at all temperature. 
Since we approximate the magnetic moments in our
scheme to be classical, we do not 
get the true quantum limit at large $U/t$.
Our theory also captures the low energy part 
of the RPA spectrum at all $U/t$ and low temperature, but
not the spin waves at zero temperature.

To set the stage for a summary of our results, 
the magnetic dynamics can be classified into three
regimes. (A).~At low temperature we observe
weakly damped dispersive
modes, with damping $\Gamma_{\bf q} \ll W_{mag}$,
where $W_{mag}$ is the magnetic bandwidth at $T=0$.
This scale is plotted in Fig.2(b).
In this regime in general $\Gamma_{\bf q} \ll 
\Omega_{\bf q}$.
(B).~Beyond a broad crossover, characterised by a scale 
$T^{cr}_1$, there is a regime of 
strongly damped but still dispersive modes,
with $\Gamma_{\bf q} \sim {\cal O}(W_{mag})$. Finally,
(C).~at even higher temperature, beyond a scale
$T_2^{cr}$, we observe spin
diffusion, with $\Omega_{\bf q} \rightarrow 0$ for all 
${\bf q}$ and $\Gamma_{\bf q} \sim {\cal O}(W_{mag})$.

An important scale in analyzing the results is 
the effective exchange $J_{eff}(U)$, 
inferred from the spin wave velocity  
computed from the  low energy spectrum. 
The spin wave velocity is the slope of 
the linear magnon branch near the Goldstone
points, $(\pi,\pi)$ and $(2\pi/3,2\pi/3)$ for the square
and triangular lattice, respectively.
$J_{eff}$ is plotted in Fig.2(a).
In terms of this scale,  our main results are the
following - first on the square lattice, and then on
the triangular lattice.\\
 
I.~For the square lattice:
\begin{itemize}
\item 
{\it Broad regimes:}
While the absolute values of the crossover temperatures increase
with decreasing $U/t$ (since the effective exchange $J_{eff}$ 
increases), the ratios  $T^{cr}_1/J_{eff}$ and $T^{cr}_2/J_{eff}$
noticeably {\it decrease} with decreasing $U/t$. This
indicator of non-Heisenberg behaviour suggests a relatively
quicker onset of mode coupling, and then diffusive behaviour,
at smaller $U/t$.
\item 
{\it Dispersion and damping:} 
The dispersion $\Omega_{\bf q}(T)$ narrows monotonically 
with increasing $T/J_{eff}$,
The onset of rapid narrowing is at $T/J_{eff} \sim 1$
when $U/t \gg 1$ and reduces to $T/J_{eff} \sim 0.5$ 
for $U \sim 6t$. 
We find that at low $T$ the thermal damping is
$\Gamma_{\bf q}(T) - \Gamma_{\bf q}(0) \propto T^{2}$ when $U/t \gg 1$ 
and $\Gamma_{\bf q}(T) - \Gamma_{\bf q}(0) \propto T$
for intermediate to small $U/t$.
The damping changes to $\sim T^{1.5}$ at higher $T$, 
and finally saturates for $T \gtrsim 2 J_{eff}$.
\item
{\it Amplitude fluctuation:} 
The amplitude fluctuations play a crucial role in broadening
the lineshape at weak coupling, where the fluctuation width
varies as $\sim \sqrt{T/U}$. While we do not capture the
real "amplitude mode" at $\omega \sim U$ we can access 
amplitude fluctuation effects
 on the spin waves at $\omega \sim J_{eff}$.
\end{itemize}


II.~On the triangular lattice:
\begin{itemize}
\item 
{\it Broad regimes:}
The triangular lattice has a finite critical interaction
for the MIT, with $U_{c} \sim 5t$. We restrict ourselves to 
$U/t$ where the 120\degree ordered state is the ground state.
The typical lineshape is two-peak in this case.
The thermal crossover scales are inferred from the behaviour 
of the peak which broadens \textit{quicker} with respect to $T$.
The behaviour of $T^{cr}_{1}$ and $T^{cr}_{2}$ with respect to $U$ 
is similar to what is observed in the square lattice, with the
distinction that their maxima occur at larger $U$ and the scales
are $\sim 0.5$ their square lattice values. 
\item
{\it Dispersion and damping:} 
Due to emergence of longer range couplings, the low
$T$ dispersion along $\Gamma-K$ shows a larger curvature 
at lower $U/t$. 
The damping is also much larger, compared to the
square lattice, at similar values of $T/J_{eff}$.
At $U/t \sim 10$, where $J_{eff}/t \sim 0.04$ 
the crossover scales are just $T^{cr}_1/J_{eff} \sim 0.4$ 
and $T^{cr}_2/J_{eff} \sim 0.8$. 
\item
{\it Fluctuation:} 
The role of amplitude fluctuations in damping the modes
is enhanced at a given $U$ and the \textit{same} $T/J_{eff}$, 
due to the finite $U_c$ and mild frustration. 
\end{itemize}

\section{Model and method}

We work with the single band, repulsive Hubbard model on square 
and triangular lattice geometries. 
The Hamiltonian reads- 
$$
H=-\sum_{<ij>\sigma}t_{ij}(c^{\dagger}_{i\sigma}c_{j\sigma} + h.c.)
+U\sum_{i}n_{i\uparrow}n_{i\downarrow} - \mu\sum_{i\sigma}n_{i\sigma}
$$
The hopping amplitude $t_{ij}$ is chosen to be non-zero only amongst 
nearest neighbours for the square case and has a uniform value $t=1.0$.
On adding the next-nearest neighbour coupling $t^{\prime}=1.0$ on top
of this \textit{along one diagonal in each square motif}, 
we get the triangular lattice. 

First, the interaction term is decoupled using a Hubbard-Stratonovich
transformation to obtain a spin-fermion model-
$$
H_{SF}=-\sum_{<ij>\sigma}t_{ij}(c^{\dagger}_{i\sigma}c_{j\sigma} + h.c.)
-U\sum_{i}{\bf m}_{i}.\sigma_{i} + U\sum_{i}|{\bf m}_{i}|^{2}
$$

We solve for the finite $T$ dynamics ${\bf m}_{i}$ using 
the following equation of motion \cite{chern}: 
\begin{equation}
\frac{d{\bf m}_{i}}{dt}=-{\bf m}_{i}\times
\frac{\partial \langle H_{SF} \rangle}{\partial{\bf m}_{i}} - 
\gamma\frac{\partial \langle H_{SF} \rangle}{\partial{\bf m}_{i}} 
+ \vec{\xi}_{i} 
\end{equation}
The noise is specified through-
\begin{eqnarray}
\langle \xi^{\mu}_{i}(t) \rangle &=& 0 \\ 
\langle \xi^{\mu}_{i}(t) \xi^{\nu}_{j}(t^{\prime}) \rangle
&=& 2\gamma k_{B}T\delta_{ij}\delta^{\mu\nu}\delta(t-t^{\prime})
\end{eqnarray}

Here $\gamma$ is a dissipation parameter. Within our 
scheme, it's value can't be determined from first principles.
To calculate it, one has to evaluate the imaginary part
of the Keldysh polarizability ($Im \Pi^{K}({\bf q},\omega)$) 
at low frequencies.
We comment that in the deep Mott phase, 
this contribution is vanishingly small 
due to the gapped single electron spectrum. 
However, on moving to lower $U$ values, 
this quantity picks up weight at finite temperature.
The evolution equation has a phenomenological justification 
as well as a a microscopic basis. We touch on these briefly. 

I.~First the 
phenomenological motivation \cite{chern,ma}. One starts from the 
Heisenberg limit with moments of fixed magnitude. 
The torque term comes from evaluating the Poisson brackets 
in the semiclassical equation of motion.
The damping is taken to be proportional to the angular momentum, 
following an analogy with the particle Langevin equation. 
Lastly, the noise is chosen so as to satisfy the 
fluctuation-dissipation relation, ensuring that one captures 
the Boltzmann distribution in the long-time limit \cite{ma,kirilyuk}.
The additive form of the damping and noise allows for 
longitudinal relaxation of the magnetic moments.
This approach does not determine the value of
the dissipation coefficient $\gamma$. 
In our treatment, we fix the $\gamma$ value by 
comparing our static results with a Monte Carlo (MC) 
method and ensuring a decent match. 
The MC strategy is briefly discussed in Appendix B.

Alternately, II.~One starts from a model of a spin coupled 
linearly to a bosonic bath and integrates out the bath degrees 
of freedom to obtain an effective equation of motion for the 
spin, it has been shown \cite{rebei1} that under certain 
conditions, a Landau-Lifshitz-Gilbert-Bloch (LLGB) equation 
\cite{brown} emerges. The derivation may also be done in presence 
of conduction electrons \cite{rebei2} or both phonons and electrons 
\cite{mera}. This equation explicitly conserves spin magnitudes. 
Our equation also reduces to the LLGB form upon 
constraining the spins on the unit sphere \cite{ma}. 

Finally, III.~One may also try to derive the present equation 
starting from the Keldysh action of the Hubbard model. First, one  
introduces auxiliary fields to decouple the interaction term and 
subsequently assumes them to be slow compared to the electrons.
This allows one to write an effective equation of motion for them.
Upon doing certain simplifications, this equation can be mapped on to Eq.1.  
We briefly allude to this in subsection E of our Discussion section.

The typical timescale for magnon oscillations is 
$\tau_{mag}\sim 1/J_{eff}$.
We set an "equilibration time" $\tau_{eq}
= 100 \tau_{mag}$ before saving data for the power spectrum.
The outer timescale, $\tau_{max} \sim 10 \tau_{eq}$.  
The "measurement time" $\tau_{meas}
= \tau_{max} - \tau_{eq}$, and the number of sites is $N$. 
Some details regarding the numerical solution of Eq.1 are given 
in Appendix A.

We calculate the 
following from the time series ${\bf m}({\bf r}_{i},t)$:
\begin{enumerate}
\item
Dynamical structure factor,
$D({\bf q}, \omega) = \vert {\bf m}({\bf q}, \omega) \vert^2$  
where
\begin{equation}
{\bf m}({\bf q}, \omega) = 
\sum_{i} \int_{\tau_{eq}}^{\tau_{max}} 
dt e^{i {\bf q}.{\bf r}_i}
e^{-i \omega t}
{\bf m}({\bf r}_i, t)
\end{equation}
\item
The instantaneous structure factor 
\begin{equation}
S({\bf q},t) =
{1 \over N^2} 
\sum_{ij} e^{i {\bf q}. ({\bf r}_i - {\bf r}_j) }
{\bf m}({\bf r}_i, t).{\bf m}({\bf r}_j, t) 
\end{equation}
The corresponding time averaged structure factor is
\begin{equation}
{\bar S}({\bf q}) = {1 \over \tau_{meas}} 
\int_{\tau_{eq}}^{{\tau}_{max}} dt S({\bf q},t) 
\end{equation}
\item
The distribution of moment magnitudes:
\begin{equation}
P(|{\bf m}|)= \frac{1}{N \tau_{meas}} \sum_{i} 
\int_{\tau_{eq}}^{\tau_{max}} dt \delta(\vert {\bf m} \vert - \vert {\bf m}_i(t) \vert)
\end{equation}
\item
Dispersion $\Omega_{\bf q}$ and damping $\Gamma_{\bf q}$: 
\begin{eqnarray}
\Omega_{\bf q} &=& \int_{0}^{\omega_{max}} d\omega 
\omega  D({\bf q}, \omega) \cr 
\Gamma_{\bf q}^{2} & = &
\int_{0}^{\omega_{max}} d\omega 
(\omega - \Omega_{{\bf q}})^2 D({\bf q}, \omega) 
\nonumber
\end{eqnarray}
\end{enumerate}


\section{Benchmarks and overall features}

\begin{figure}[t]
\centerline{
\includegraphics[height=4.25cm,width=4.3cm]{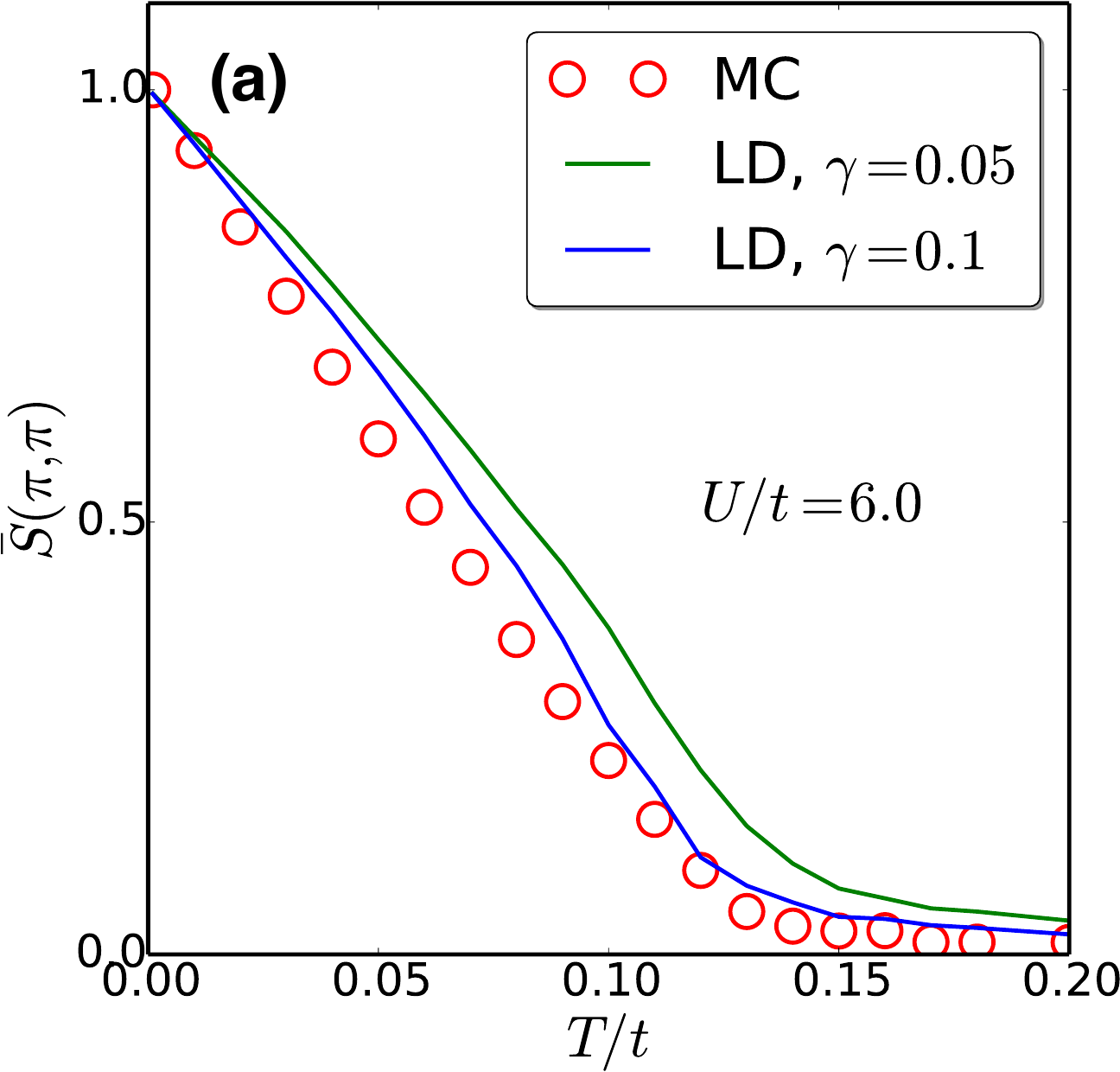}
\includegraphics[height=4.25cm,width=4.3cm]{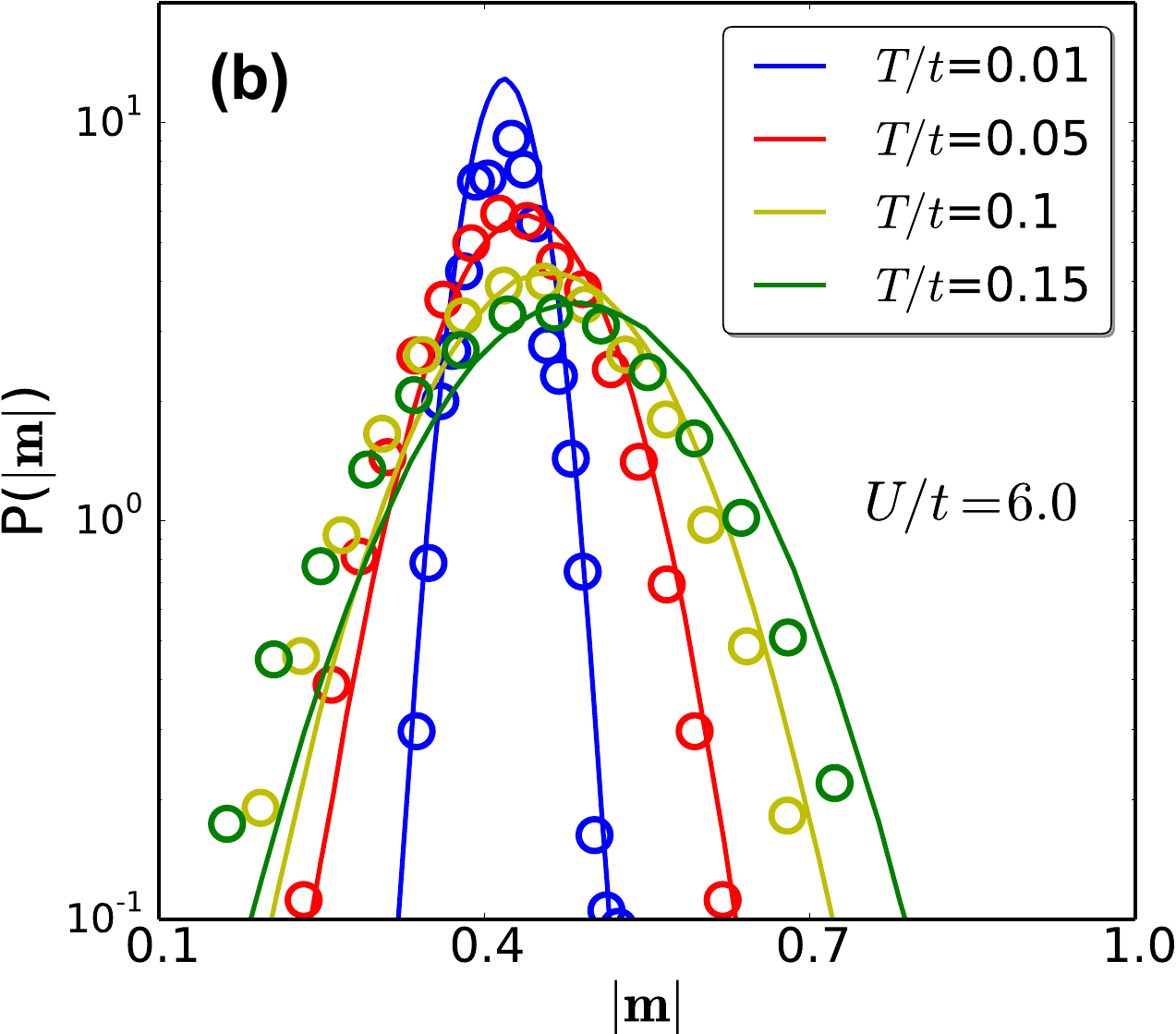}
}
\caption{$\bar{S}(\pi,\pi)$ (a) and $P(|{\bf m}|)$ (b)
for the square lattice Hubbard model at $U/t=6.0$.
Solid lines denote answers obtained using the present LD
method and open circles indicate MC data. We observe a 
reasonable agreement between the two methods.
}
\end{figure}

\subsection{Fixing the Langevin parameters}

We do a bechmarking of the Langevin scheme using the square lattice
as a test case. Three coupling regimes are explored- weak ($U/t=3.0$),
intermediate ($U/t=6.0$) and strong ($U/t=10.0$).
The statics is quantified through two quantities- the
structure factor $S(\pi,\pi)$ and the moment magnitude distribution
$P(|{\bf m}|)$. The former shows the correlation temperatures 
($T_{corr}$), below which the correlation length approaches the system size.
The latter details the longitudinal fluctuations of local moments.
The alternate technique used to compute these quantities is a
Monte Carlo calculation done assuming the auxiliary ${\bf m}_{i}$
field to be classical and using the sum of electronic free energy and
the stiffness cost (last term in $H_{SF}$) as the sampling weight
\cite{swain} (see Appendix B for more details).

The method of fixing $\gamma$ was the following. We started with
a low value (motivated by its vanishing magnitude at strong coupling, 
and the fact that we should get undamped spin waves at low enough $T$)
at a fixed coupling and run length. Next, we increased the $\gamma$ 
at that coupling in steps till the match with MC results on temperature
dependence became reasonable, while ensuring that the low $T$ spin 
waves remain sharp enough. Results for a typical coupling are quoted above.

Fig.1(a) shows a comparison
of $\bar{S}(\pi,\pi)$ at $U/t=6$, with a reasonable match.
The dissipative coefficients are $\gamma=0.05$ and $\gamma=0.1$.
In Fig.1(b), the $P(|{\bf m}|)$ distributions also show
reasonable agreement (for $\gamma=0.05$).
We've used $\gamma=0.025$ to generate the bulk of 
our final dynamics results, which roughly corresponds to 
a relaxation timescale $\tau_{rel} \sim 40\tau_{mag}$.
We will later quantify the increasing relevance of magnitude 
fluctuations on decreasing coupling, which is an important piece 
of the non-Heisenberg physics.

\subsection{Magnetic scales for varying $U/t$}

At low temperature, our dynamical equation (Eq.1) gives rise to 
weakly damped, dispersive spin wave excitations. From the
obtained spectrum, we extract two scales- (i)~the spin-wave 
stiffness, $J_{eff}$, and (ii)~the magnon bandwidth, $W_{mag}$. 
The first is computed from the spin wave velocity of the linear
branch near the respective Goldstone modes on the square
and triangular lattice. The latter requires knowledge of 
the full magnon band structure.  
We plot these quantities 
for both the square and triangular lattice in Fig.2.

In Fig.2(a), we find a monotonic decrease of $J_{eff}$ with
$U/t$ in the square lattice case, 
with a $1/U$ asymptote at strong coupling. The value at $U/t=20.0$
matches the expected $J_{eff}=4t^{2}/U$, indicating
that one has reached the Heisenberg limit. On the triangular lattice, 
the stiffness goes to zero for $U/t=6.0$, indicating a breakdown of
the $120\degree$ ordered state. The scale then rises and finally
falls as $\sim 1/U$ at strong coupling.
In Appendix C, we compare the extracted spin wave velocities 
with those obtained from RPA \cite{singh1}.

The magnon bandwidths Fig.2(b) feature a \textit{non-monotonicity}
in the square case, with a maximum around $U/t=6.0$. $W_{mag}$ 
increases on lowering $U$ on the triangle, rising to $0.6t$
before the ordered state breaks down. 

\begin{figure}[t]
\centerline{
\includegraphics[height=4.5cm,width=4.5cm]{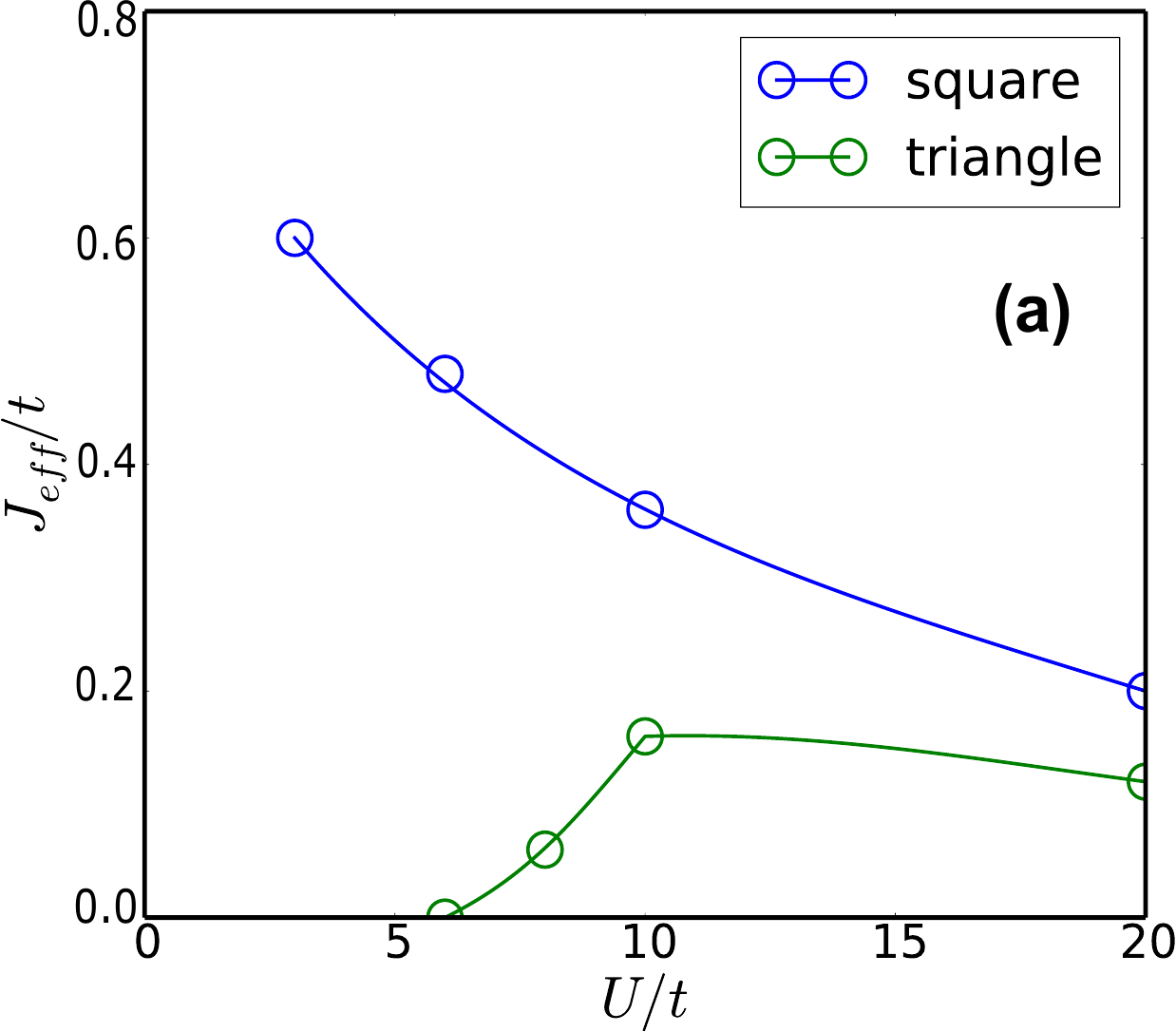}
\includegraphics[height=4.5cm,width=4.5cm]{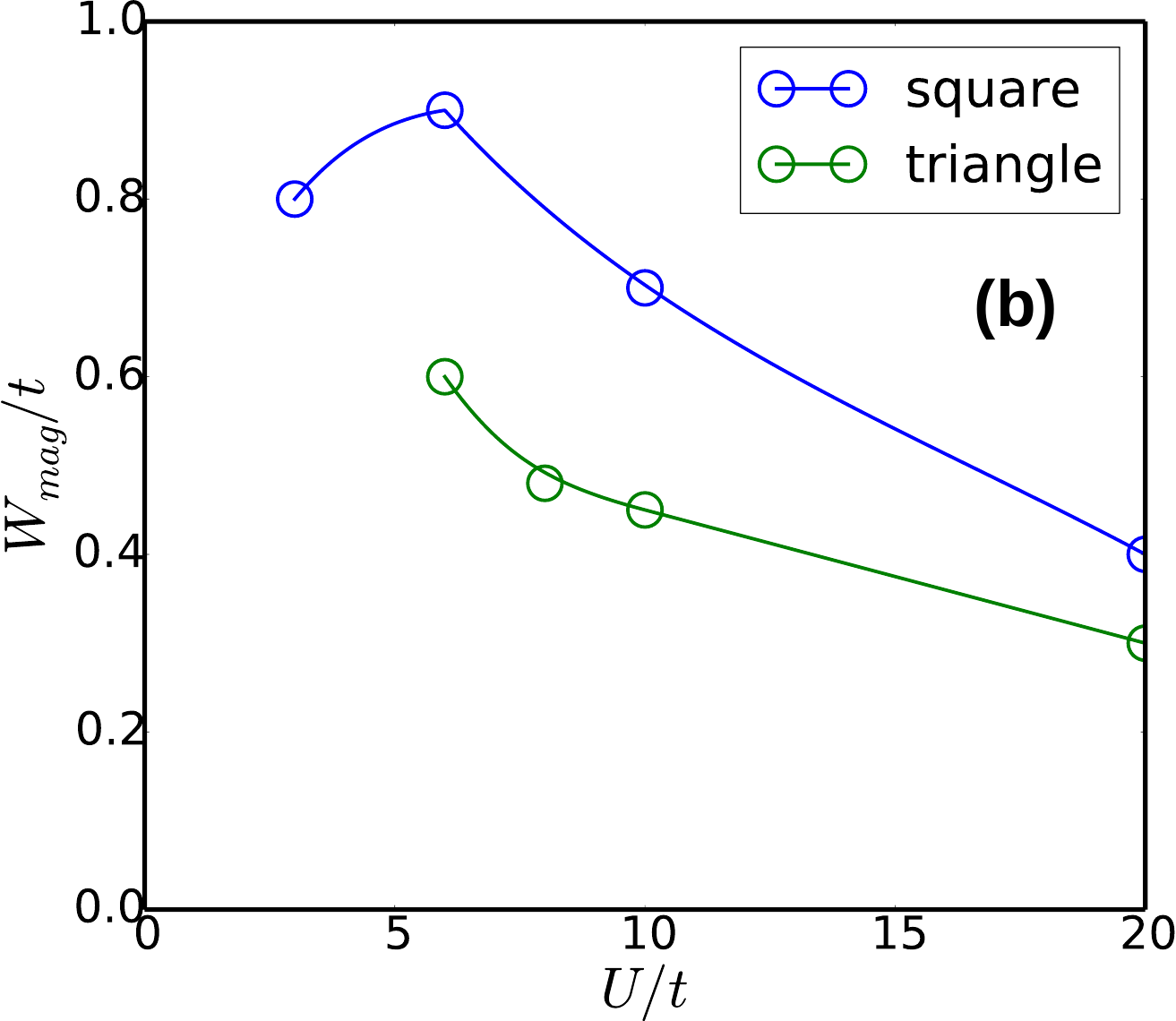}
}
\caption{(a): The dimensionless effective exchange ($J_{eff}/t$), 
calculated from the spin wave velocity, for the square and triangular 
lattice Hubbard models at various $U/t$ values. 
We see a monotonic behaviour for the square lattice and a 
non-monotonic behaviour for the triangular lattice case.
Moreover, the scale vanishes around $U/t=6.0$ for the latter,
signalling a breakdown of $120\degree$ order.
(b):The spin wave bandwidth ($W_{mag}$), calculated from the
full magnon dispersion, for the square and triangular cases.
Here, we see a non-monotonicity in the square lattice, and
a gradual decrease in the triangular lattice.}
\end{figure}

\subsection{Comparison with Heisenberg as $U/t \rightarrow \infty$} 

We compare the Hubbard results at $U/t =20$ on the square
lattice with the Heisenberg model with $J=1$.
The former effectively reduces to the latter with
$J_{eff}=4t^{2}/U$ and $|{\bf m}_{i}|=1/2$.
First, in 3(a), the low $T$ dispersions are compared, 
with both being scaled by $W_{mag}$, the spin wave
bandwidth. There's a nearly perfect agreement.
 
The Heisenberg model features three broad thermal regimes.
These are- (i)~weakly damped ($T \ll J$), where we obtain 
dispersive excitations with low damping, 
(ii)~strongly damped ($T \sim {\cal O}(J)$), where there's significant
mode coupling among spin waves, but dispersion is still discernable, 
and (iii)~diffusive ($T \gg J$), where mode frequencies collapse to zero
and the dampings are comparable to $W_{mag}$.
In these regimes, we compare the lineshapes of the Heisenberg model 
at ${\bf q}=(\pi/2,\pi/2)$ with those of the large $U$ Hubbard model
in Fig.3(b). 
In regime (i), a sharp lineshape centered around $\Omega_{\bf q}=4J$
is seen, which picks up significant damping in regime (ii), before
becoming diffusive in (iii).
A quantitative agreement is seen between the Hubbard and
Heisenberg results.
The frequencies are scaled by $J_{eff}$ in the Hubbard
case, and $J$ in the Heisenberg one. 
\begin{figure}[t]
\centerline{
\includegraphics[height=4cm,width=4.5cm]{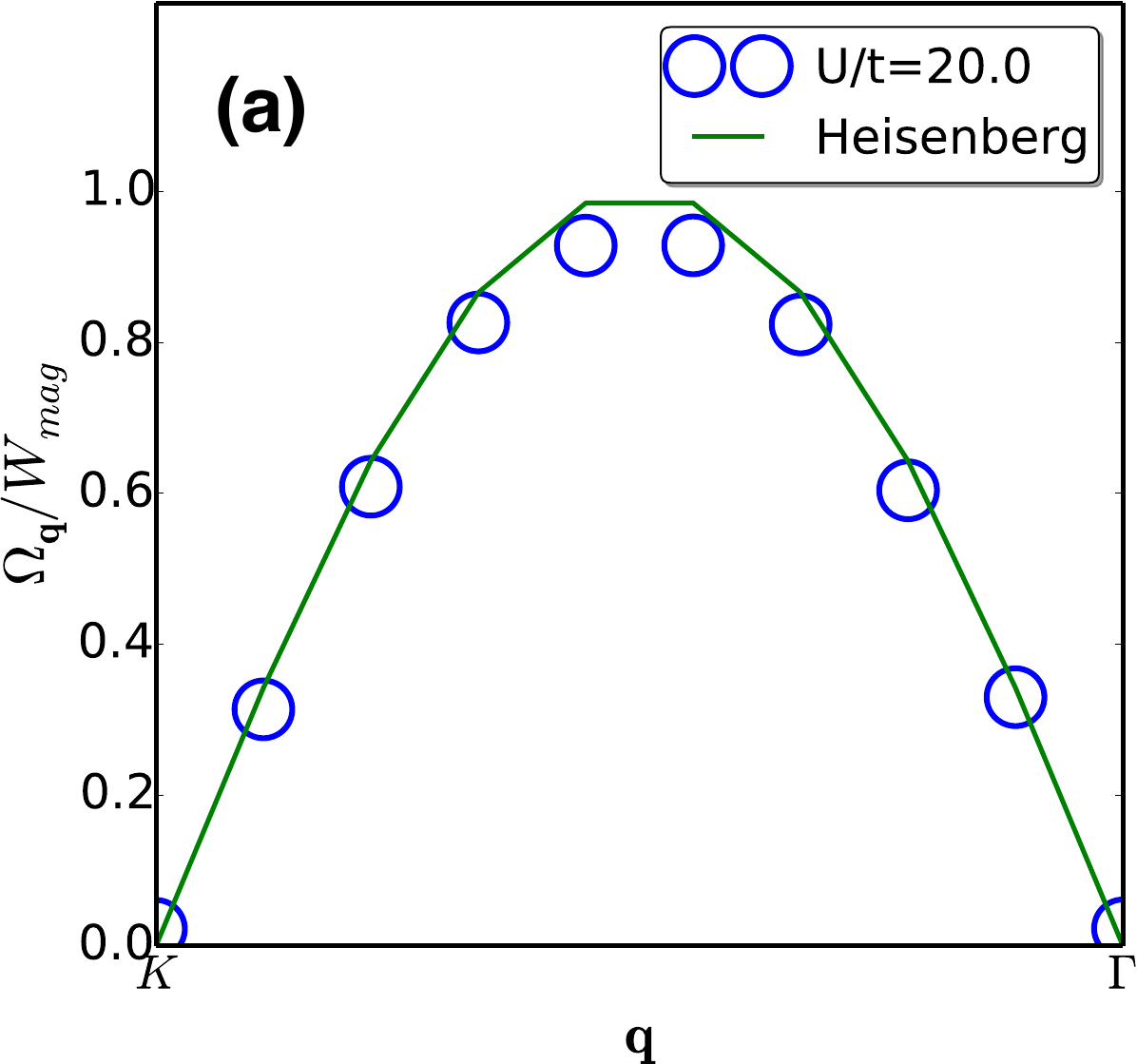}
}
\hspace{0.2cm}
\centerline{
\includegraphics[height=4.1cm,width=4.3cm]{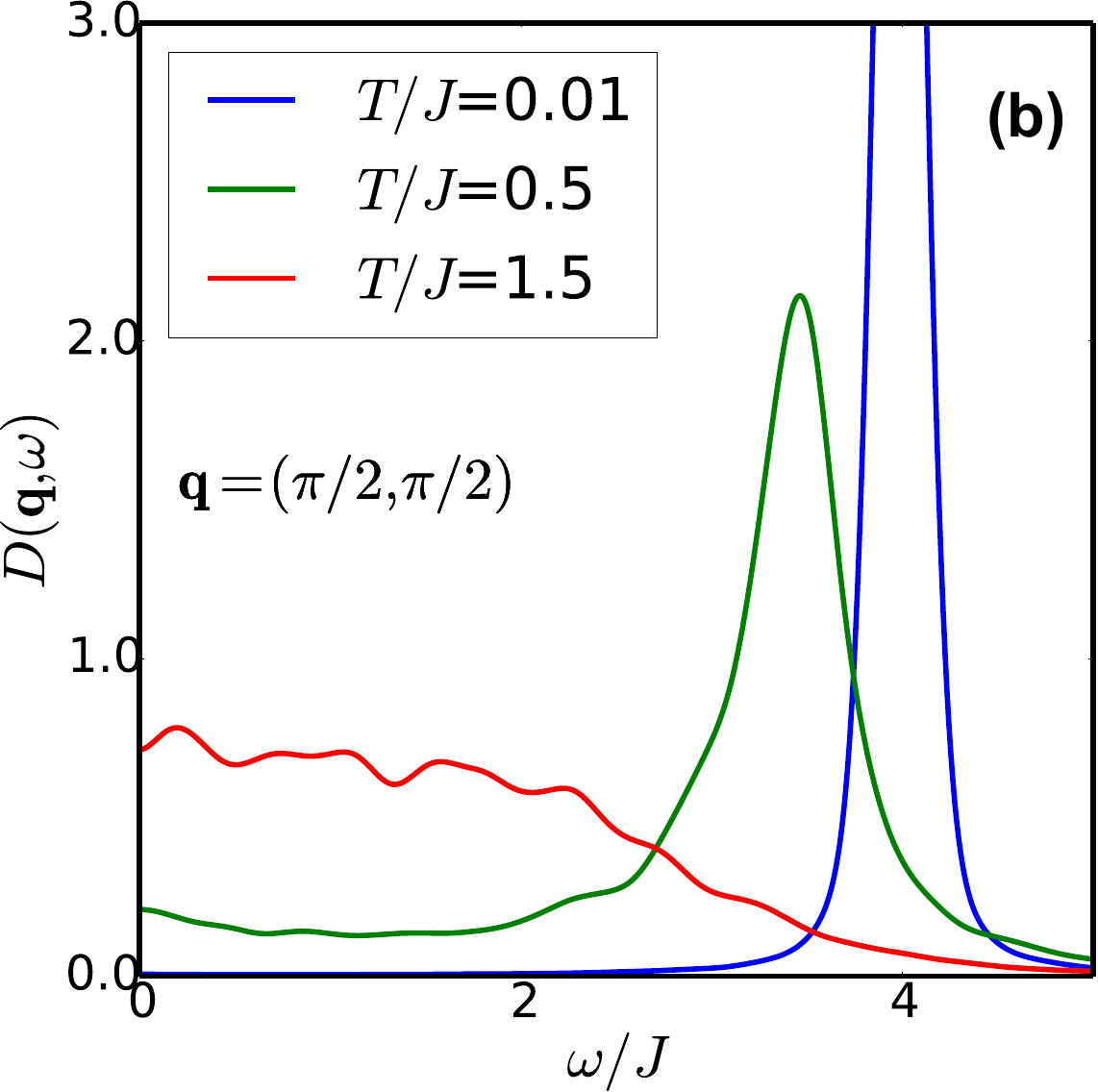}
\includegraphics[height=4.1cm,width=4.3cm]{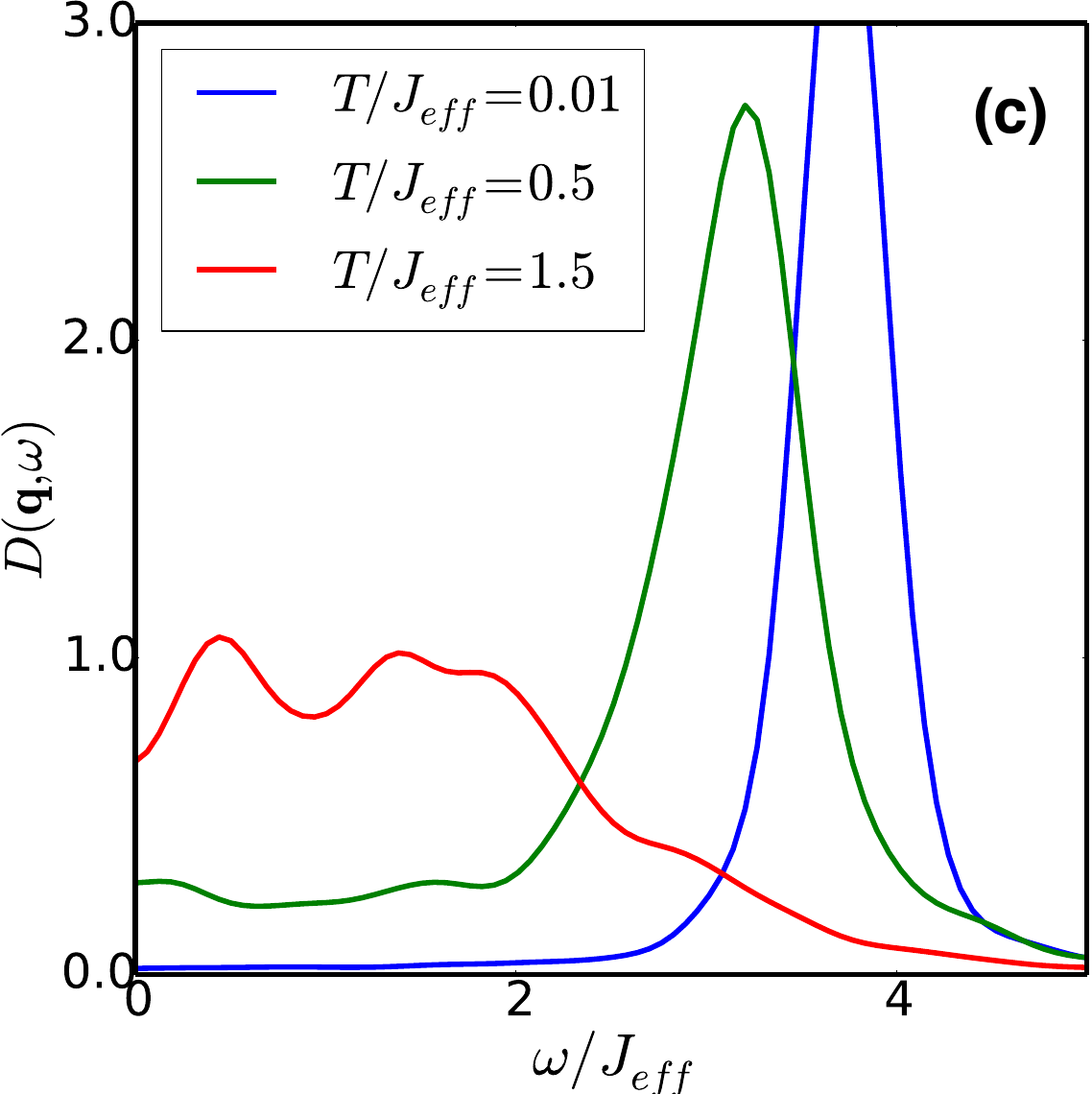}
}
\caption{(a): Comparison of dispersions $\Omega_{\bf q}$ along
the $K-\Gamma$ direction of the Brillouin Zone (BZ) between
the square lattice Hubbard model at $U/t=20.0$ and the Heisenberg
model with $J=1$. One gets a near perfect agreement on scaling
the former by $J_{eff}=4t^{2}/U$. 
(b),(c): Lineshapes at three characteristic temperatures
$T/J=0.01,0.5,1.5$ for the Heisenberg model (in (b)) and
the $U/t=20.0$ Hubbard model (in (c)). There's again 
a marked agreement.
}
\end{figure}

\begin{figure}[t]
\centerline{
\includegraphics[height=4.5cm,width=4.3cm]{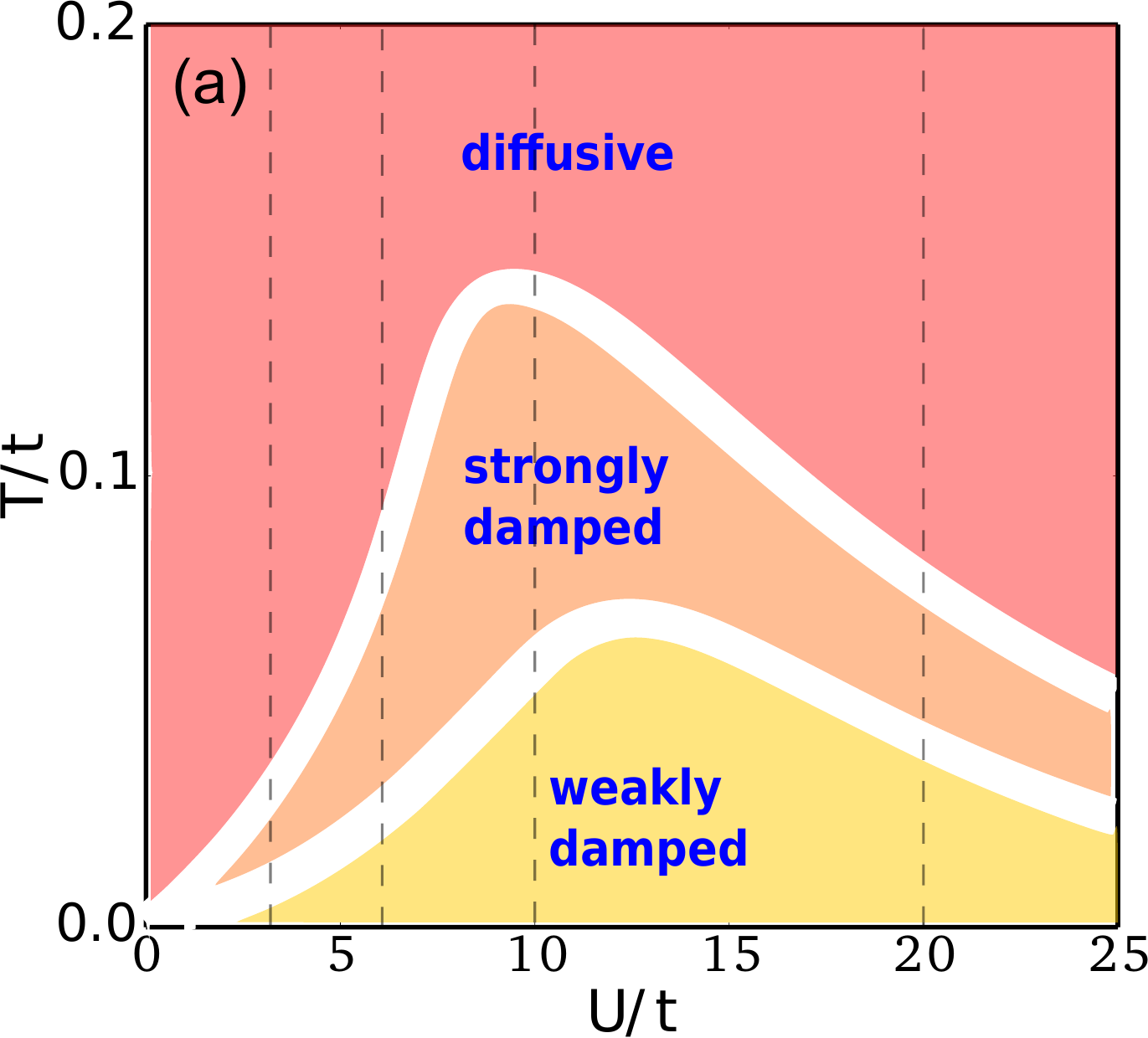}
\includegraphics[height=4.5cm,width=4.3cm]{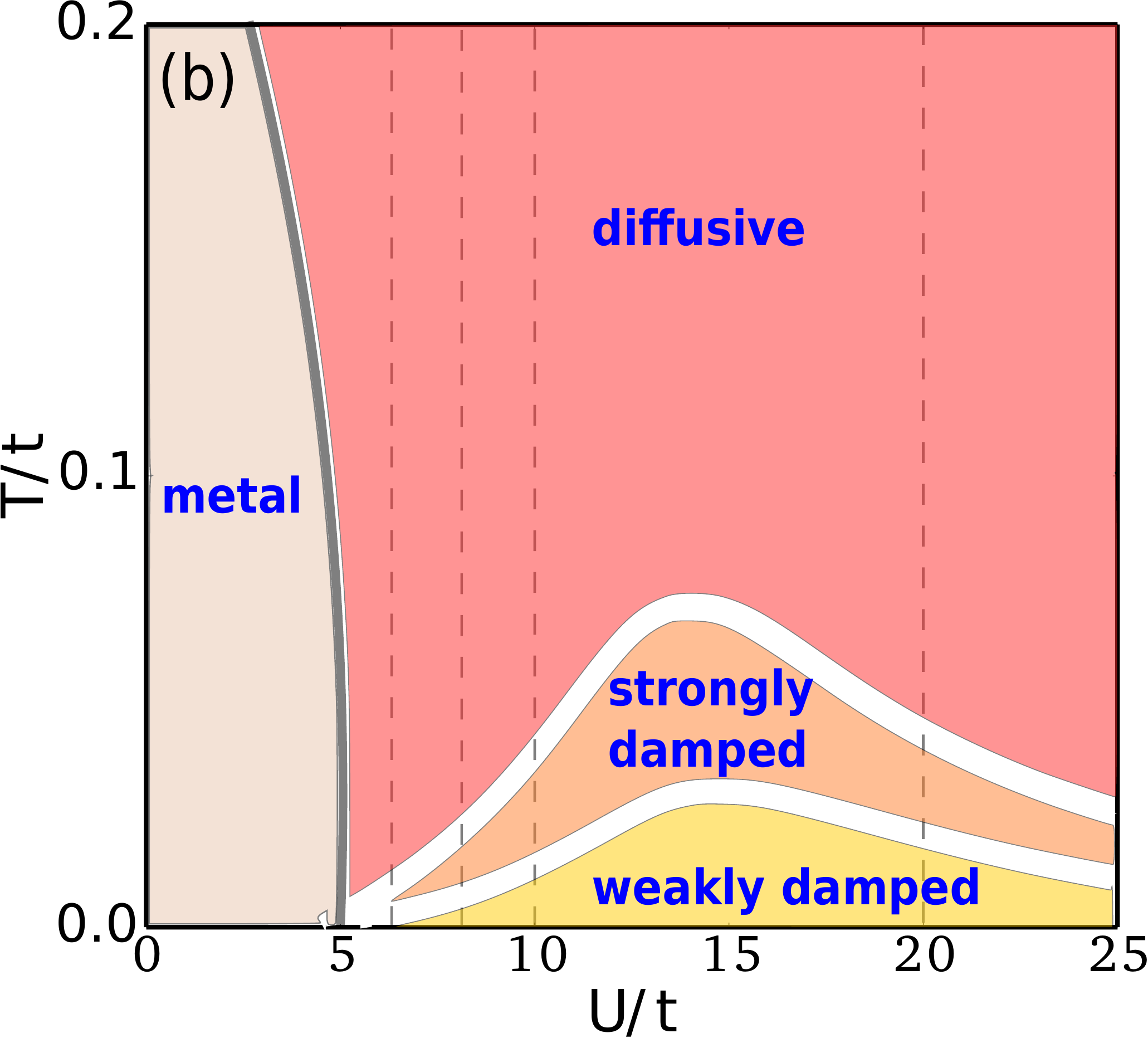}
}
\centerline{
\includegraphics[height=4.5cm,width=4.3cm]{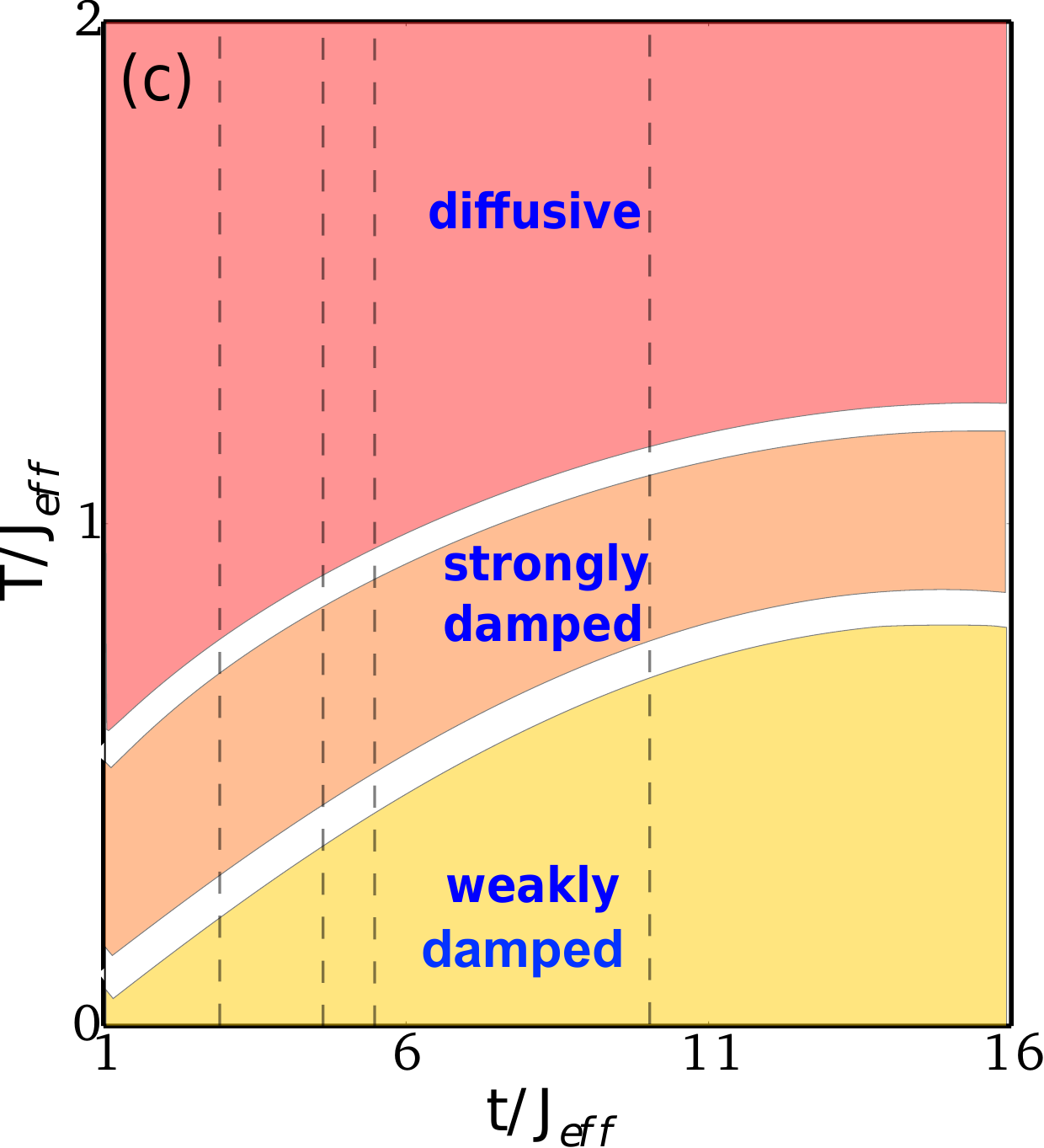}
\includegraphics[height=4.5cm,width=4.3cm]{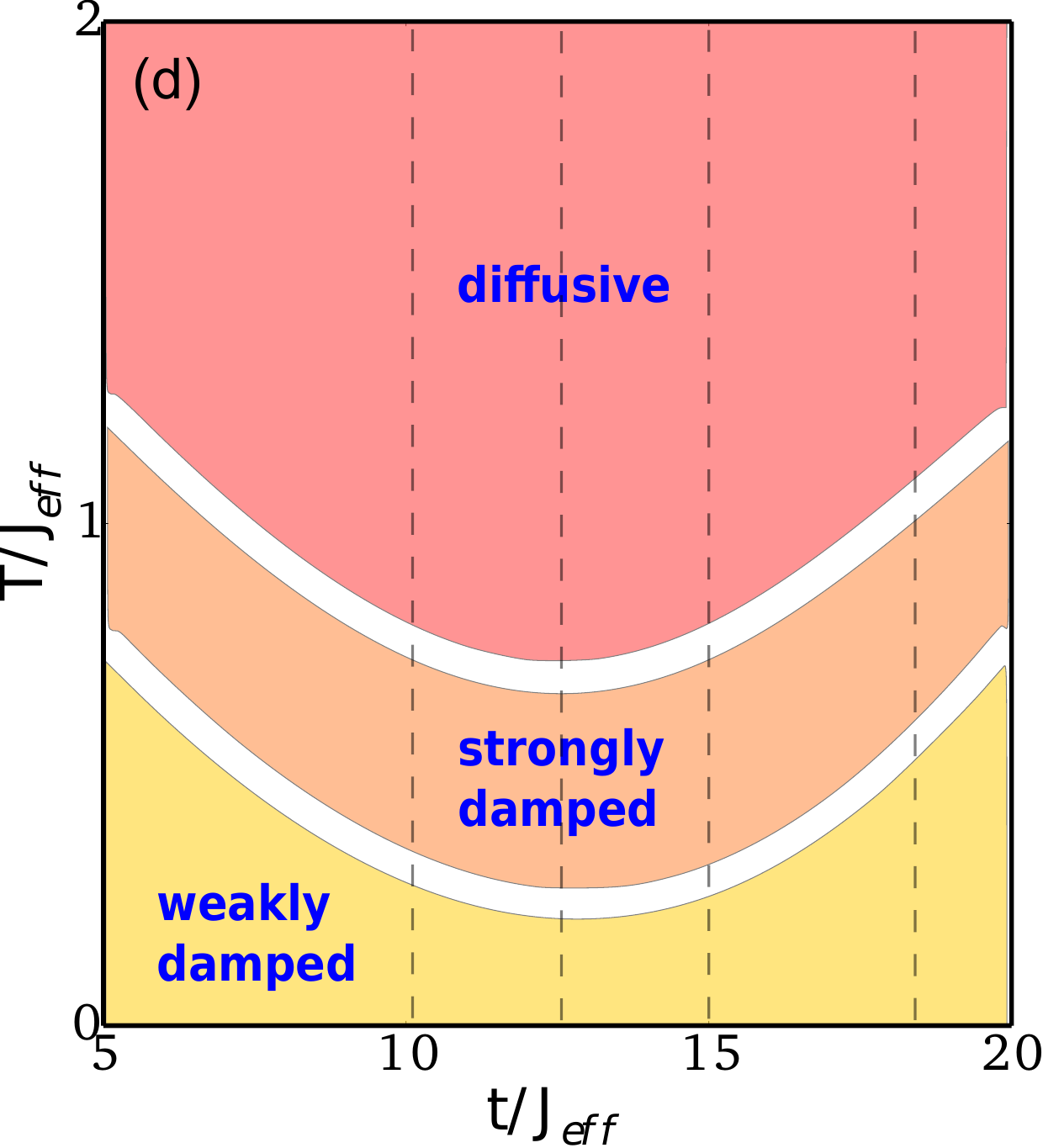}
}
\caption{Magnon phase diagrams for square ((a) and (c)) and
triangular ((b) and (d)) lattice Hubbard models at half-filling.
The top row features the $U/t-T/t$ phase diagrams, while the
bottom one exhibits the $T/J_{eff}-t/J_{eff}$ plots.  We broadly
observe three "dynamical regimes"- (i) weakly damped (where
$\Gamma_{\bf q} \lesssim 0.2 W_{mag}$), (ii) "strongly damped"
(where $\Gamma_{\bf q} \sim {\cal O} (W_{mag})$) and (iii) "diffusive"
(where $\Gamma_{\bf q} \sim {\cal O} (W_{mag})$ and $\Omega_{\bf q}
\rightarrow 0$).  The metallic region in (b) is not tackled by our
approach.  Vertical sections indicate couplings used in actual
simulations.
}
\end{figure}

\subsection{General features of dynamics in the Mott phase}

We first comment on the broad dynamical regimes obtained on the
square and triangular lattice problems. This is characterized by
the the number of peaks, their location, and width.

As mentioned earlier, we find three broad dynamical regimes on 
analyzing the data-
(i)~weakly damped, where
the linewidth for a generic momentum $\Gamma_{\bf q} \ll W_{mag}$, 
(ii)~strongly damped, where
$\Gamma_{\bf q} \sim {\cal O} (W_{mag})$ and 
(iii)~diffusive, where
$\Gamma_{\bf q} \sim {\cal O} (W_{mag})$ and $\Omega_{\bf q} \rightarrow 0$. 

On the square lattice (Figs.4(a) and 4(c)), the
low $T$ lineshapes are unimodal. 
There is a gradual crossover to regimes (ii) and (iii) 
at $T^{cr}_{1}(U)$ and $T^{cr}_{2}(U)$ respectively.
The window of regime (ii) is maximum around $U/t=6.0$. 
The crossover lines behave $\sim 1/U$ asymptotically, 
but have a maximum around $U/t=10.0$. Below this coupling,
the amplitude fluctuation effect dominates and consequent
excess thermal dampings cause a downward trend. 
This non-Heisenberg feature is much better highlighted
in 2(c), where both $T^{cr}_{1}/J_{eff}$ and $T^{cr}_{2}/J_{eff}$
decrease markedly on lowering $U$.
At weak coupling, both these scales collapse quickly. 

The loss of antiferromagnetic correlations 
at finite temperature is characterized through a temperature scale 
$T_{corr}$, extracted from $S(\pi,\pi)$.
The crossover lines have a similarity to the 
locus of this $T_{corr}(U)$ \cite{tiwari}, which 
also coincides with the metal-insulator transition line at weak coupling.
However, there are quantitative differences.
the peak location in our dynamical phase diagram Fig.4(a) 
is at $\sim U/t=10$, a \textit{higher} coupling compared to the
peak location in $T_{corr}$ at $\sim U/t=4$.
We emphasize that our focus is on the
"local moment" regime, {\it i.e}, 
intermediate to strong coupling.
Our method can address the weak coupling Slater regime as well
but that regime is dominated by amplitude fluctuations and
also requires larger system size.

In Section VI (subsection C), we discuss an effective 
\textit{classical} moment model which actually interpolates 
between the Heisenberg and Slater limits, 
borrowing a few parameters from the Hubbard mean field and RPA results.
This captures the low temperature dynamics of the Hubbard problem
fairly well at all $U/t$, and the Heisenberg limit at all temperatures. 
Moreover, the non-monotonicity of $T_{corr}$
as a function of $U/t$ and the qualitative behaviour of the 
thermal regimes are also captured by the effective model.

\begin{figure*}[t]
\centerline{
\includegraphics[height=3.6cm,width=15.2cm]{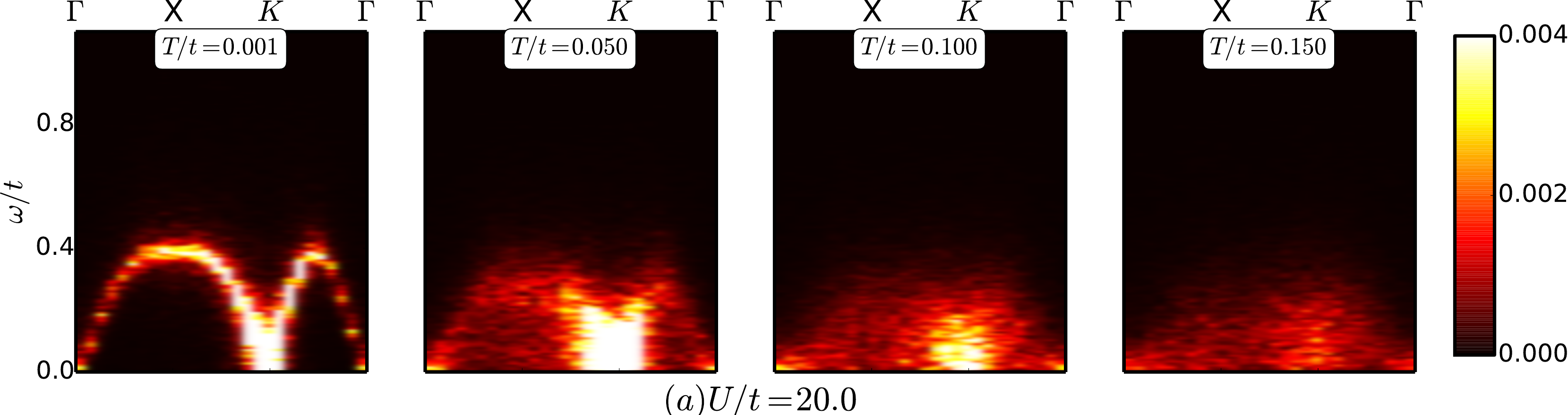}
}
\vspace{0.2cm}
\centerline{
\includegraphics[height=3.6cm,width=15.5cm]{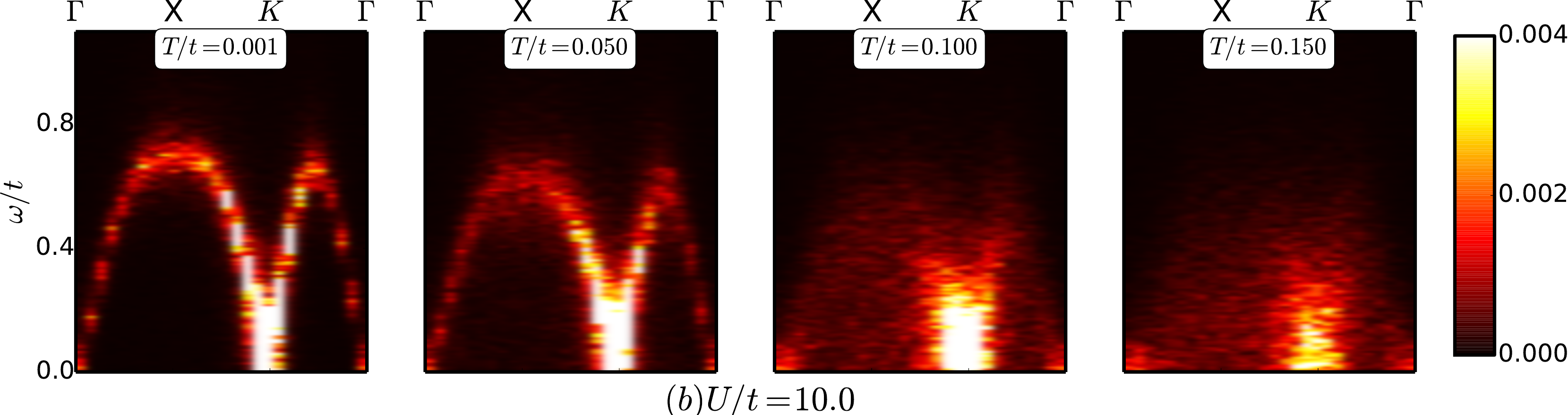}
}
\vspace{0.2cm}
\centerline{
\includegraphics[height=3.6cm,width=15.5cm]{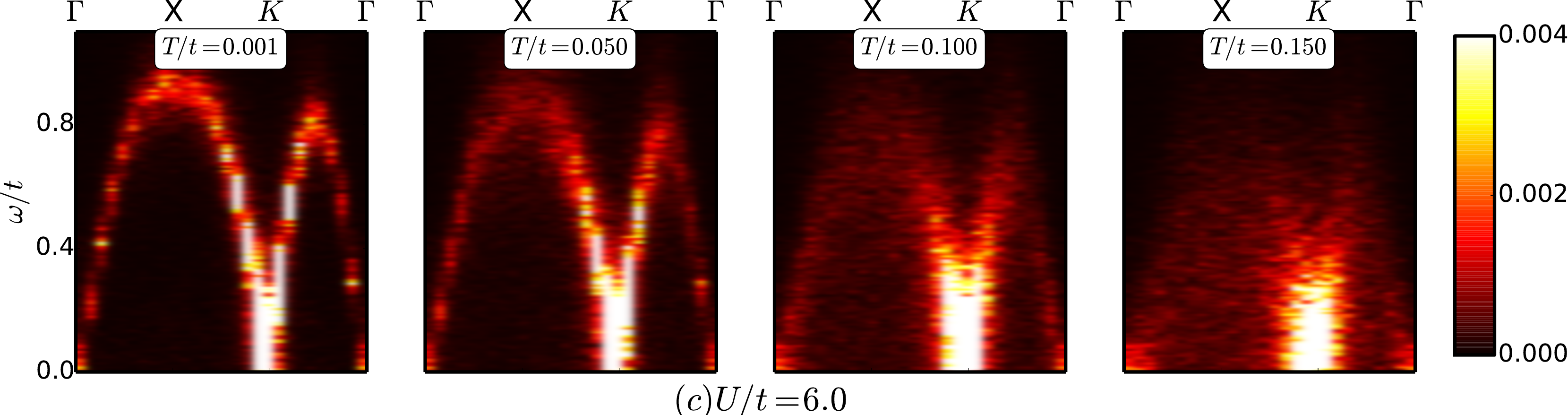}
}
\vspace{0.2cm}
\centerline{
\includegraphics[height=3.6cm,width=15.5cm]{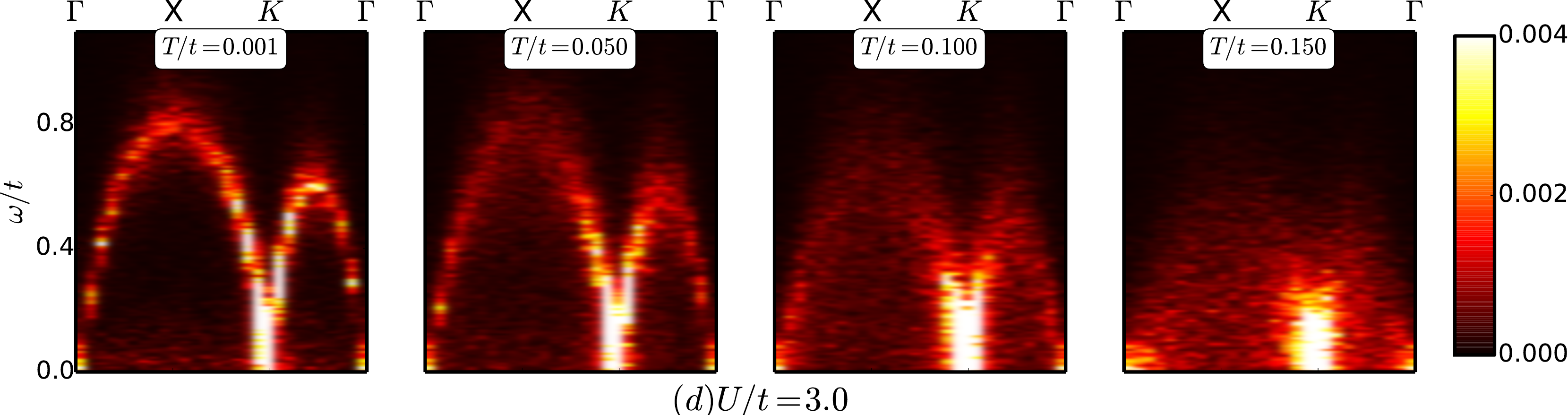}
}
\caption{Power spectrum of magnetization field $D({\bf q},\omega)$ for the 
Hubbard model on the square lattice for $U/t=20,10,6,3$ respectively. 
The trajectory chosen in Brillouin Zone is $\Gamma-X-K-\Gamma$. 
Temperatures are scaled by electron hopping $t$. 
We observe a resemblance of the strong coupling Hubbard spectrum 
with that of the Heisenberg model with $J_{eff}=4t^{2}/U$. 
At lower couplngs, the dispersion changes at low $T$, 
owing to longer-range spin couplings. Thermal damping is more prominent 
at weaker couplings, as the stiffness for amplitude fluctuation decreases.
}
\end{figure*}

In the triangular case (Fig.4(b) and 4(d)), 
the generic low $T$ lineshapes is two-peak. 
The crossover regimes (ii) and (iii) 
occur at much lower temperatures compared to the square case, 
owing to mild geometric frustration and consequently fragile
magnetic order. 
The fall of the crossover scales on decreasing $U$ (below
$U/t=10.0$, say) is 
also sharper than the former.
Close to the transition ($U/t \sim 6$)
the lineshapes become diffusive
even at very low temperatures ($T/t \sim 0.01$).
The scaled phase diagram (4(d)) reveals a minimum in
the crossover scales around $t/J_{eff} \sim 12.5$. 
This is related to the
non-monotonic behaviour of $J_{eff}$ itself, shown in Fig.2.

We comment that our scheme at weak coupling generates
a peak centered at zero frequency for all momenta, exclusively 
due to amplitude fluctuations. This arises from an oversimplification
of our equations of motion. 
However, the fraction of this weight isn't
visible on a linear scale above $U/t \sim 4$ on the square.
Moreover, if we ignore the near-zero energy part of the
magnon spectrum (upto some cutoff $\sim 0.05 W_{mag}$), the
rest of it doesn't have any \textit{spurious} features.
We still capture the impact of magnitude fluctuations on the
damping of spin waves, which reside at higher energies.

Next, we present detailed numerical results on the dynamics
of square and triangular lattice Hubbard models found using our scheme.
The focus is on deviations from the Heisenberg limit, 
quantified through finite temperature behaviour of the damping
of spin waves. 

\section{Dynamics on the square lattice}

In this section, we first show the spectral maps of $D({\bf q},\omega)$
across a section of the Brillouin Zone (BZ) for four representative
couplings, starting from the Heisenberg limit. Next, we extract the
mode energies and magnon damping from the data and plot their
variation with respect to $T$ and ${\bf q}$ respectively.
Finally, a comparison of actual lineshapes for a generic
wavevector ${\bf q}=(\pi/2,\pi/2)$ is featured.

\subsection{Spectral maps for varying $U/t$ and temperature}

The dynamical structure factor maps are exhibited in Fig.5. The top
row shows results for a $U/t=20.0$ Hubbard model (the Heisenberg limit) 
in various temperature regimes. The first column corresponds to the lowest $T$.
Here, we see sharply defined spin waves, with Goldstone modes at
both $(0,0)$ and $(\pi,\pi)$ and a characteristic antiferromagnetic
dispersion. At intermediate temperatures ($T/t=0.05$), 
the bandwidth reduces and the spin waves broaden. 
On further increase in $T$,  
the correlations weaken to give a diffusive spectrum, with prominent
low-energy weight close to $(\pi,\pi)$. 
Ultimately, the momentum dependence is also lost for $T/t=0.15$.

The lower panels show results on the Hubbard model for
three successively lower couplings- 
strong ($U/t=10.0$), intermediate ($U/t=6.0$) and
weak ($U/t=3.0$) respectively. At strong coupling, the behaviour is
Heisenberg-like, with $J_{eff} \sim t^2/U$, with small deviations. 
The spectrum remains mostly coherent till $T \sim J_{eff}$, 
with momentum dependent thermal damping. 
The Goldstone mode at $(\pi,\pi)$ survives as 
a broad low-energy feature till $T \sim 2J_{eff}$.

\begin{figure}[b]
\centerline{
\includegraphics[height=4.2cm,width=4.3cm]{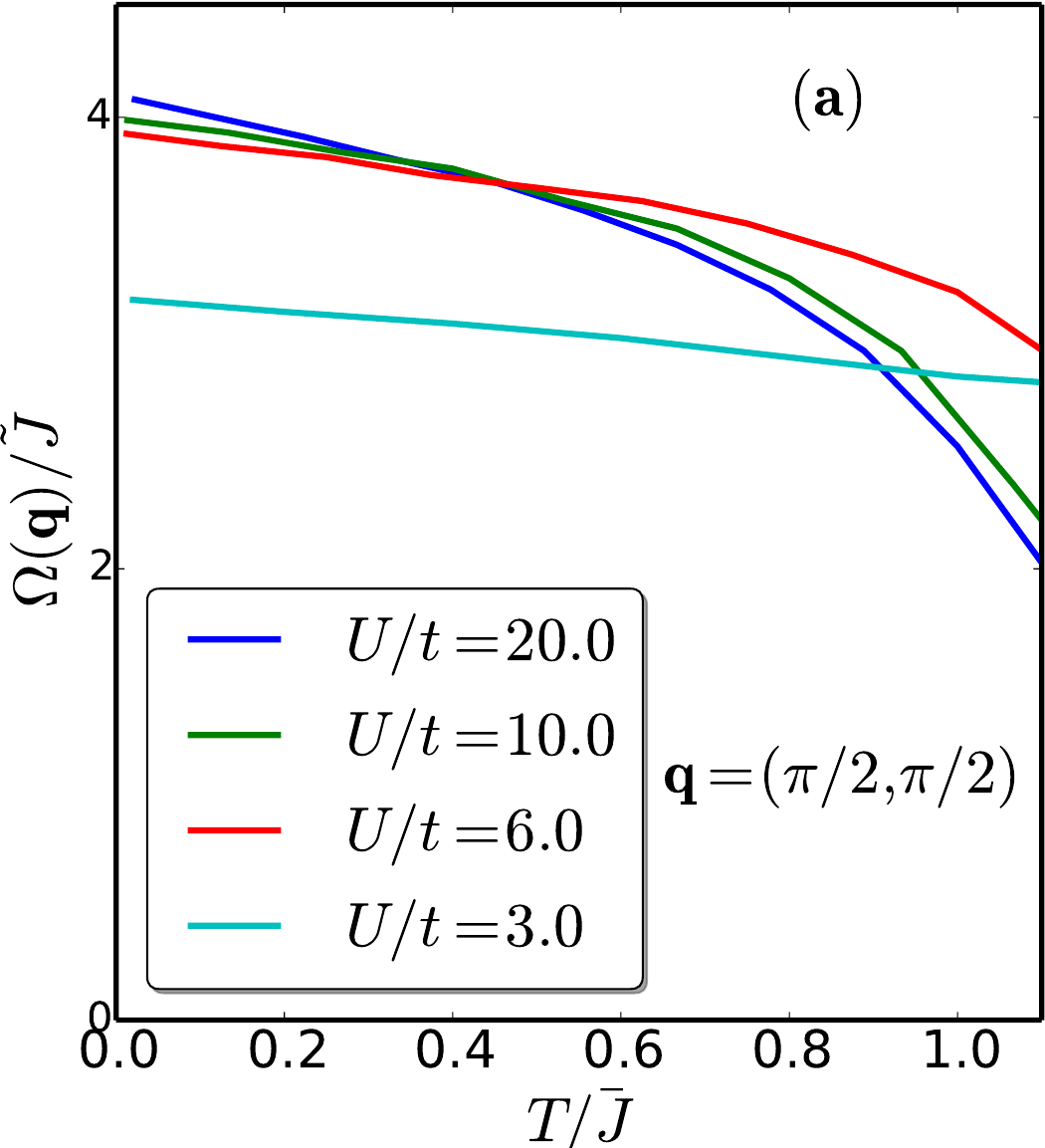}
\includegraphics[height=4.2cm,width=4.3cm]{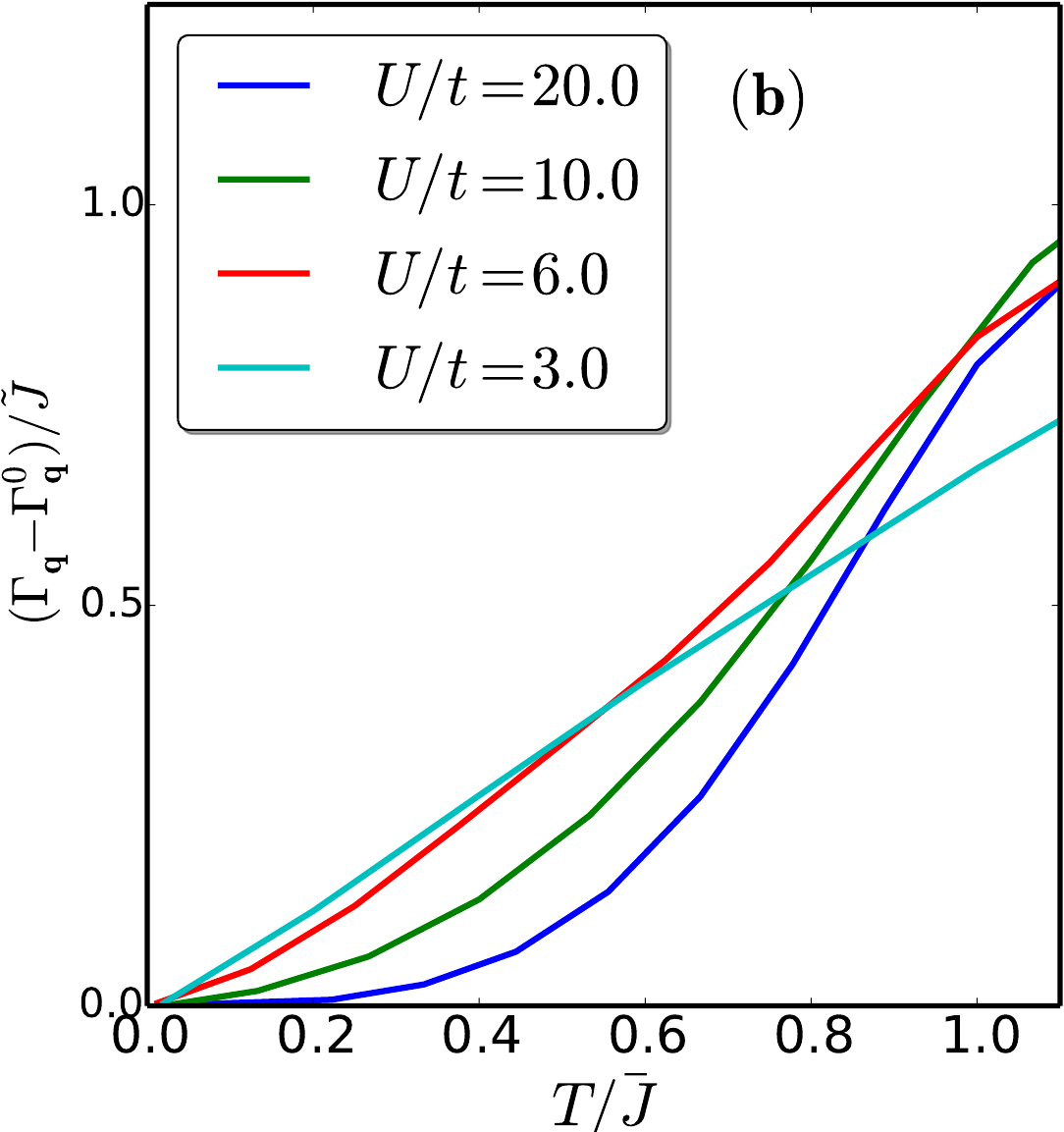}
}
\caption{Fitted dispersions ($\Omega_{\bf q}$) and intrinsic thermal dampings 
($\Gamma_{\bf q}-\Gamma^{0}_{\bf q}$) as functions of $T$, extracted from the 
dynamical spectra in the square lattice. The temperature axes are scaled by 
${\bar J}=J_{eff}|{\bf m}_{HF}|^{2}$, while the frequencies are scaled by 
${\tilde J}=J_{eff}|{\bf m}_{HF}|$ 
values for the various couplings studied.  
The dispersions soften slowly with increasing $T$, 
while one clearly observes the onset of non-Heisenberg behaviour 
in (b) for lower $U$ values, with large dampings showing up much below 
$T/{\bar J}=1$.  }
\end{figure}

At intermediate coupling ($U/t=6.0$), the bandwidth increases
compared to the earlier case and the low $T$ dispersion changes 
in shape. This owes its origin to the emergence of multi-spin
couplings. There's also a faint, momentum-independent low-energy
band, more clearly visible in a logarithmic color scale. 
This band arises from longitudinal fluctuations of moments 
within our scheme, which is controlled by the local stiffness. 
Thermal fluctuations broaden the spin waves gradually, 
with the dispersion being discernable even at $T \sim 0.1t$. 

The bottom row features weak coupling ($U/t=3.0$) results, 
where the low energy band gains more weight 
(now visible on a linear scale) and the bandwidth
shortens again. Thermal effects are stronger, as amplitude 
fluctuations are more prominent here.

\begin{figure}[b]
\centerline{
\includegraphics[height=3.2cm,width=2.9cm]{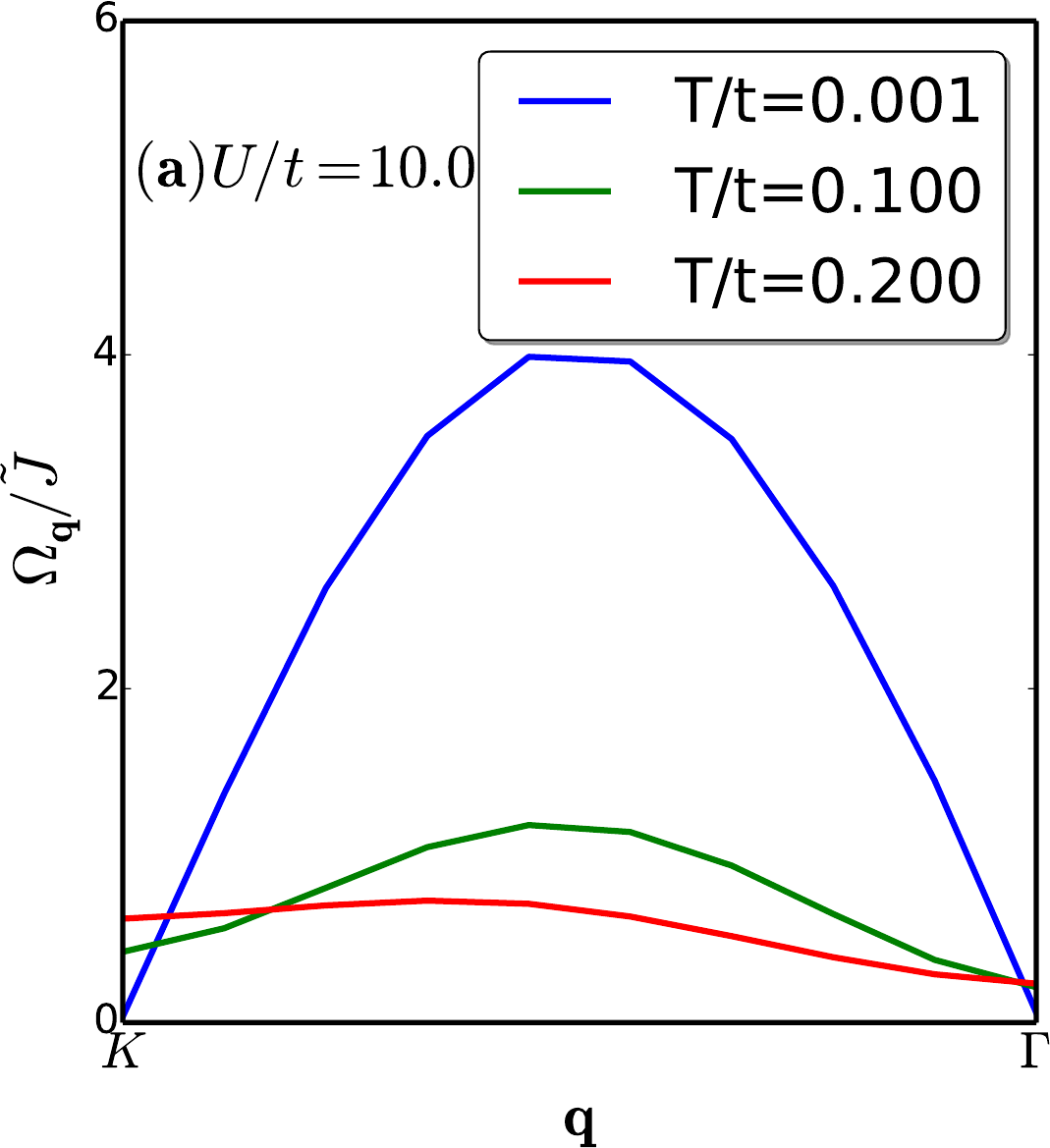}
\includegraphics[height=3.2cm,width=2.9cm]{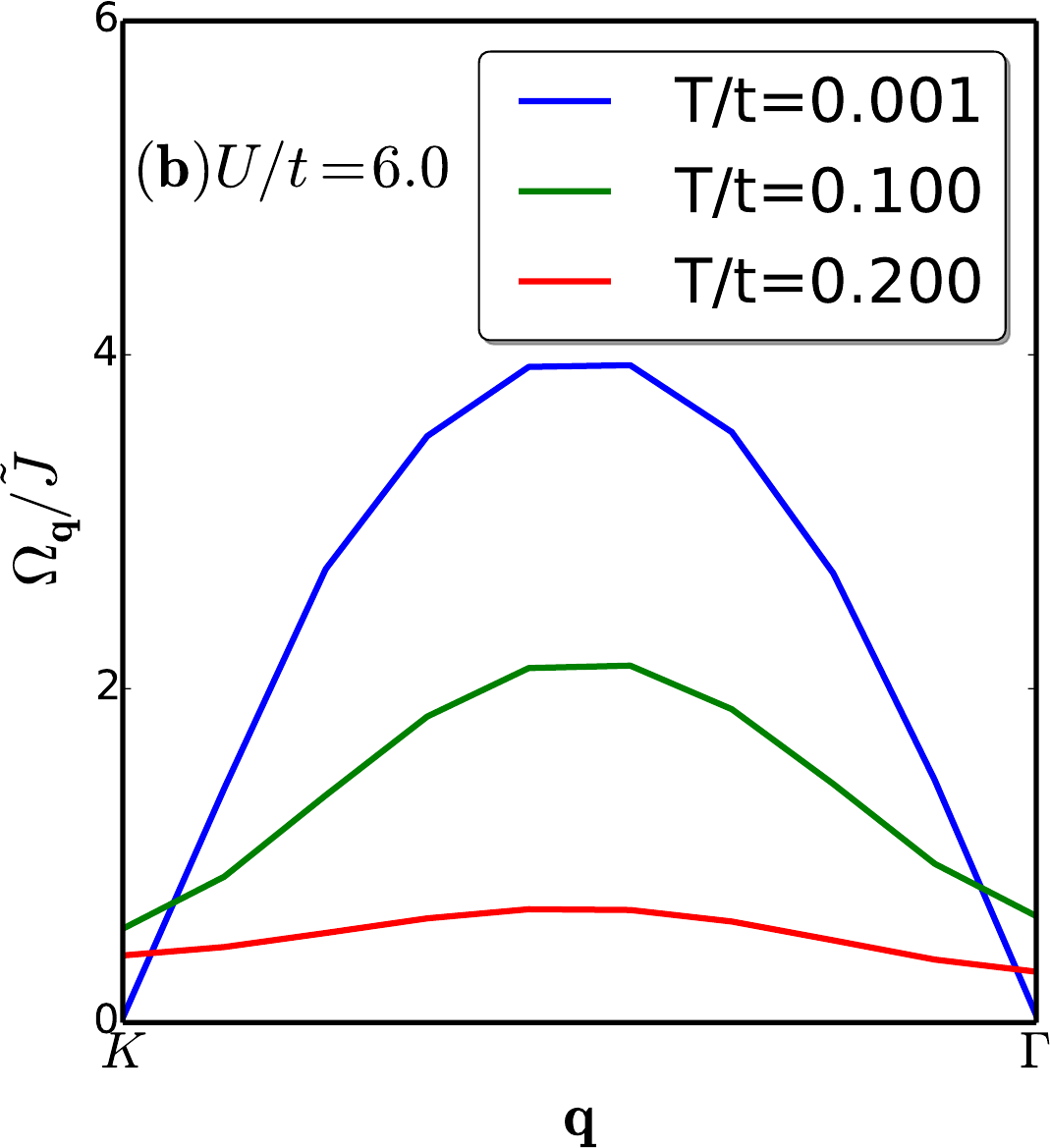}
\includegraphics[height=3.2cm,width=2.9cm]{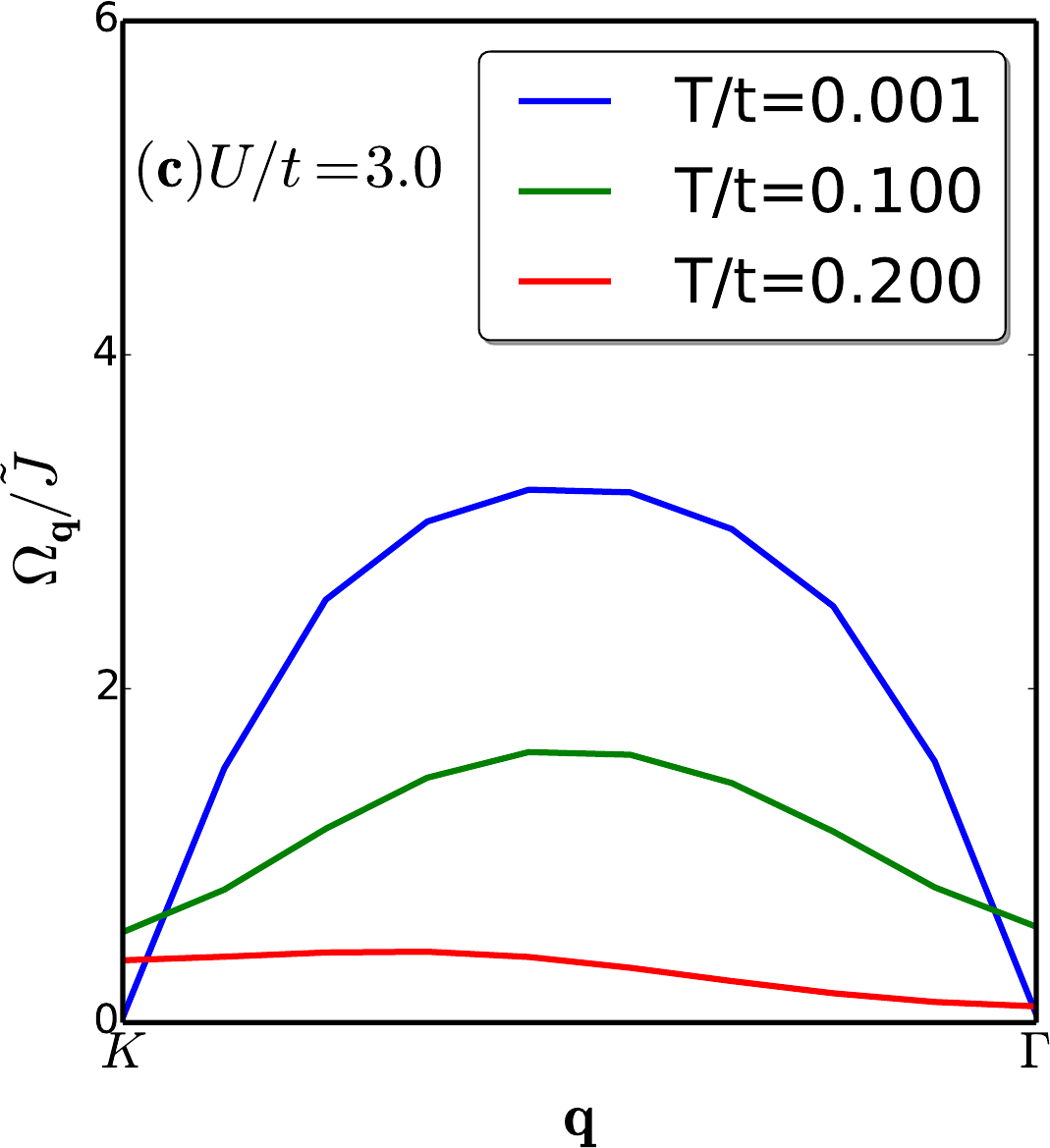}
}
\vspace{0.2cm}
\centerline{
\includegraphics[height=3.2cm,width=2.9cm]{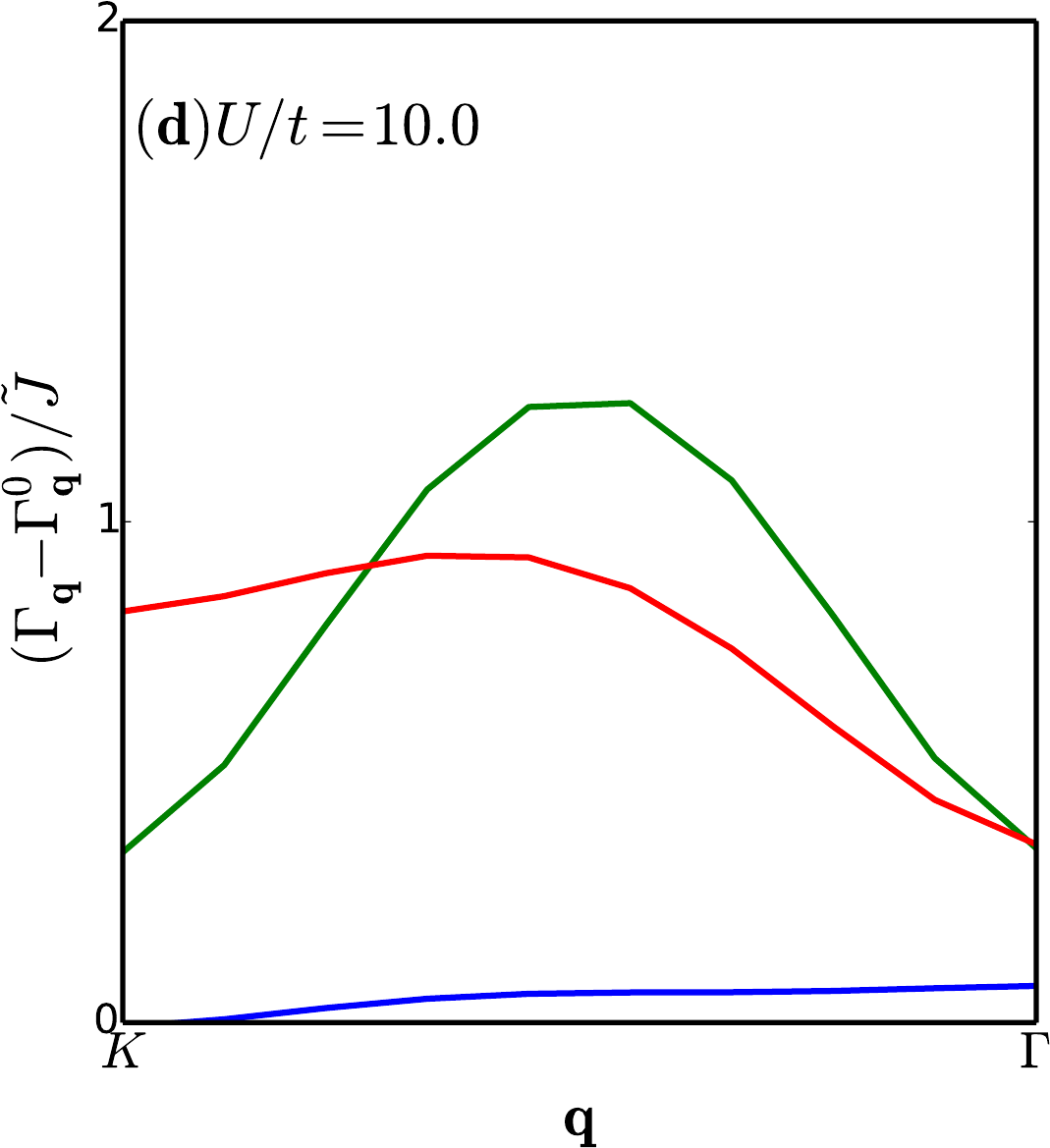}
\includegraphics[height=3.2cm,width=2.9cm]{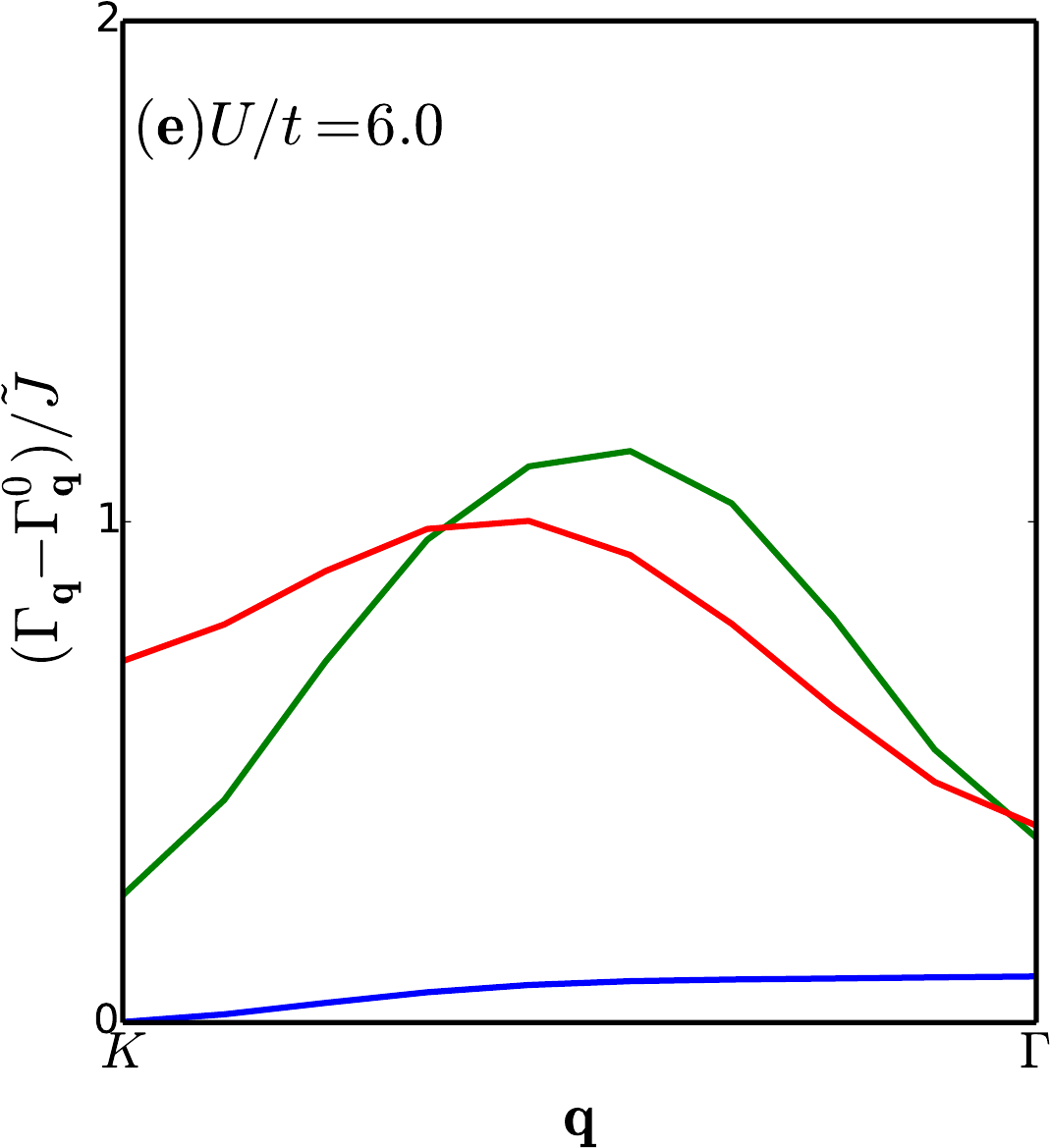}
\includegraphics[height=3.2cm,width=2.9cm]{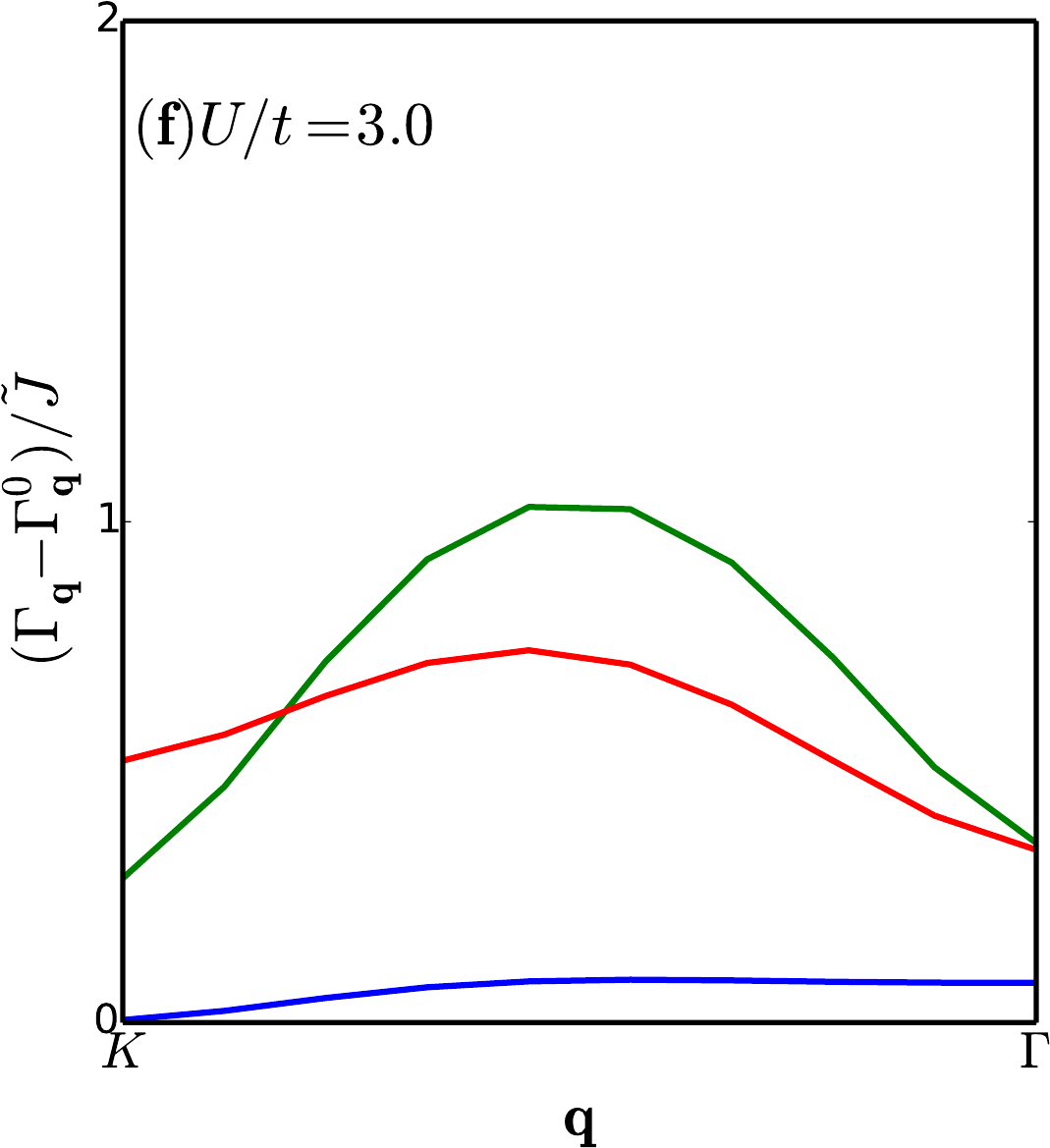}
}
\caption{Fitted dispersions ($\Omega_{\bf q}$) in (a)-(c) and intrinsic thermal 
dampings ($\Gamma_{\bf q}-\Gamma^{0}_{\bf q}$) in (d)-(f), plotted against ${\bf q}$ 
along the $K-\Gamma$ trajectory in three thermal regimes- (i)~weakly damped, 
(ii)~strongly damped and (iii)~diffusive. The couplings chosen are $U/t=3,6,10$ and
the absolute temperatures are $T/t=0.001,0.1,0.2$. We observe a non-monotonicity in 
the peak frequency, and a mild shift of this peak to lower ${\bf q}$ on heating up. The 
bottom row reveals a residual momentum dependence of magnon damping even in the 
diffusive regime.  }
\end{figure}

\begin{figure*}[t]
\centerline{
\includegraphics[height=4.5cm,width=5.5cm]{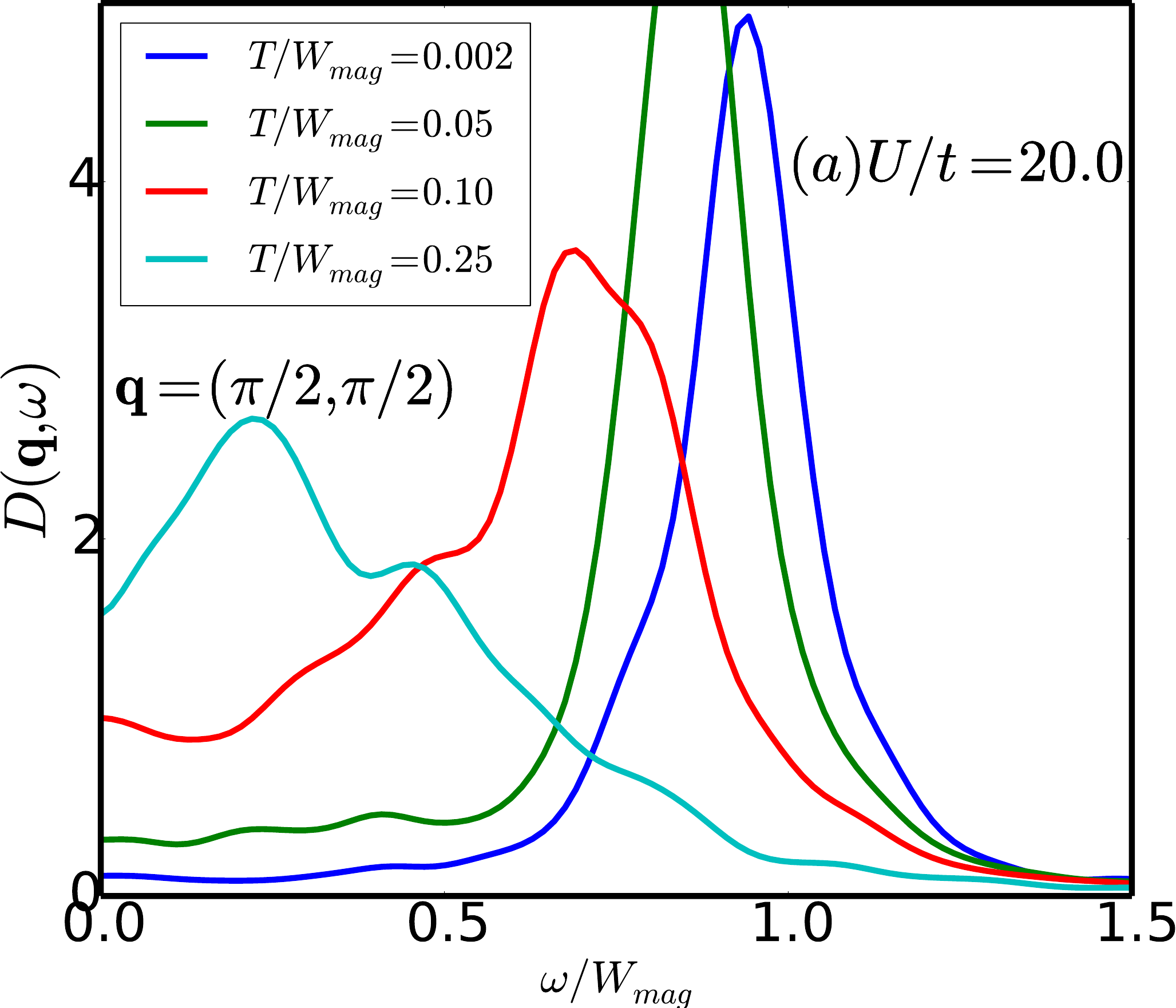}
\includegraphics[height=4.5cm,width=5.5cm]{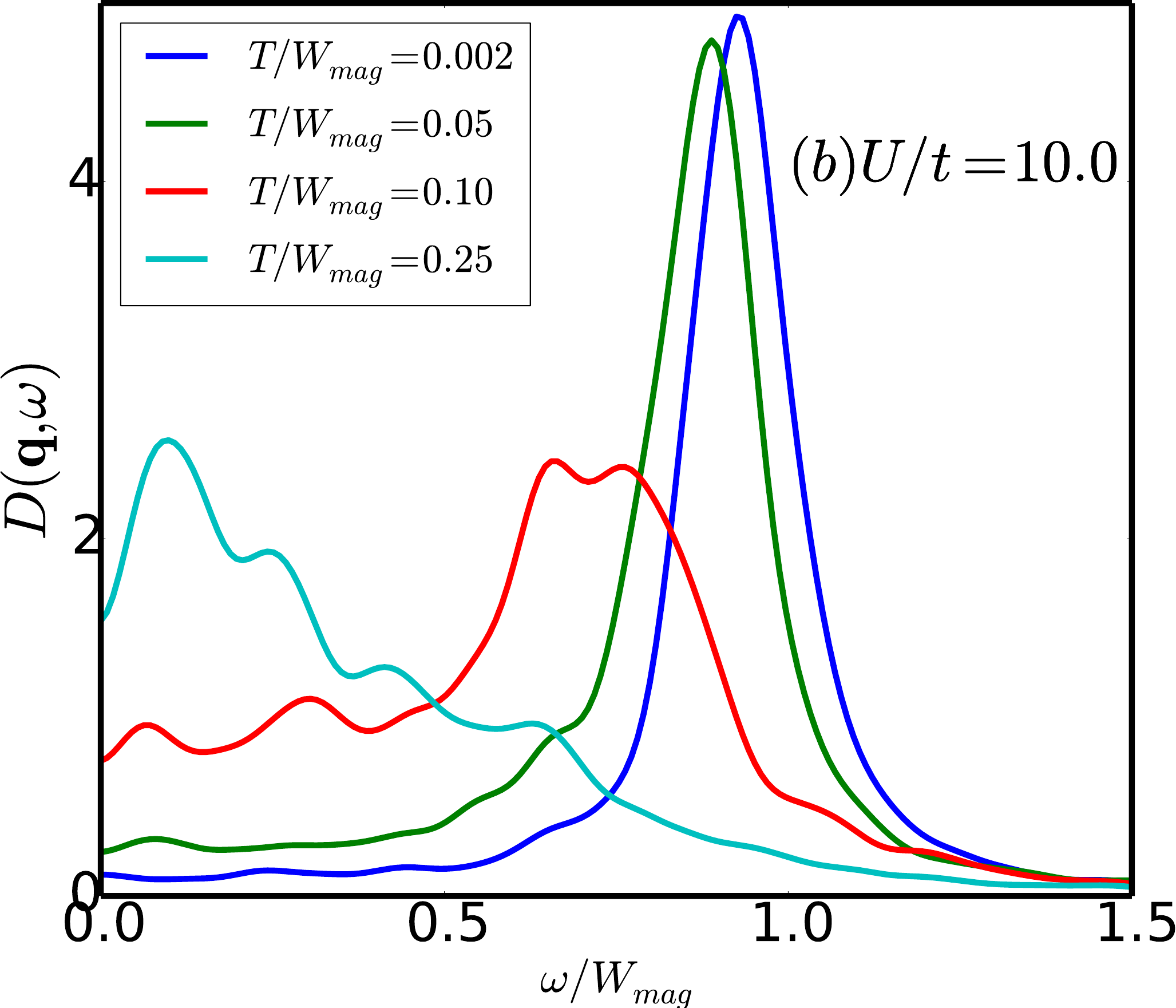}
}
\vspace{0.2cm}
\centerline{
\includegraphics[height=4.5cm,width=5.5cm]{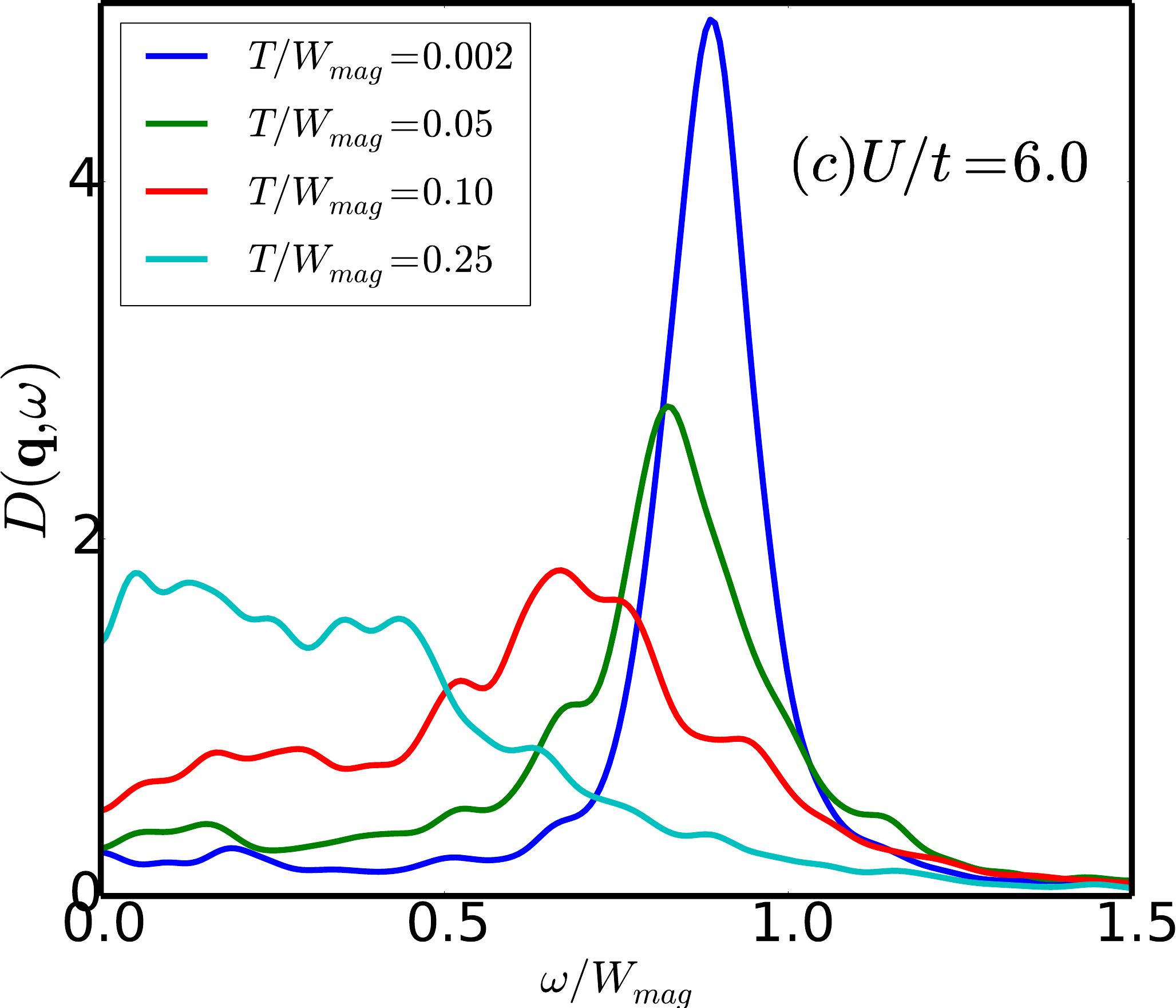}
\includegraphics[height=4.5cm,width=5.5cm]{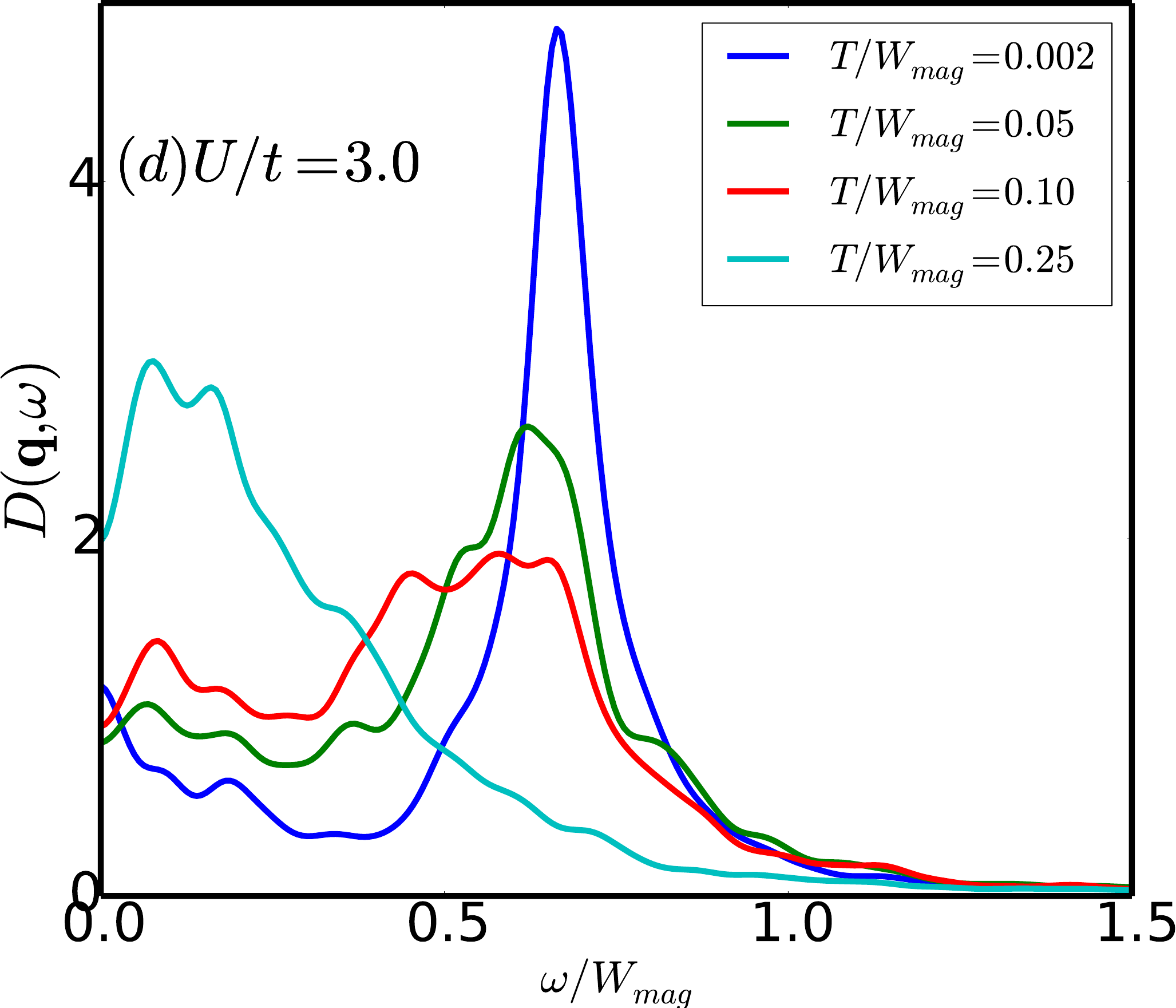}
}
\caption{Lineshapes at ${\bf q}=(\pi/2,\pi/2)$ for the Hubbard model 
(a-d) for $U/t=20,10,6,3$ respectively. 
We see a clear deviation from Heisenberg-like
behaviour in the thermal trends on decreasing coupling. Frequencies
and temperatures are scaled by the respective bandwidths ($W_{mag}$)
of the magnetization spectrum.
}
\end{figure*}

\begin{figure*}[t]
\centerline{
\includegraphics[height=3.6cm,width=15.2cm]{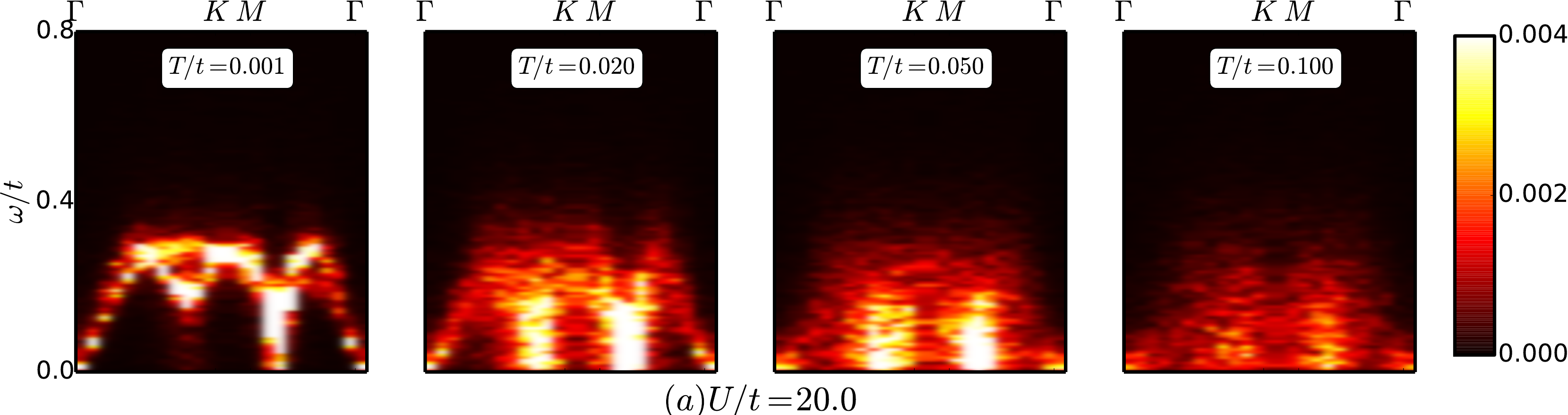}
}
\vspace{0.2cm}
\centerline{
\includegraphics[height=3.6cm,width=15.5cm]{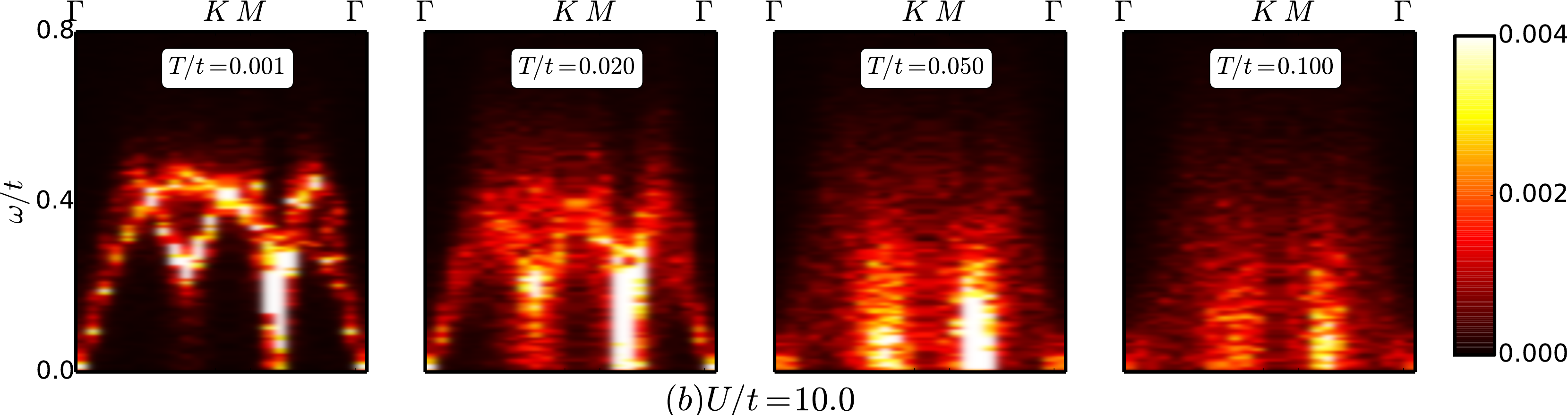}
}
\vspace{0.2cm}
\centerline{
\includegraphics[height=3.6cm,width=15.5cm]{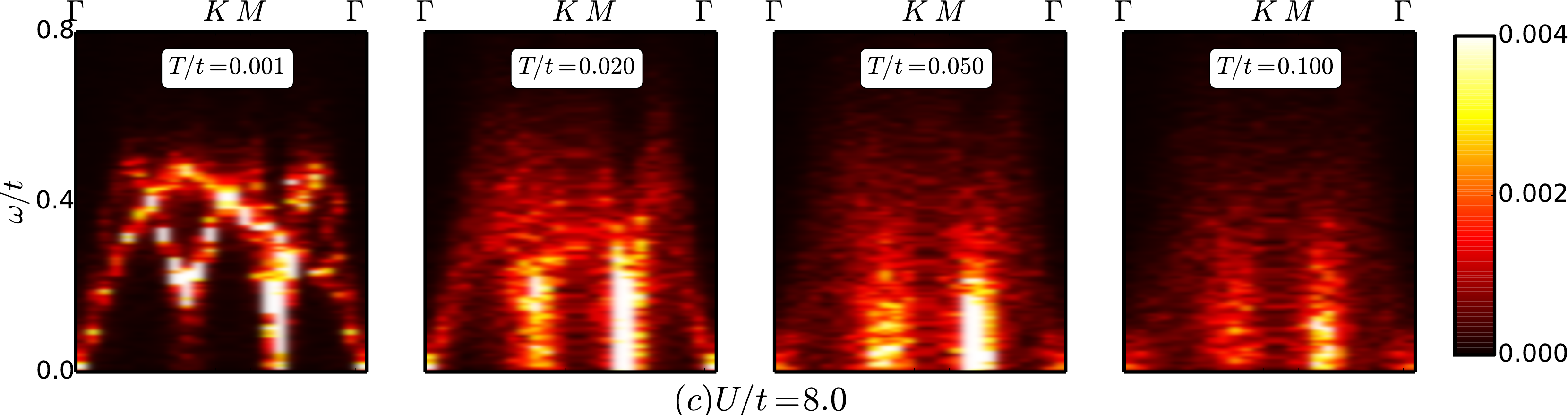}
}
\vspace{0.2cm}
\centerline{
\includegraphics[height=3.6cm,width=15.5cm]{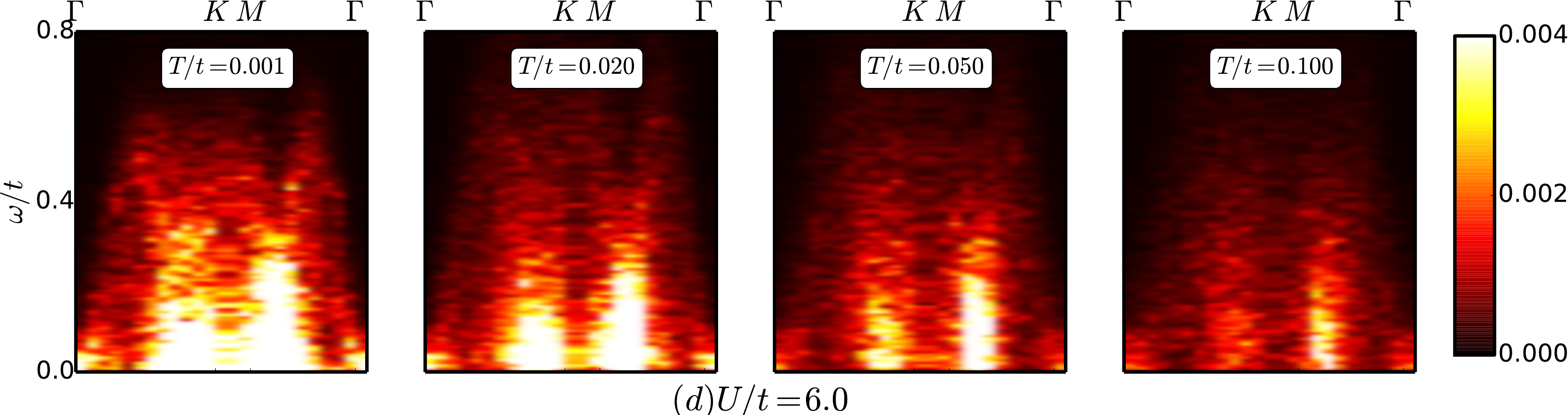}
}
\caption{Power spectrum of magnetization field $D({\bf q},\omega)$
for the Hubbard model on the triangular lattice for $U/t=20,10,8,6$ 
respectively. The trajectory chosen
in Brillouin Zone is $\Gamma-K-M-\Gamma$. Temperatures are scaled by
electron hopping $t$. Again, we observe a similarity
of the strong coupling Hubbard spectrum with the Heisenberg case. 
The lower branch between $\Gamma-K$ in the Heisenberg limit develops 
a prominent dip for lower $U$ values. The thermal dampings are stronger 
on moving to weaker couplings compared to the square case.
}
\end{figure*}

\subsection{Variation of mode energy and damping with $T$}

Fig.6 highlights the evolution of mean frequency ($\Omega_{\bf q}$)
and thermally induced linewidth ($\Gamma_{\bf q}-\Gamma^{0}_{\bf q}$)
with temperature at a generic wavevector ${\bf q}=(\pi/2,\pi/2)$. 
The former monotonically falls with increasing $T$, as seen in 6(a). 
The rate of decrease speeds up around successively 
lower fractions of ${\tilde J}=J_{eff}|{\bf m}_{HF}|$ on moving to 
lower couplings. 
In 6(b), we see that the rise in thermal damping has an initially
quadratic trend at large $U$ and low $T$, 
which then changes to a linear one one moving to lower couplings, 
and becomes $T^{\alpha}$ with $1<\alpha<2$ on raising $T$.
A somewhat sharper fall is seen in the "onset temperature" for
strongly damped behaviour on lowering $U/t$, compared to the trend
followed by the mean. 

\begin{figure*}[t]
\centerline{
\includegraphics[height=4.2cm,width=4.2cm]{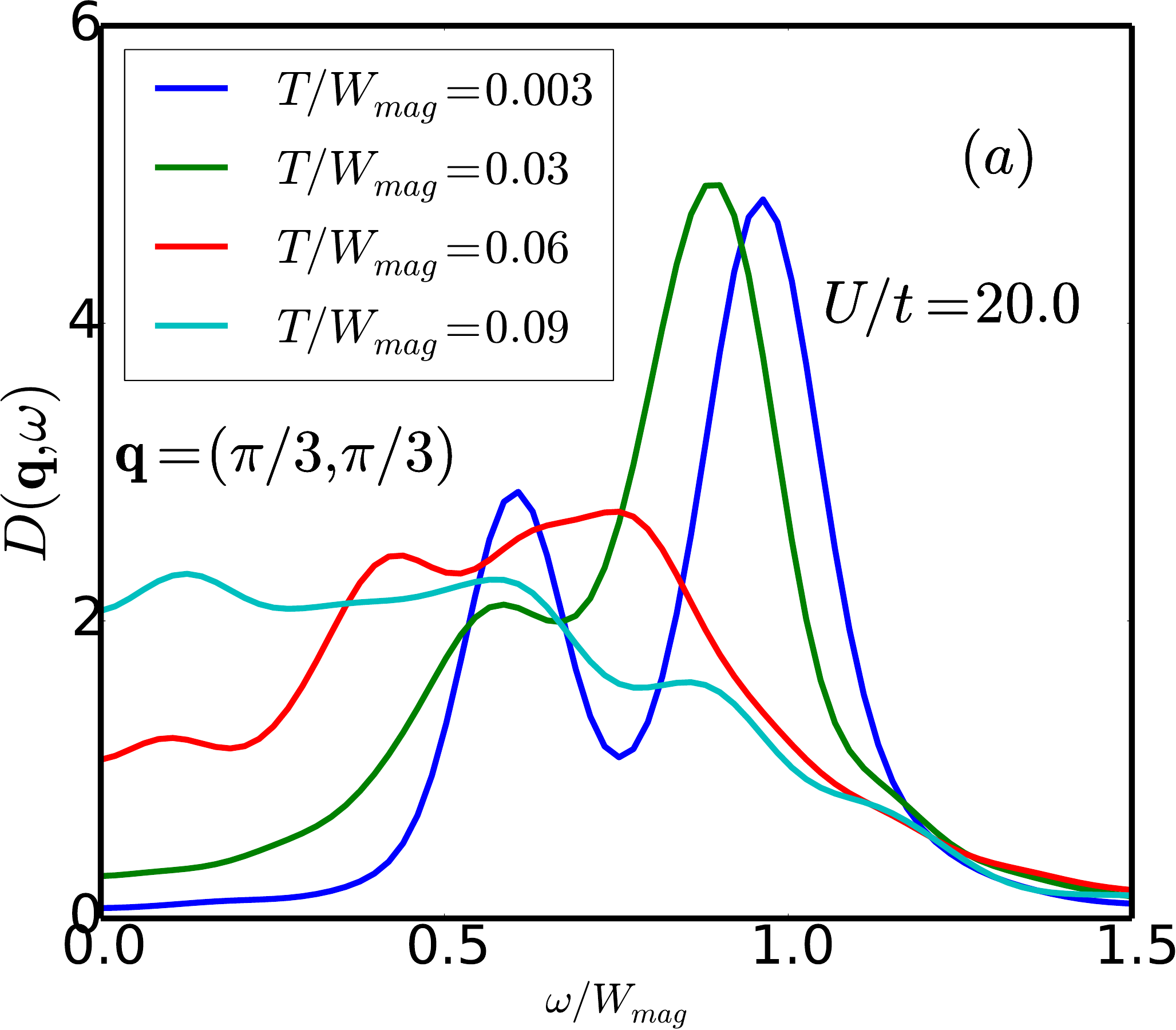}
\includegraphics[height=4.2cm,width=4.2cm]{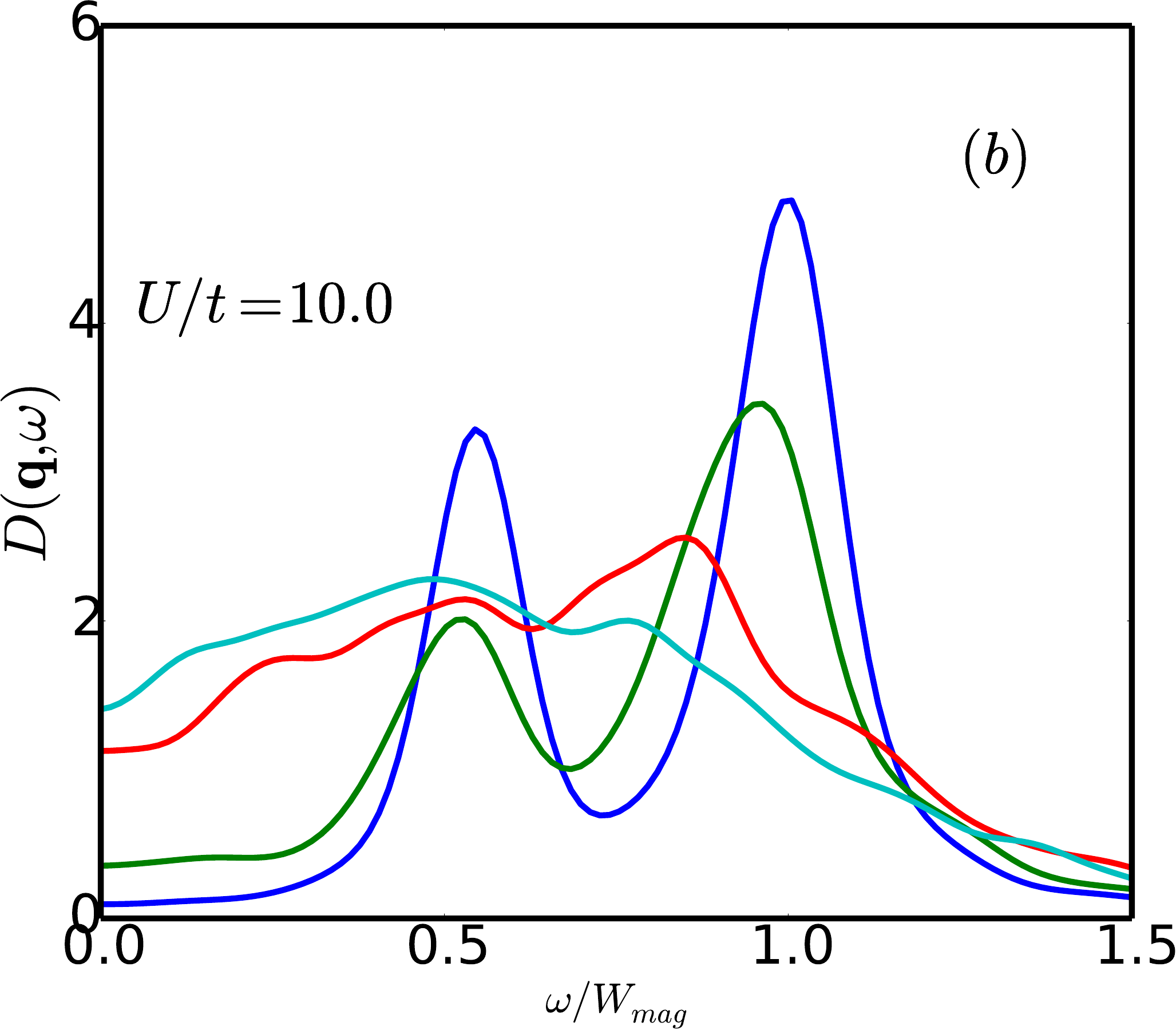}
\includegraphics[height=4.2cm,width=4.2cm]{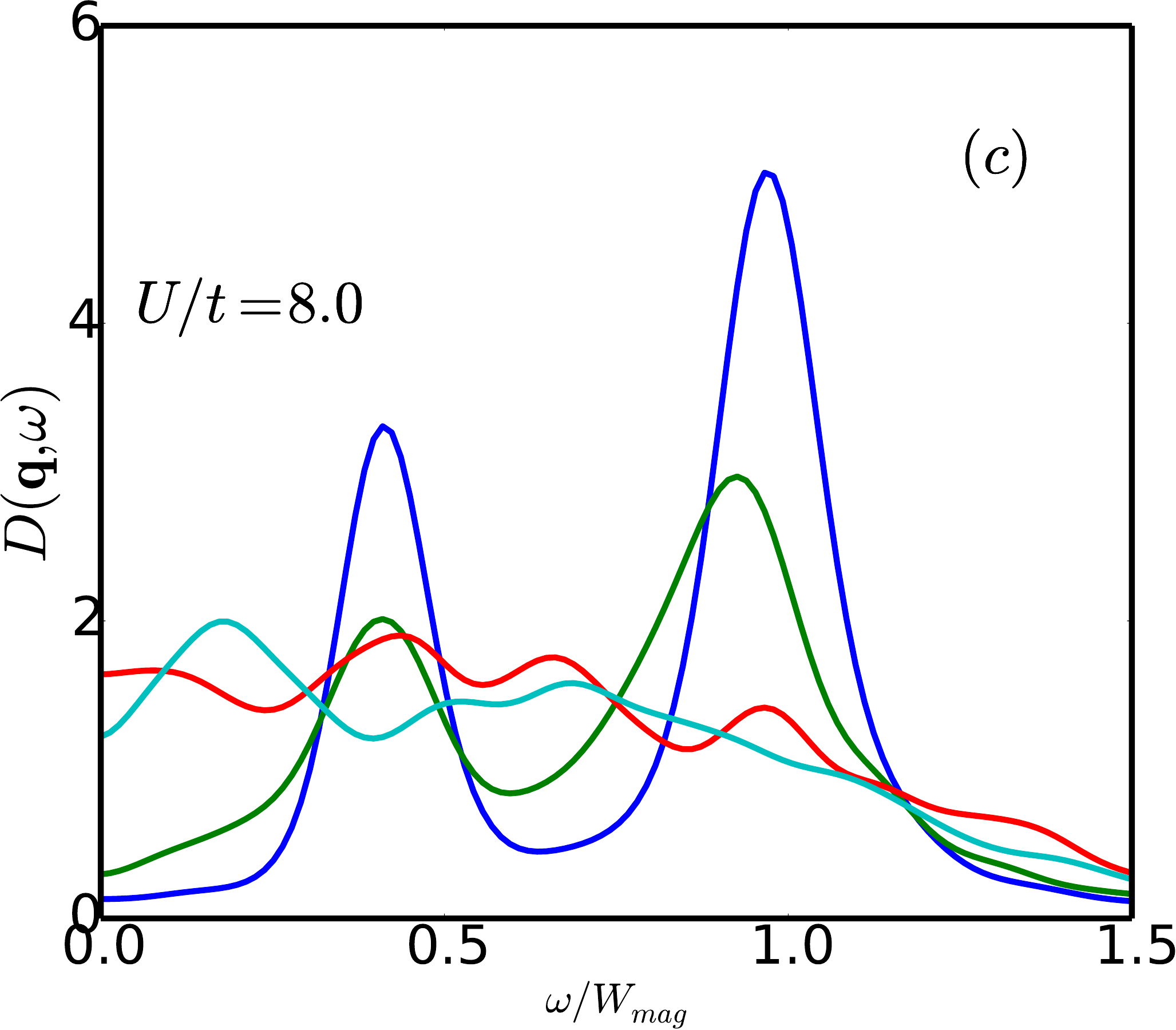}
\includegraphics[height=4.2cm,width=4.2cm]{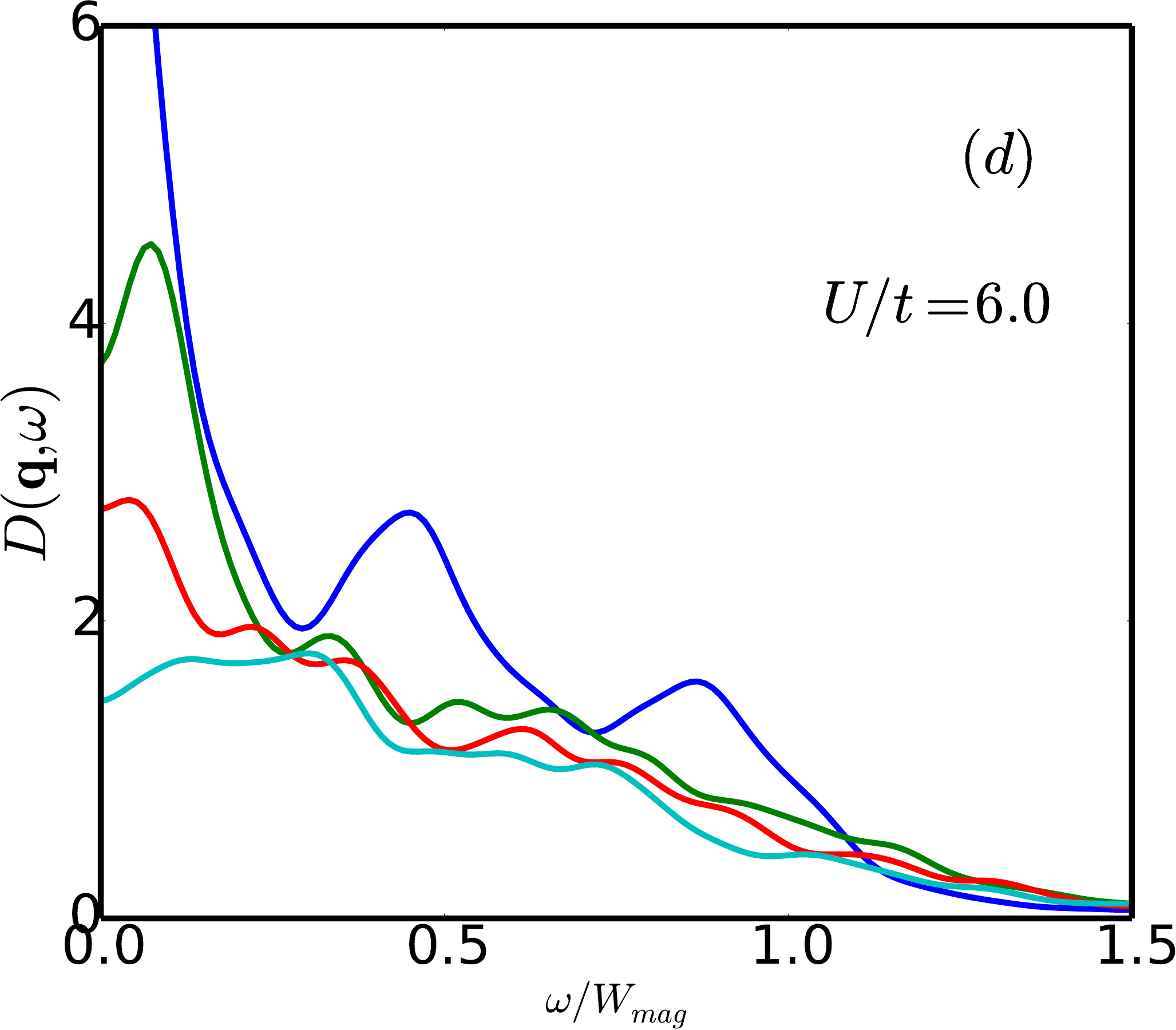}
}
\vspace{0.2cm}
\centerline{
\includegraphics[height=4.2cm,width=4.2cm]{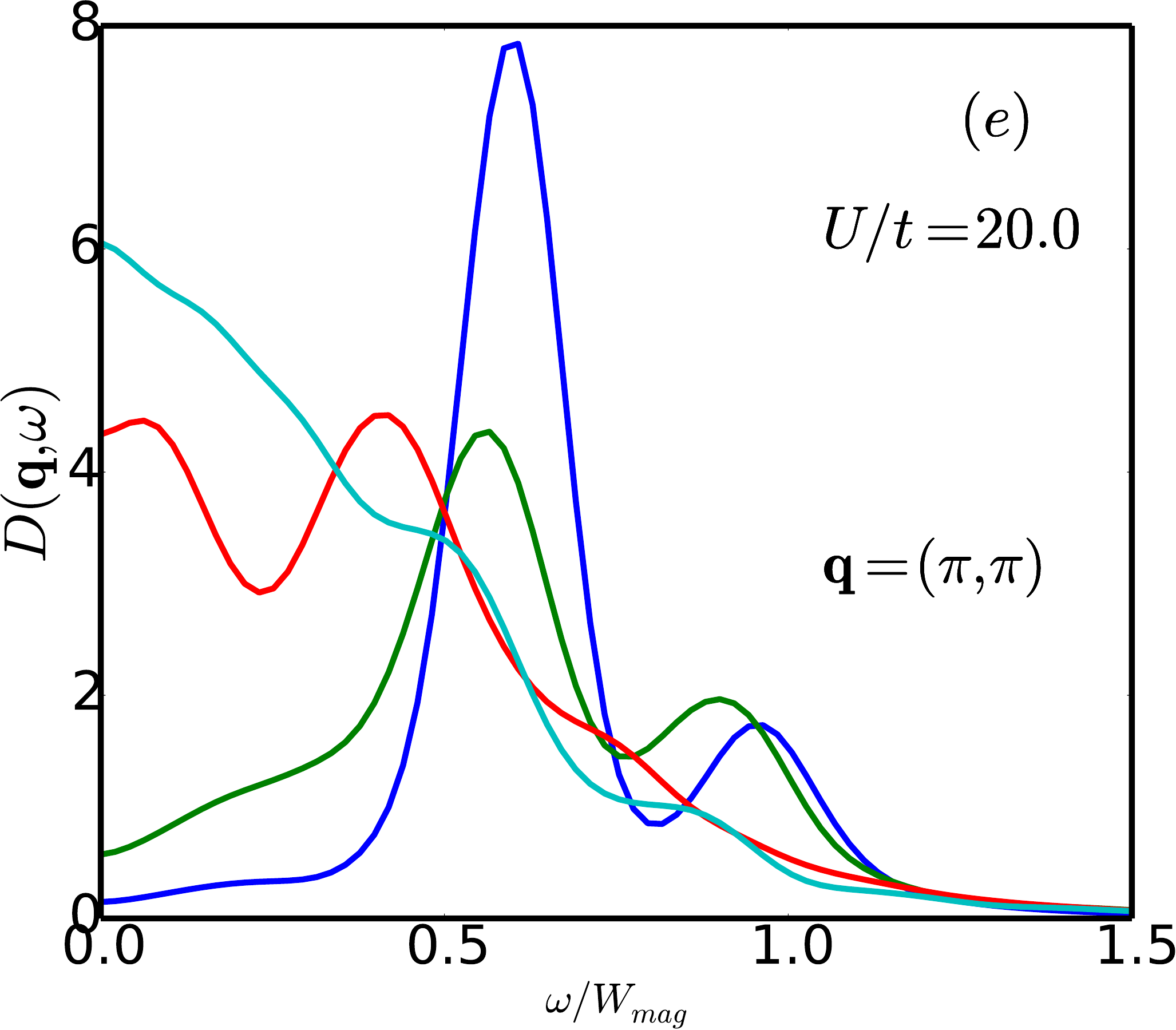}
\includegraphics[height=4.2cm,width=4.2cm]{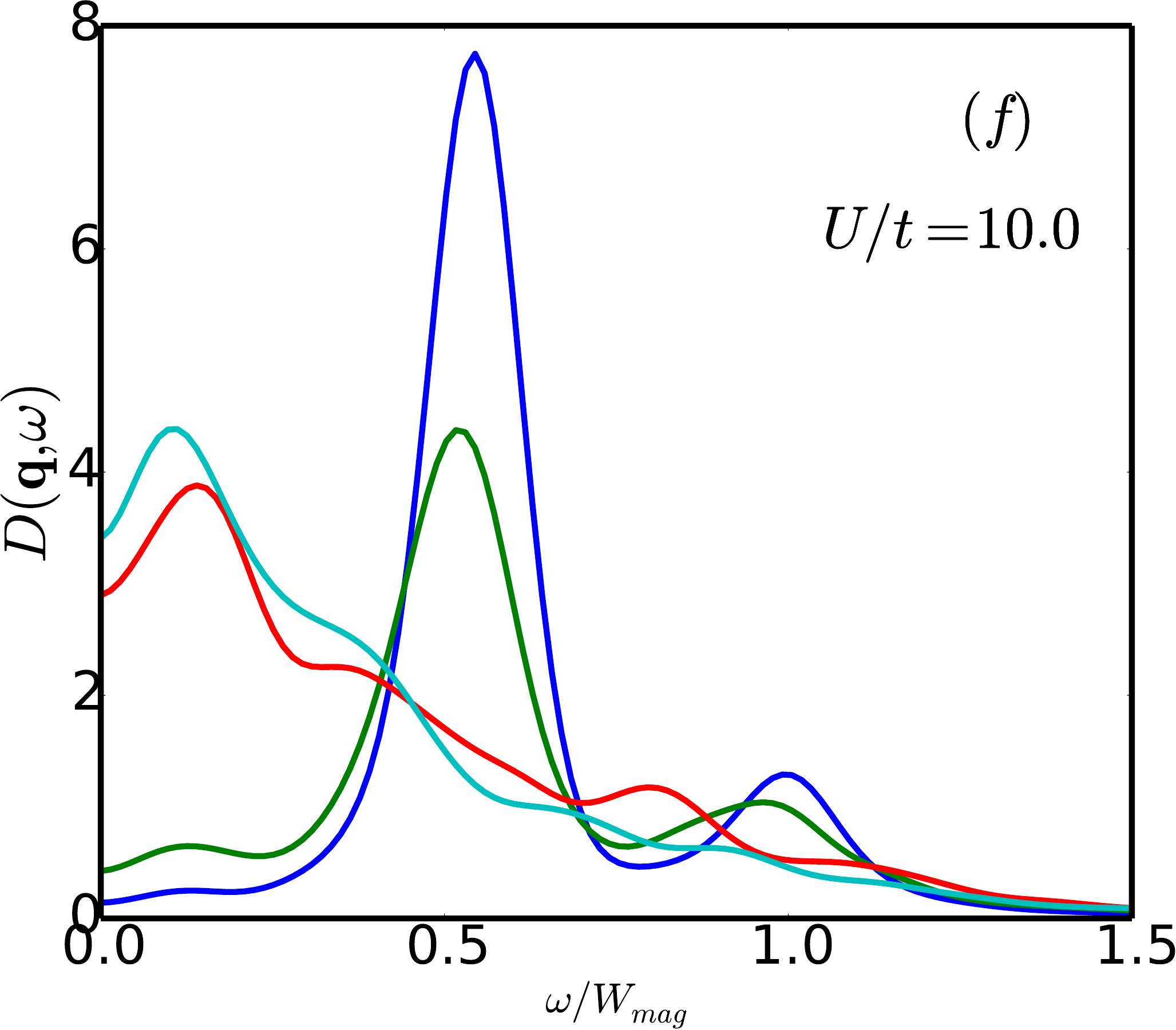}
\includegraphics[height=4.2cm,width=4.2cm]{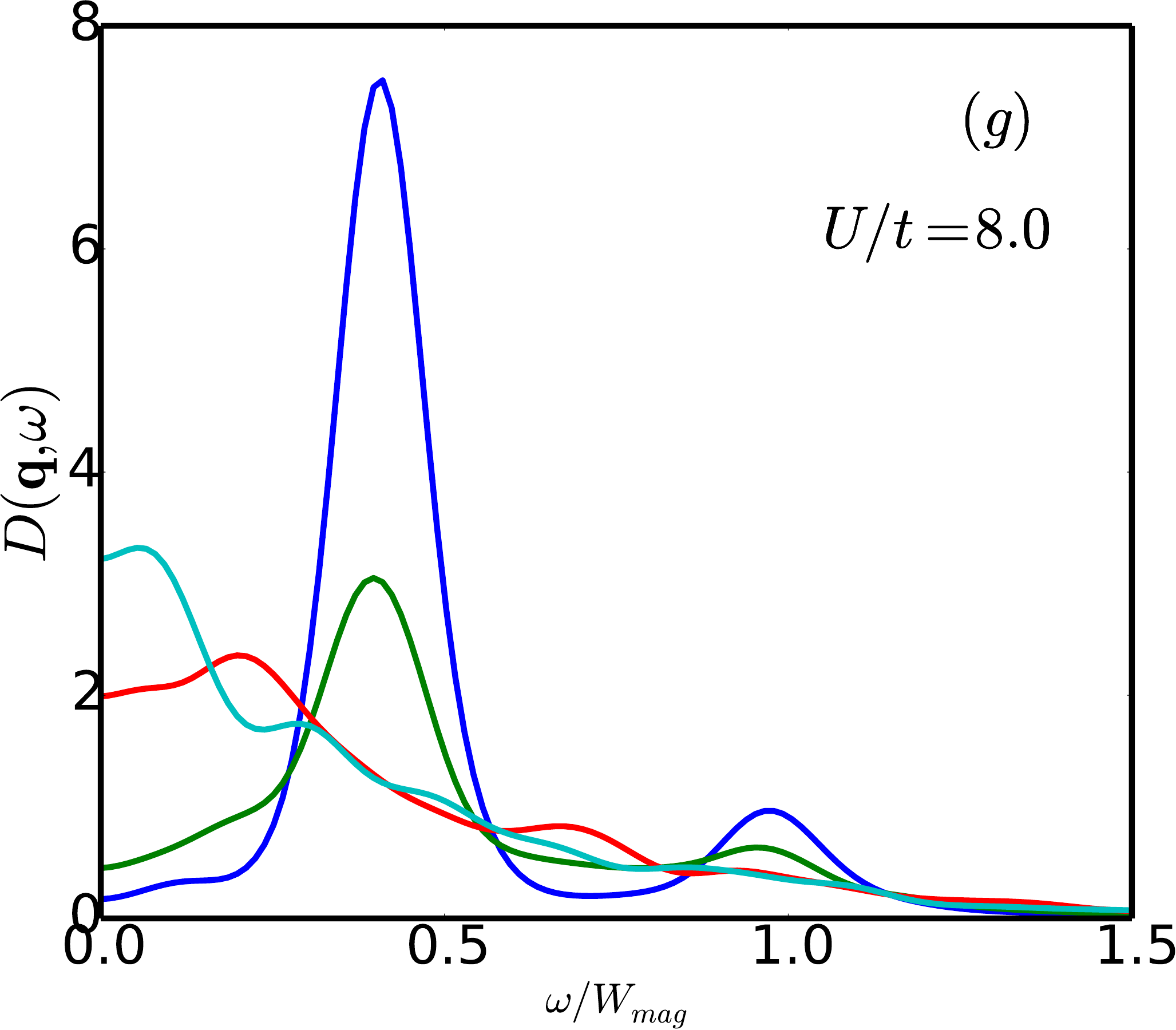}
\includegraphics[height=4.2cm,width=4.2cm]{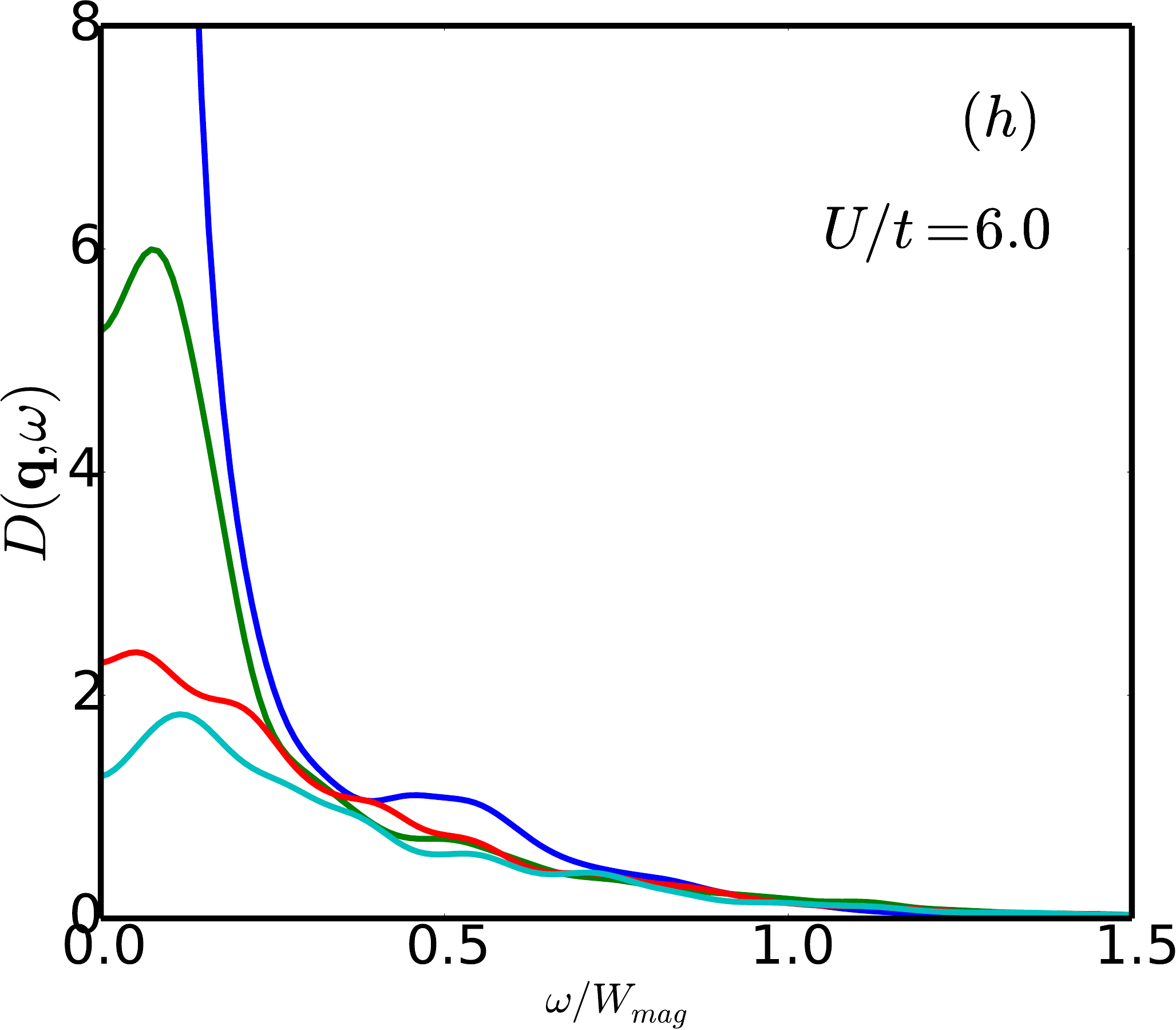}
}
\caption{Triangular lattice: 
lineshapes at ${\bf q}=(\pi/3,\pi/3)$ (a-d)
and ${\bf q}=(\pi,\pi)$ (e-h) for the Hubbard model for $U/t=20,10,6,3$
respectively. We see a clear deviation from Heisenberg-like
behaviour in the thermal trends on decreasing coupling. Frequencies
and temperatures are scaled by the respective bandwidths ($W_{mag}$)
of the magnetization spectrum.
}
\end{figure*}

\subsection{Momentum dependence of energy and damping with
changing temperature}

In Fig.7, we concentrate on the momentum dependence of the same
two quantities in the three broad thermal regimes, discussed before.
We firstly see a monotonic behaviour of the peak frequency
(at ${\bf q}=(\pi/2,\pi/2)$), as
well as the finite $T$ bandwidth (scaled by $\tilde{J}$), 
on lowering $U$ in the weakly damped regime. 
The linewidths here are very small.
In the strongly damped regime (green curves), the peak
location of mean frequency shifts to slightly lower ${\bf q}$ at
weak coupling, while the peak in magnon damping shifts
towards higher ${\bf q}$ values. Finally, even in the
diffusive regime, a residual momentum dependence can be
observed in the linewidth plots ((d)-(f)).

\subsection{Lineshapes on the square lattice}

Fig.8 highlights the behaviour of a specific high-momentum
lineshape (at ${\bf q}=(\pi/2,\pi/2)$) as a function of frequency
for several temperatures. Fig.8(a) is the 
Heisenberg limit ($U/t=20.0$) result.
We see sharp mode gradually broadening and developing 
a tail-like feature upto $T/W_{mag}=0.1$ on increase in $T$.
Finally, a diffusive lineshape emerges at high temperature ($T/W_{mag}=0.25$).
The plots for $U/t=10.0$ shares most of these
qualitative features. However, the extent of broadening at intermediate
temperatures is much more at the same scaled temperatures for 
$U/t=6.0$. There's a zero frequency feature for weaker couplings, 
most prominent for $U/t=3.0$. As discussed already, this is an 
artifact of the present method and shouldn't be taken seriously.

We next move on to an example of a weakly frustrated system, the 
Hubbard model on the isotropic triangular lattice. 
This system has a \textit{finite} $U_{c} \sim 4.5t$ and features
120\degree ordered ground states for $U \gtrsim 6t$. 
We focus our attention to the latter coupling regime. 
First, the spectral maps are exhibited, followed by lineshapes 
at two specific momenta.

\section{Dynamics on the triangular lattice}

\subsection{Spectral maps for varying $U/t$ and temperature}

Fig.9 exhibits the spectral maps for the triangular lattice, in 
the same layout as in the square case. The four couplings 
represent "Heisenberg" ($U/t=20.0$), "strong" ($U/t=10.0$), 
"intermediate" ($U/t=8.0$) and "close to the transition" ($U/t=6.0$) regimes. 
The non-Heisenberg features like amplitude fluctuations 
and multi-spin couplings increase column-wise. 

The spectrum in the Heisenberg limit is much more
complicated than in the square case, as the background order
corresponds to ${\bf q}=(2\pi/3,2\pi/3)$ due to the effect of mild
frustration. We plot the spectrum along $\Gamma-K-M-\Gamma$
trajectory in the Magnetic Brillouin Zone (MBZ). 
There are two bands at a generic wavevector.
The magnetic order is fragile, as indicated by the reduced
bandwidth compared to the square case. 
Even on mild increase in $T$
($T/W_{mag}=0.2$), the multi-band structure becomes fuzzy and
large linewidths develop in the $M-\Gamma$ region. 
Further increase in $T$ 
makes most of the spectrum incoherent, apart from the
Goldstone mode at the ordering wavevector. 

Moving to the lower coupling counterparts, the strong 
coupling spectrum at low $T$ is similar to the Heisenberg result, 
with $J_{eff} \sim t^2/U$.
The dip near $M$ point is more prominent.
Thermal effects are also Heisenberg-like. 
On decreasing the coupling to $U/t=8.0$, 
the curvature of the $\Gamma-K$ branch increases at low $T$, as does the dip.
Amplitude fluctuations induce more dramatic damping of the spin-wave
modes at comparable temperatures. Finally, close to the Mott
transition ($U/t=6.0$), even the low-$T$ spectrum is incoherent.
Soft modes are visible in a wide region of momentum space.
In Appendix D, we show the gradual evolution of the low temperature
spectrum as one approaches the Mott transition, staying within the 
120\degree ordered family of states.  

\subsection{Lineshapes on the triangular lattice}

Fig.10 elaborates the comparison of detailed lineshapes of the Hubbard
model with those of the Heisenberg in the triangular case. The two
rows feature lineshapes for ${\bf q}=(\pi/3,\pi/3)$ and 
${\bf q}=(\pi,\pi)$ respectively. Once again, the frequencies
and temperatures are scaled with respect to the low $T$ bandwidth.
The leftmost columns represent the Heisenberg limit 
($U/t=20$) results. 
We observe that for both wavevectors, a bimodal spectrum is obtained
at low $T$, which gradually broadens on increasing temperature.
Even upto $T/W_{mag} \sim 0.1$, the spectra retain two distinct peaks.

Moving to the Hubbard results, we see that the strong coupling results
($U/t=10.0$) bear a striking resemblance to the Heisenberg case, as
expected. However, even at moderately high coupling ($U/t=8.0$), the
thermal damping results in diffusive behaviour even at
$T/W_{mag} \sim 0.05$. On going closer to the Mott transition
($U/t=6.0$), even the low $T$ lineshapes significantly change
their character, with prominent zero frequency weights cropping
up in both the wavevectors. Diffusive behaviour sets in immediately
on increasing $T$.

\section{Discussion}

We have tried to organise the results in this paper in terms of
three dynamical regimes and then quantified the detailed response
on these regimes in terms of the lineshape, the mode energy and
the damping. In what follows we shall try to provide the analytic
basis of some of the results seen in the Langevin simulations,
also point out some of the limitations of our approach. 
The main effect observed in this paper is the enhancement 
of thermal damping of magnons as one 
moves away from the Heisenberg limit. 
We argue this effect maybe 
minimally captured by a simpler classical toy model, 
which allows for amplitude fluctuations and approaches 
the classical Heisenberg limit upon tuning a single parameter.

\subsection{Classification of non-Heisenberg effects at finite $U/t$}

We first comment that there exists a two-particle 
continuum of excitations, originating from particle-hole processes, 
missed out by the present scheme. 
This is accessed by a quantum RPA calculation done on the 
mean-field ordered states on square and triangular geometries. 
However, this continuum is energetically well separated from the 
spin wave spectrum at strong coupling and hence don't influence
each other at the temperature scales of interest. But, this
argument breaks down at weak coupling (e.g. $U/t=3.0$), where
indeed there's appreciable mixing even at low temperature, 
and our dynamical results are indeed 
imperfect, except near special, symmetry-protected wavevectors 
like $(0,0)$ or $(\pi,\pi)$. In what follows, we only underline the 
non-Heisenberg features observed in the spin wave part.  

In the full Hubbard problem, at intermediate $U/t$ values, 
there are two main non-Heisenberg features- (i)~the ordered state
and the low $T$ dispersion are modified, and 
(ii)~the moment magnitudes are no longer fixed but
are reduced at low $T$ and also fluctuate thermally. We'll discuss
the impact of the second class of features in detail in the 
upcoming subsections. To obtain the effects of the first class 
systematically at low $T$, 
one does an expansion about the mean-field state, which may 
(as in the square lattice case) or may not (as in the triangular one) have
the same ordering as in the Heisenberg limit, with a reduced moment value. 
The effective Hamiltonian for ${\bf m}_{i}$'s, 
obtained through integrating out the electrons perturbatively 
in $t/U$, now involves longer range, multi-spin terms \cite{capriotti,yang}.
The couplings are decided by the electronic band structure 
on the mean field state. 
However, we should remember that our model is composed of 
\textit{classical} moments. Hence, the coefficients 
don't match with those in the actual quantum model. 

These coefficients depend non-trivially on $U/t$. 
As a result, the crossover lines 
between the thermal regimes are \textit{modified}
with respect to the Heisenberg case.  
 
To lowest order, a \textit{linear} theory 
maybe written down for the fluctuations, 
which has an analytic solution.
We'll discuss this subsequently in subsection C.
The contribution to the effective field 
($\frac{\partial <H>}{\partial {\bf m}_{i}}$) 
coming from the leading non-Heisenberg term, 
expanded upto ${\cal O} (\delta{\bf m}_{i})$ in fluctuations, looks like-
$$
\sum_{ijkl} K_{ijkl} ({\bf m}^{0}_{j}
({\bf m}^{0}_{k}.\delta{\bf m}_{l} + \delta{\bf m}_{k}.{\bf m}^{0}_{l})
+\delta{\bf m}_{j}({\bf m}^{0}_{k}.{\bf m}^{0}_{l}))
$$

The coupling $K_{ijkl}$ has a lowest order contribution of
${\cal O} (t^{4}/U^{3})$, as maybe motivated from a perturbative
argument, starting from the strong coupling limit. 
One now puts this expression back in the first and second terms of
Eq.1, along with the Heisenberg term $4t^{2}/U\sum_{<j>}{\bf m}_{j}$
and the stiffness contribution ($U(|{\bf m}_{i}|-1/2)^{2}$), 
and solves the resulting equation via Fourier transformation.
From the poles of the ensuing power spectrum, 
one gets the low $T$ dispersion, which contains the leading 
non-Heisenberg effects.

\begin{figure}[t]
\centerline{
\includegraphics[height=4.5cm,width=4.35cm]{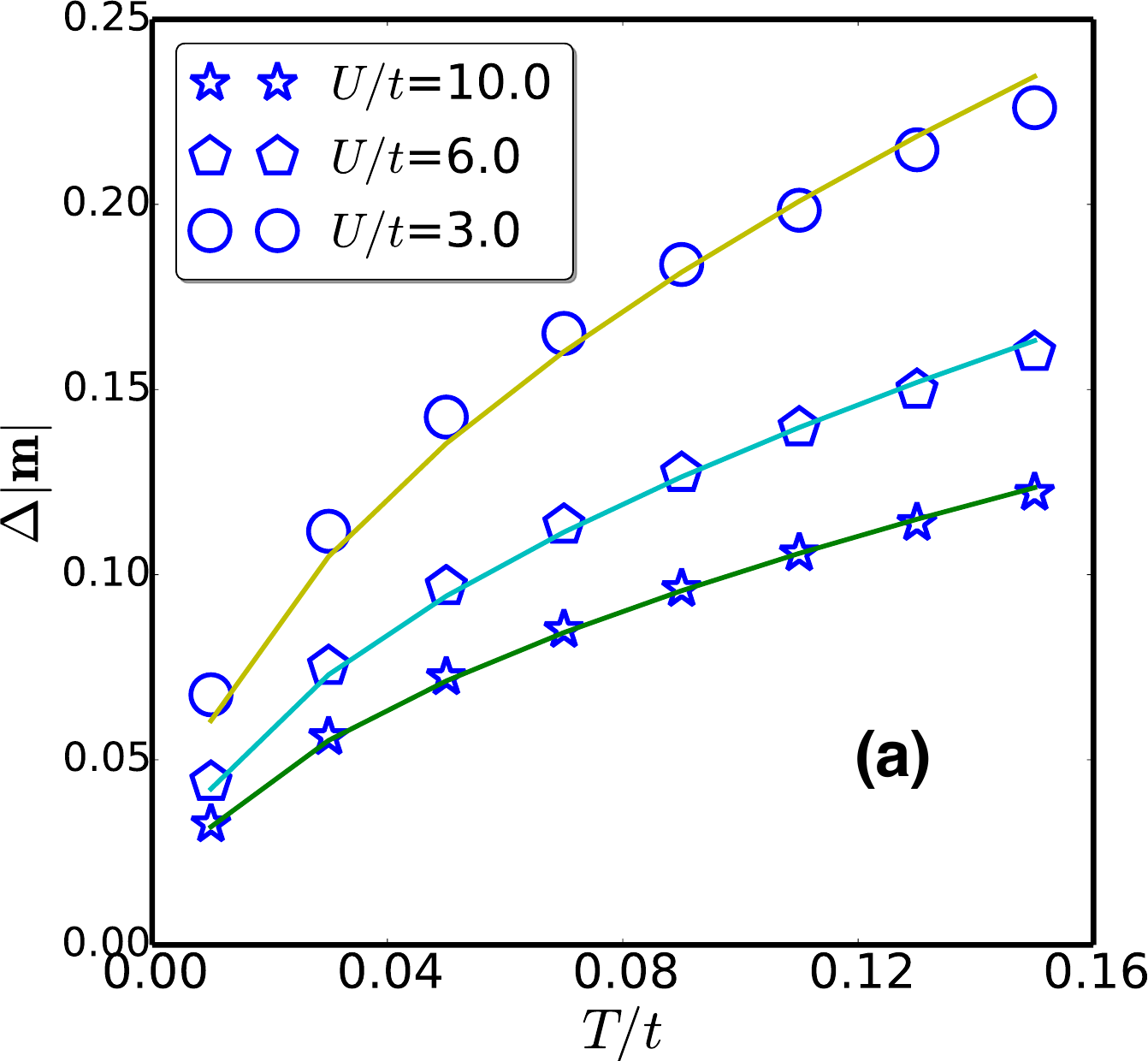}
\includegraphics[height=4.5cm,width=4.35cm]{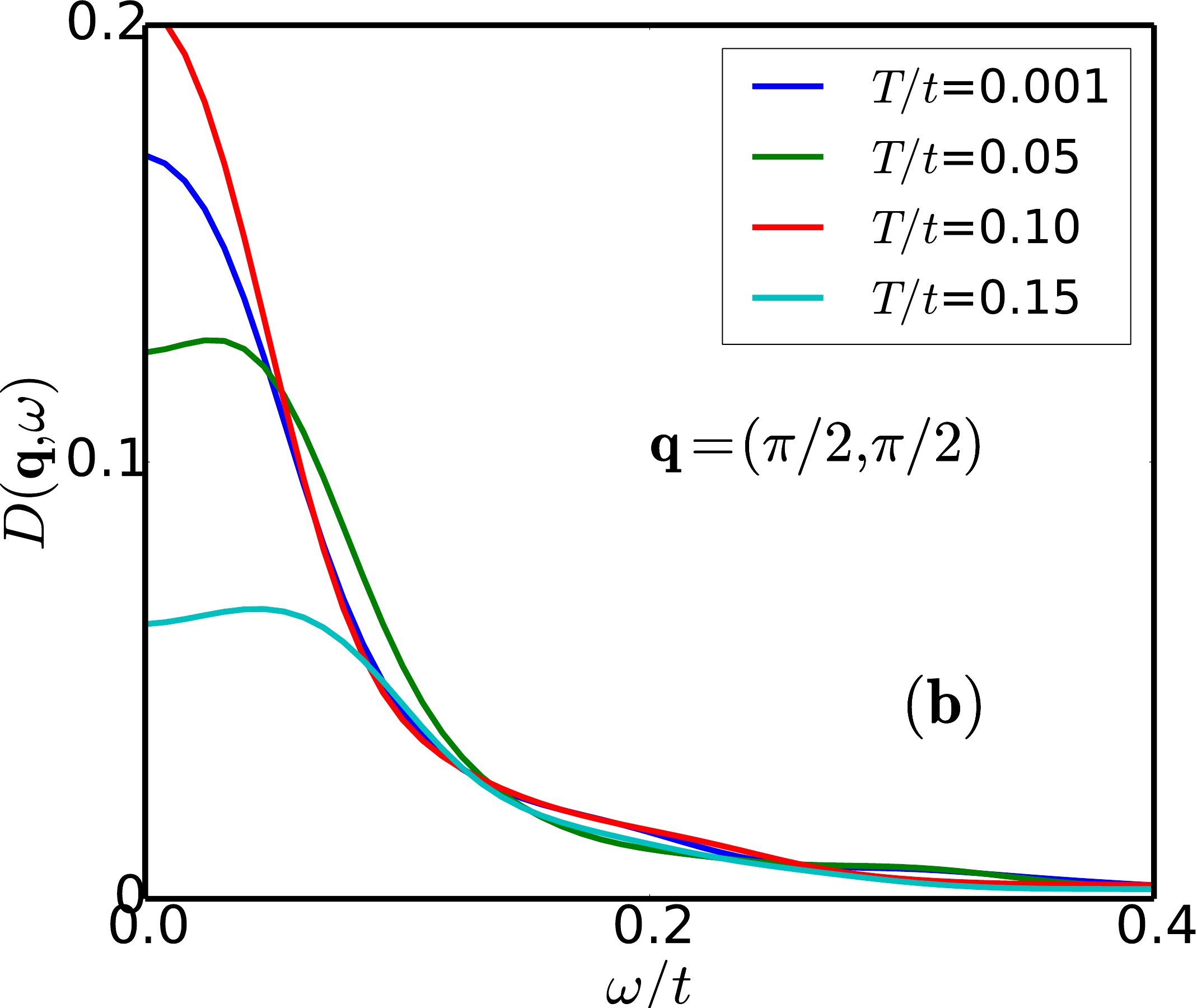}
}
\caption{(a): Fitted standard deviations ($\Delta|{\bf m}|$)
from $P(|{\bf m}|)$ distributions, plotted against temperature
for three couplings in the square lattice case.
Blue open circles denote actual data points, while
solid lines are fits using a square root function.
The trends indicate the increasing importance of amplitude
fluctuations at weaker couplings and a square root dependence,
expected of a "soft spin" Heisenberg model.
(b): Lineshapes at ${\bf q}=(\pi/2,\pi/2)$ for the
amplitude fluctuations at $U/t=6.0$, indicating a diffusive mode
centered at zero energy.
}
\end{figure}

\subsection{Quantifying amplitude fluctuations}

In this subsection, we quantify the extent and \textit{intrinsic}
dynamical signature of fluctuations in the moment magnitude, before 
launching into the construction of an effective model to describe them.
Fig.11(a) focusses on the longitudinal fluctuations of the magnetic
moments. These are, of course, frozen in the Heisenberg limit. 
We fit the $P(|{\bf m}|)$ distributions, shown earlier in Fig.1,
to Gaussians and extracted the corresponding standard deviations.
These are plotted as functions of temperature for various
coupling values in the square lattice case. In a "soft spin"
Heisenberg model, where the intersite term is Heisenberg but
longitudinal fluctuations are allowed, the behaviour should
be $\sim\sqrt{T}$. However, we observe deviations from this
trend at lower $U$ values. The coefficient of the
square root fits is exactly $1/\sqrt{U}$ at strong 
coupling. Even at weaker couplings, the deviations are
small. Hence, the amplitude fluctuations can be effectively 
captured by a local term $H_{amp}=\sum_{i}U(|{\bf m}_{i}|-1/2)^{2}$.

The spectral signature of these fluctuations is a diffusive mode
centered at zero frequency, shown in Fig.11(b). This is obvious 
from the locality of $H_{amp}$, which deactivates the torque term in Eq.1. 
The width is regulated by $\gamma$. Interestingly, the weight at
low frequency shows a non-monotonic behaviour with $T$. 
This behaviour, however, doesn't capture the true physics 
of the amplitude mode, which should have a signature at
$\omega \sim U$. For that, one needs to incorporate 
\textit{quantum} fluctuations of the magnetization field in
the effective equation of motion. We'll discuss this briefly
in subsection E.

\subsection{Construction of an effective model}

In the following, we describe the construction of an effective 
"classical moment" model, which essentially captures the 
qualitative features of the full Hubbard model calculation at all $U/t$. 
The model reads-
\begin{eqnarray}
H_{eff}&=&J_{eff}\sum_{<ij>}{\bf m}_{i}.{\bf m}_{j} + \frac{K_{eff}}{2}\sum_{i}
(|{\bf m}_{i}|-|{\bf m}_{HF}|)^2 \nonumber \\ 
&& - 2J_{eff}\sum_{i}|{\bf m}_{i}|^2
\end{eqnarray}

\begin{figure*}[t]
\centerline{
\includegraphics[height=4.25cm,width=5.5cm]{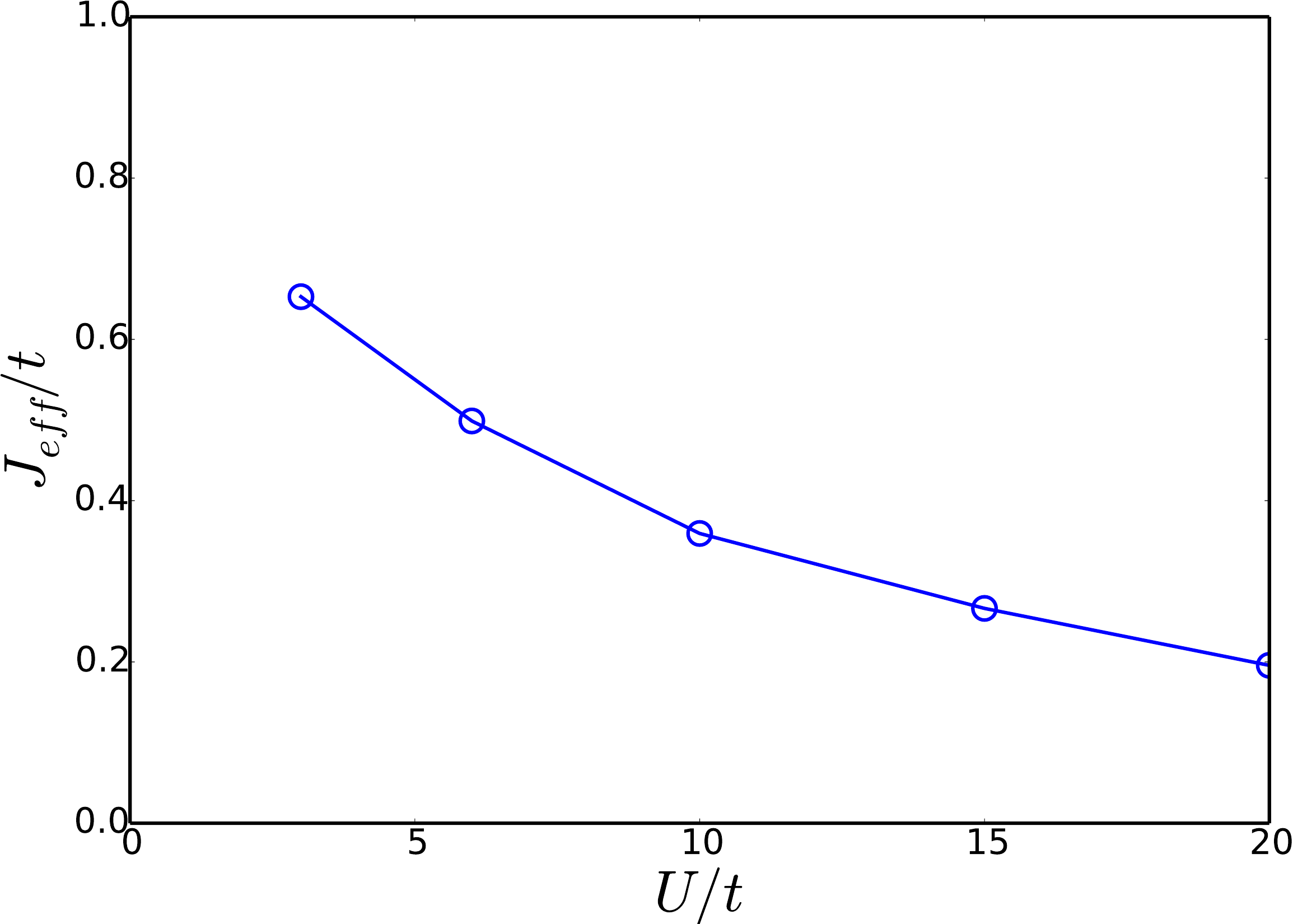}
\includegraphics[height=4.25cm,width=5.5cm]{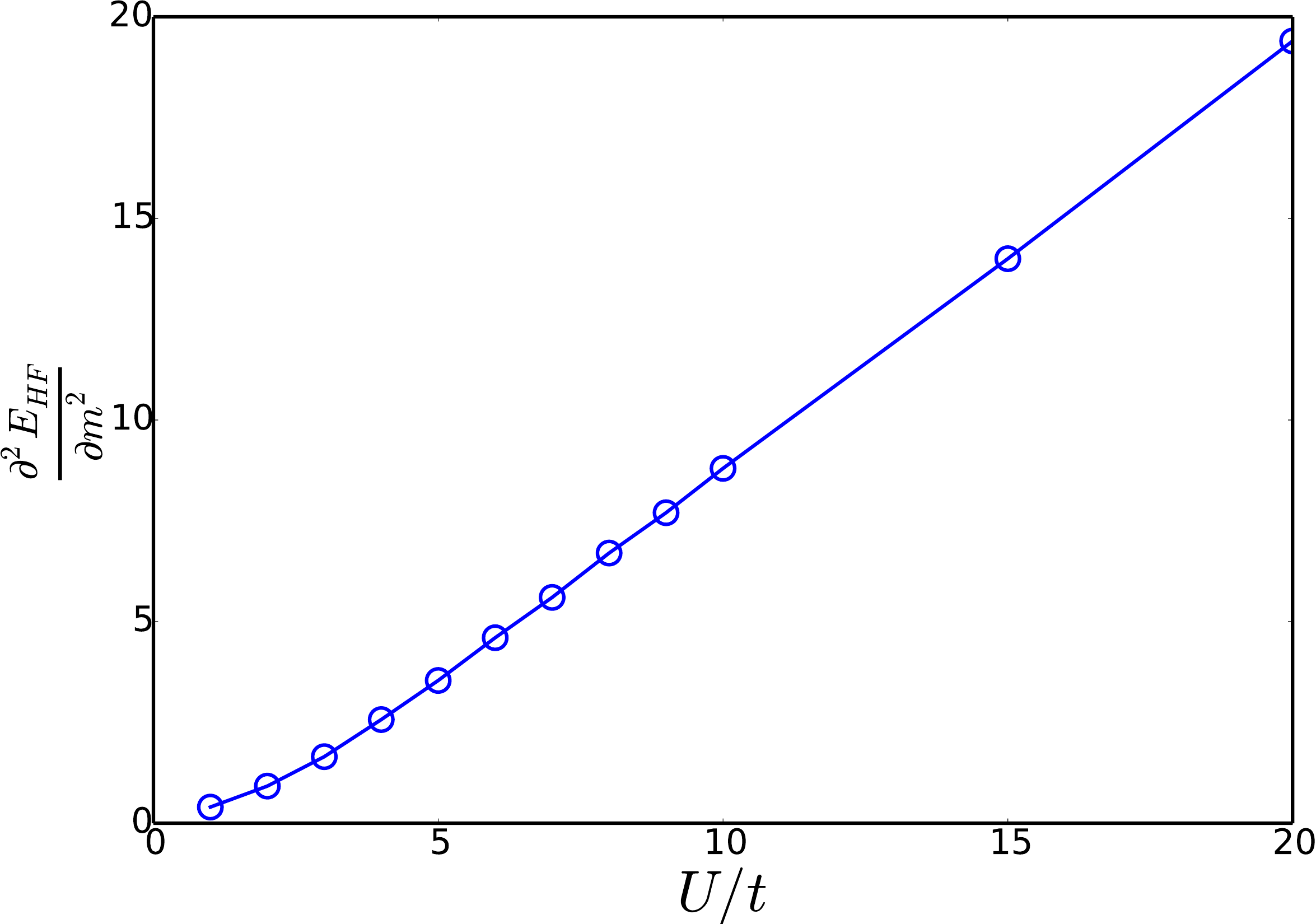}
\includegraphics[height=4.25cm,width=5.5cm]{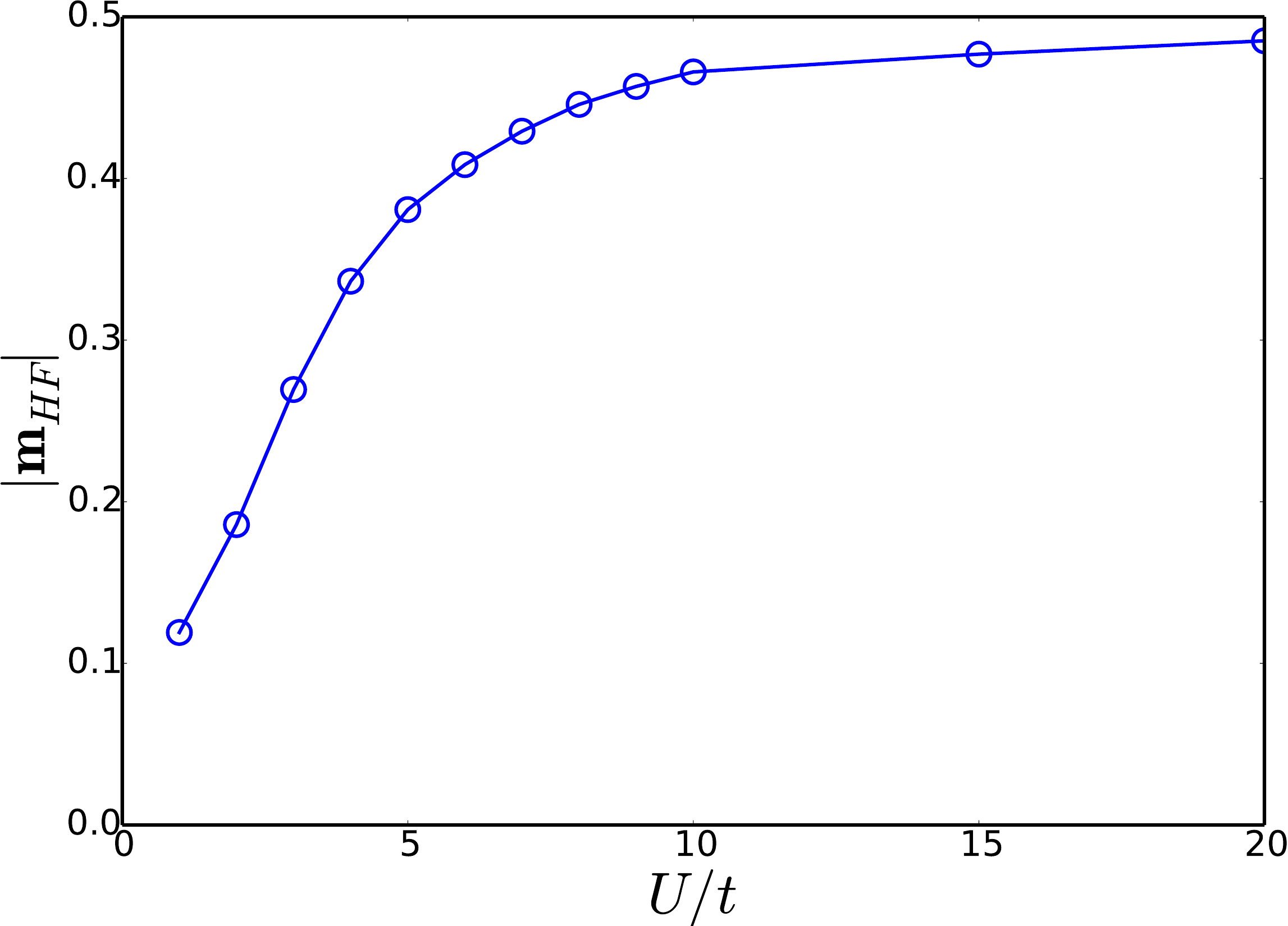}
}
\caption{The effective exchange $J_{eff}$, second derivative of Hartree-Fock 
energy with respect to moment magnitude 
($\frac{\partial^{2}E_{HF}}{\partial{m}^2}$), which is proportional to 
the amplitude stiffness $K_{eff}$ and 
Hartree-Fock moment value ($|{\bf m}_{HF}|$), 
as determined from HF and RPA calculations, for 
various $U/t$ values on the square lattice Hubbard model.}
\end{figure*}

The first term encapsulates an "effective" nearest neighbour exchange between 
the local moments ${\bf m}_{i}$, the second term is an amplitude stiffness 
which regulates the thermally induced fluctuations of the moment magnitude 
and the third term is a counterterm that fixes the low $T$ moment size to 
exactly $|{\bf m}_{HF}|$, the Hartree-Fock value. The parameters $J_{eff}$ and 
$K_{eff}$ are extracted, respectively, 
from the low $T$ RPA spin wave velocity (fitted to a 
nearest-neighbour Heisenberg model) and the ``curvature'' of the 
Hartree-Fock energy,  ${\partial^{2}E_{HF}}/{\partial{m}^2}$.
Fig.12 illustrates the behaviour of the above parameters for various 
$U/t$ values.

\begin{figure}[b]
\centerline{
\includegraphics[height=4.2cm,width=4.4cm]{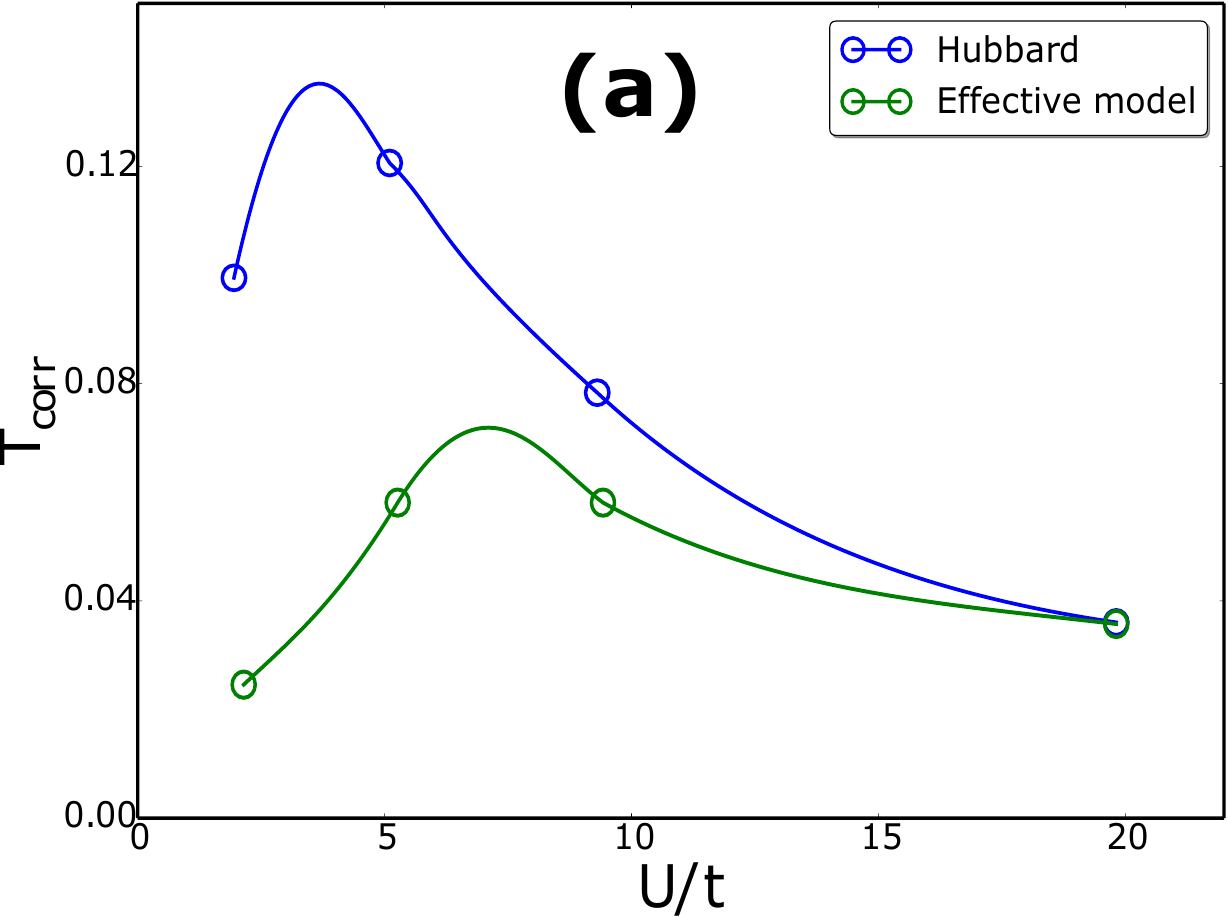}
\includegraphics[height=4.2cm,width=4.2cm]{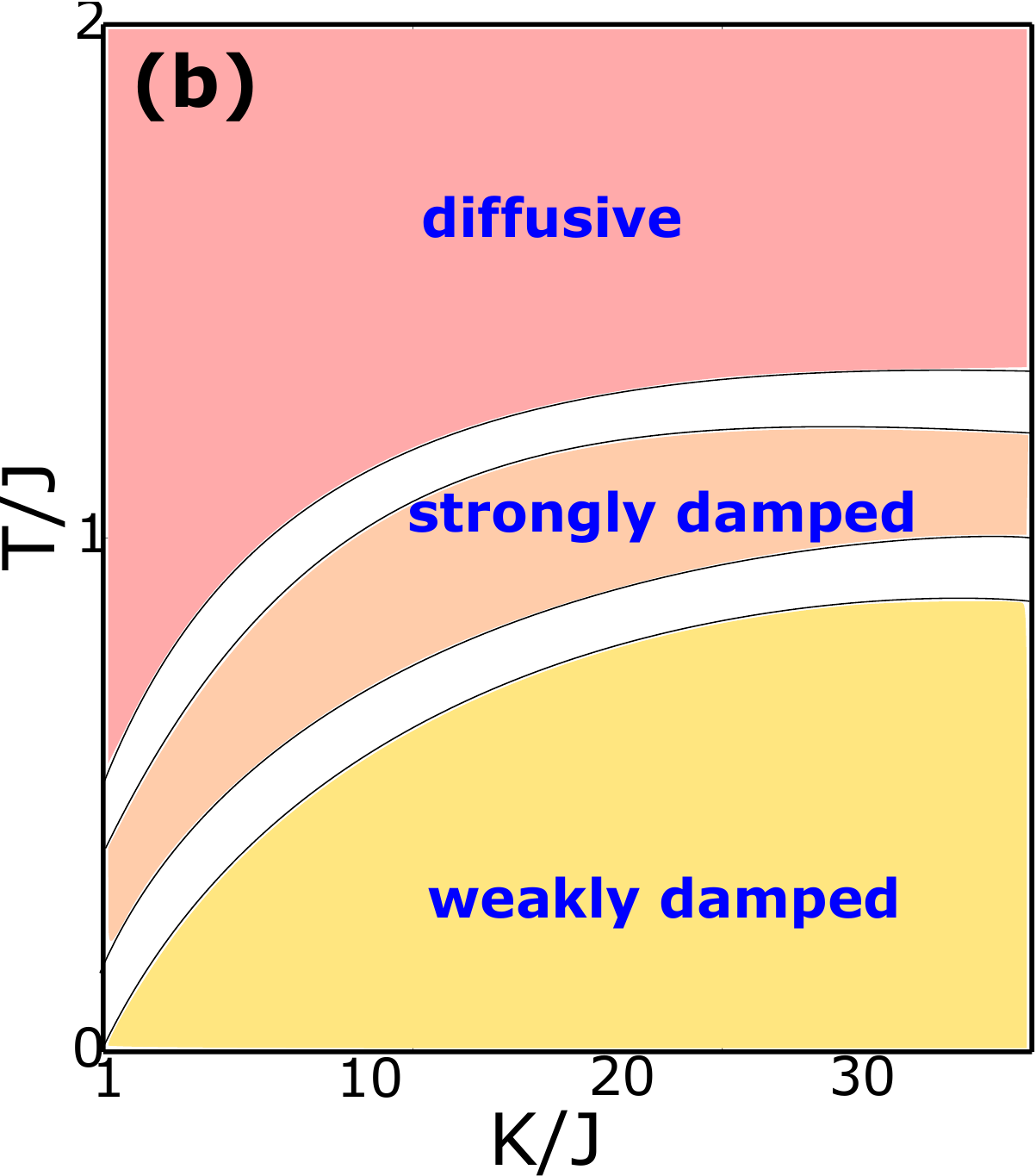}
}
\caption{Left: Comparison of the correlation temperatures ($T_{corr}$), 
extracted from the respective structure factors $S(\pi,\pi)$ of the 
full Hubbard (blue curves) and effective model (green curves) 
obtained using Monte Carlo (MC) method described in the paper. 
One observes that the non-monotonicity is well captured by the former
model. Right: Thermal regimes obtained using Langevin dynamics of 
the effective model (Eq.10) with varying $K/J$. 
A qualitative resemblance with the square lattice Hubbard 
results (Fig.4(c)) is apparent.}
\end{figure}

The model is constructed based on a strong coupling expansion 
argument.
At large $U/t$, the Hubbard model reduces to a spin model of the 
following form-
\begin{eqnarray}
H_{eff} &=& H_{loc} + H_{coup} \cr
\cr
H_{loc}~ &=& U(|{\bf m}_{i}|-\frac{1}{2})^2 + ... \cr
\cr
H_{coup} &=& J_{2}\sum_{<ij>}{\bf m}_{i}.{\bf m}_{j} + 
J_{4}\sum_{ijkl}f[{\bf m}_{i},...{\bf m}_{l}] + ... 
\end{eqnarray}

$H_{loc}$ is basically the HF energy
in terms of moment magnitude, expanded to quadratic order in the deviations.
$H_{coup}$ reduces to the first term with $J_{2}=4t^{2}/U$ as 
$U/t\rightarrow\infty$.
This can be shown explicitly by expanding 
about the $U/t \rightarrow \infty$ local limit. 
On including further terms 
in the expansion (subleading in $t/U$), 
one gets longer range, multi-spin couplings.
We lump the effect of all non local terms into an equivalent 
nearest neighbour coupling $J_{eff}$ and retain the local amplitude stiffness
in our simplified model. The strong coupling limit is also correctly 
recovered as $J_{eff} \rightarrow 4t^2/U$, $K_{eff} \rightarrow 2U$ and 
$|{\bf m}_{i}| \rightarrow 1/2$ as $U/t \rightarrow \infty$ in our model. 
The result of the aforesaid construction is that it reproduces the thermal
physics of the \textit{classical} Heisenberg model at all $T/t$ 
for large $U/t$. At weaker couplings, the $T=0$ 
state is captured with the correct (mean-field) moment value and the 
low-energy spin wave excitations (in particular their velocity $v_{SW}$) 
are also correctly captured by construction.

\begin{figure*}[t]
\centerline{
\includegraphics[height=4.7cm,width=5cm]{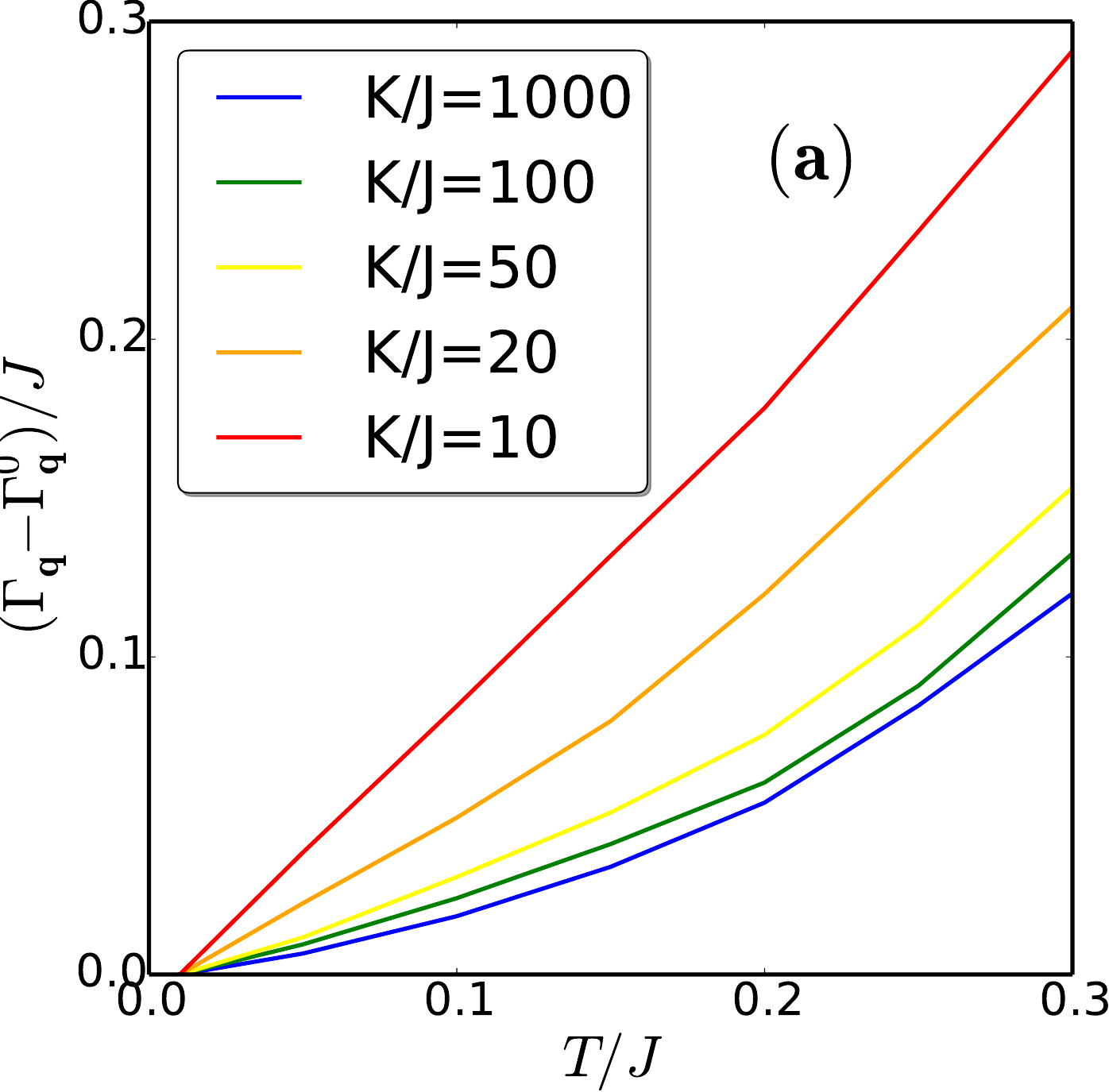}
\hspace{0.2cm}
\includegraphics[height=4.7cm,width=5cm]{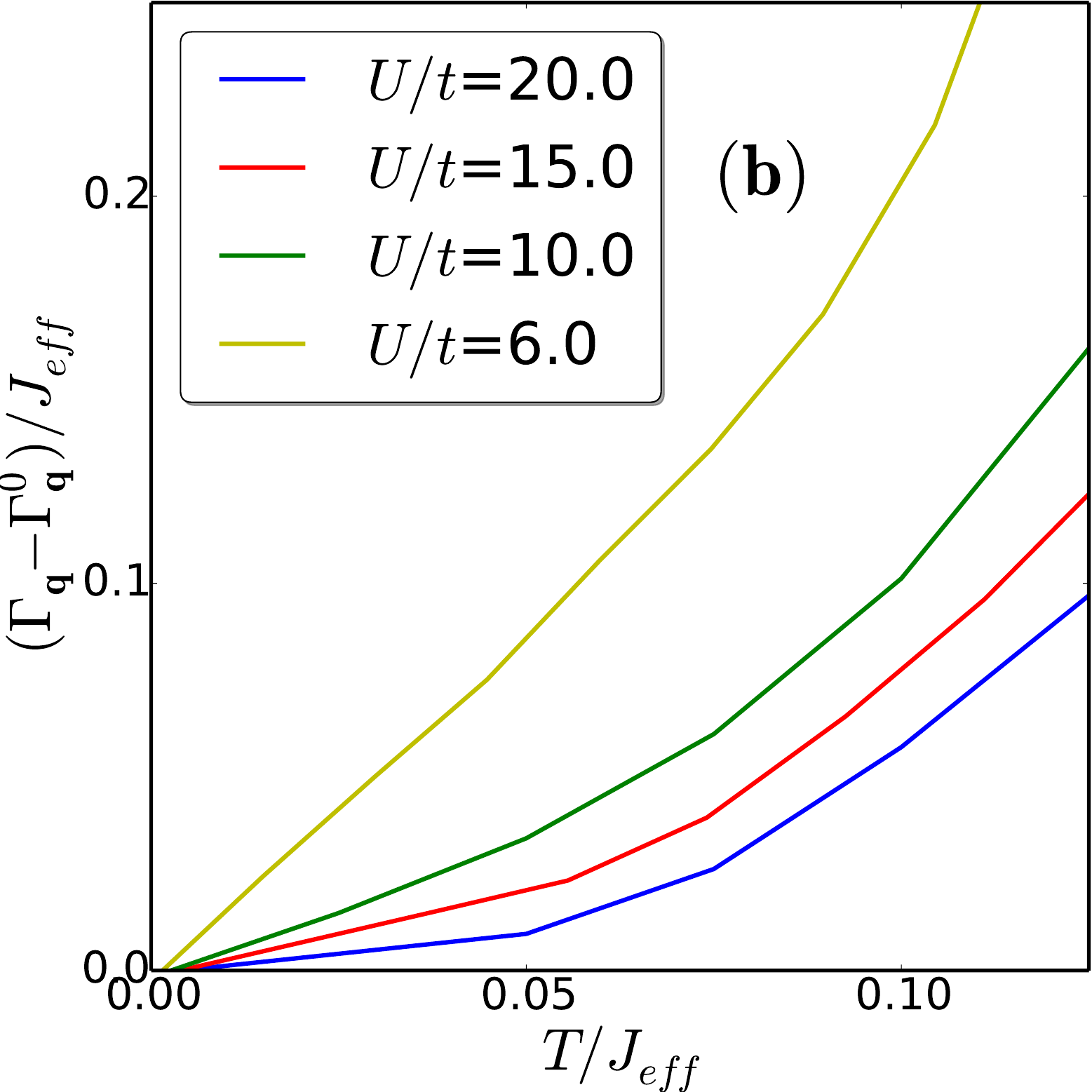}
\hspace{0.2cm}
\includegraphics[height=4.7cm,width=5cm]{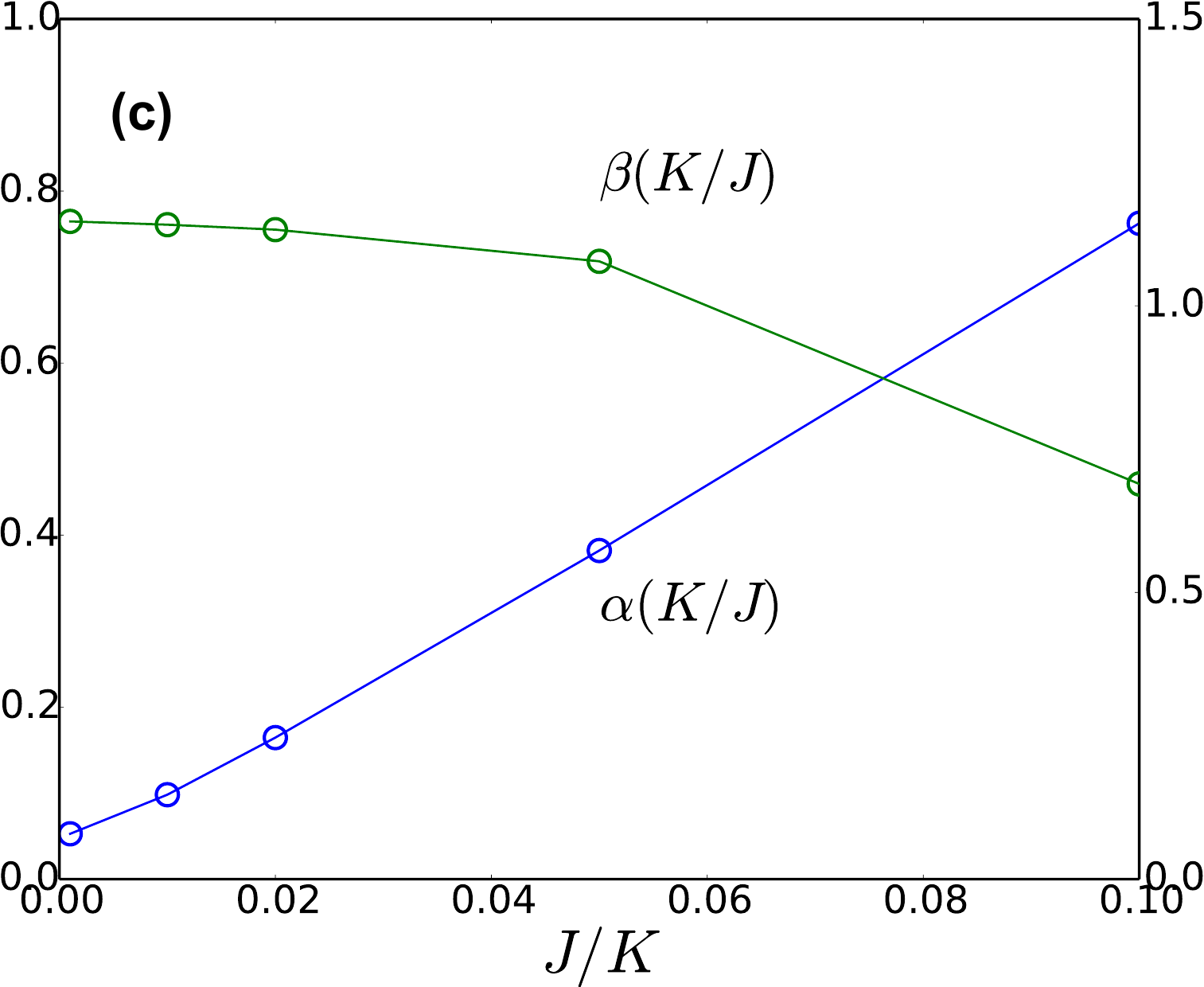}
}
\caption{(a): The excess thermal damping ($\Gamma_{\bf q}-\Gamma^{0}_{\bf q}$), 
plotted as a function of $T/J$ for various stiffness values in the 
approximate $J-K$ model for ${\bf q}=(\pi/2,\pi/2)$. 
One notes that the low $T$ linear regime shrinks on increasing $K/J$
and the behaviour turns to parabolic.
(b): The same quantity extracted from the full Hubbard model calculation
at various $U/t$ values. Similar qualitative features are observed.
(c): Plot of fitting parameters $\alpha$ and $\beta$ for the approximate 
model, showing the quadratic to linear crossover on decreasing $K/J$.
}
\end{figure*}

As regards the results obtained using the above model, 
we first compare the static indicators, in particular the low temperature 
structure factor $S(\pi,\pi)$ between the original Hubbard model and 
this effective model at various $U/t$ values. 
To minimize parametric dependencies, the comparison 
was done using the Monte Carlo technique, elaborated in Appendix B. 
The results for the correlation temperatures ($T_{corr}$) are shown in 
Fig.13(a). The basic observation is 
that the non-monotonicity of this scale as a function of $U/t$, is 
succesfully captured by the effective model, albeit the maximum is 
slightly shifted to higher $U/t$.
The $T_{corr}$ within the effective
model scales roughly as $\sim |{\bf m}|_{HF}^2 J_{eff}$ for large $U/t$, 
but crashes faster at lower $U$ due to the effect of $K_{eff}$.

To further simplify the three parameter effective model of Eq.8, we scaled 
the effective couplings $J_{eff}$ and $K_{eff}$ by the moment value 
$|{\bf m}_{HF}|$ appropriately and reduced Eq.8 to an 
"equivalent one-parameter" model of the following form-

\begin{equation}
H_{1par}=J\sum_{<ij>}{\bf m}_{i}.{\bf m}_{j} + \frac{K}{2}\sum_{i}
(|{\bf m}_{i}|-1)^2 - 2J\sum_{i}|{\bf m}_{i}|^2
\end{equation} 

where $J$ is set to 1 and $K/J$ is varied to mimic the behaviour of the 
earlier model. The moment magnitudes fluctuate about unity for all couplings
in this model. The results obtained using Eq.10 agree quantitatively 
with those originating from Eq.8, which is formally equivalent. 

Next, we move to the dynamics. The thermal regimes in the dynamics of
the effective model (Eq.10) are depicted in Fig.13(b). They qualitatively
resemble the scaled phase diagram (Fig.4(c)) of the full Hubbard problem. 
This corroborates the usefulness of the effective model, not only to 
understand the static properties, but also dynamical features.

After comparing the gross features of the dynamics, we also examined
whether the same effective model (Eq.10) can mimic the changing 
low $T$ behaviour of the damping in the full Hubbard problem. 
We extracted the excess damping at finite $T$ and 
plotted it for the generic ${\bf q}=(\pi/2,\pi/2)$ as a function of $T/J$.
One finds that \textit{empirically} one may fit
this excess damping $\Gamma_{\bf q}-\Gamma^{0}_{\bf q}$ 
to a polynomial of the form $\alpha T + \beta T^{2}$, 
with the coefficients depending on $K/J$.
 
Upon examining the fitting parameters, one observes that the 
$\alpha \propto 1/K$ at low $K$ and decreases to zero in the fixed 
moment limit ($K/J \rightarrow \infty$).
The quadratic coefficient $\beta$ is roughly constant at large $K$. 
The results are shown in Fig.14(a) and 14(c).
Such features are also observed qualitatively 
in the full Hubbard calculation, where the normalizing energy scale 
is chosen as $J_{eff} = 4t^2/U$. These results are shown in Fig.14(b).

We next try to find an \textit{a posteriori} justification for
the rising linear coefficient and rise in damping as
on reduces the amplitude stiffness by imagining
undamped spin wave modes getting affected by amplitude disorder. 
If one is at sufficiently low temperature, the equation of motion 
(Eq.1) maybe linearized in terms of deviation from the ground state 
configuration. On the square lattice, for instance, 
one simply expands the ${\bf m}_{i}$ as
\begin{eqnarray}
{\bf m}_{i}&=&{\bf m}^{0}_{i}+\delta{\bf m}_{i} \cr
{\bf m}^{0}_{i}&=&(-1)^{i_{x}+i_{y}}\hat{z}
\end{eqnarray}
Keeping upto the linear order in fluctuations $\delta {\bf m}_{i}$
gives us an analytically solvable starting point.
The effective equation is- 
\begin{eqnarray}
\frac{d \delta {\bf m}_{i}}{dt}&+&
J({\bf m}^{0}_{i}\times \sum_{<j>}\delta{\bf m}_{j}
-\sum_{<j>}{\bf m}^{0}_{j}\times \delta{\bf m}_{i}) \cr
&+&\gamma (J\sum_{<j>}\delta{\bf m}_{j}
+K\sum_{i}(-1)^{i}\delta m^{z}_{i}\hat{z})
={\vec \xi}_{i}
\end{eqnarray}

The transverse and longitudinal modes gets 
decoupled at this order. On Fourier transforming this equation and
solving for the power spectrum, one finds the usual dispersion 
of the antiferromagnetic classical Heisenberg model, while the damping 
of transverse spin wave modes is limited by $\gamma J$. 
The longitudinal modes generally give rise to a diffusive lineshape, 
and freeze for $K/J \rightarrow \infty$. 
On top of this low temperature, purely transverse theory, 
one may switch-on amplitude fluctuations perturbatively. 
The width of these fluctuations is $\propto 1/K$. 
On treating them as static, uncorrelated disorder, 
they cause the eigenmodes of the linear theory to scatter.
In the lowest order Born approximation, this generates 
a self-energy, whose imaginary part translates to an 
additional contribution to the magnon linewidth. 
This has a prefactor $T$ coming from the propagator of transverse fluctuations. 
In the static limit, the coefficient of this correction is thus 
proportional to $T/K$. Hence as $K$ is reduced from infinity, 
the linear $T$ correction to spin wave damping increases as $1/K$, 
as is seen in the numerical data. 

The aforesaid argument doesn't include the effect of non-linear 
interactions among the transverse fluctuations. 
To evaluate their effect, one expands upto second order in 
the deviation field, which generates a $\delta {\bf m}_{\bf q} \times 
\delta {\bf m}_{\bf q^{\prime}}$ contribution in the equation of motion. 
If one substitutes the lowest order solution in this and 
averages over the noise, this correction term vanishes, 
owing to the fact that the noise is uncorrelated between 
different Cartesian axes.
Hence, no $O(T)$ contribution is found for 
the damping of transverse fluctuations. 
The lowest order correction is of ($O(T^2)$), 
as is found in the extensive literature \cite{harris,tyc}. 
This becomes the leading term when amplitude fluctuations 
are completely restricted (in the $K/J \rightarrow \infty$ limit).

\subsection{Computational issues for frustrated systems}

One would want to ultimately apply this formalism to study the Hubbard
model on fully frustrated geometries (e.g. Kagome in 2d and 
pyrochlore in 3d). The rich spin dynamics, with the moment 
softening and multipsin coupling effects present beyond the
Heisenberg limit, should be accessible at finite temperature.
However, there are some tough computational difficulties 
associated with this attempt. Briefly, the issues are- 

\begin{itemize}
\item Extracting even the static properties correctly 
(vis-a-vis Monte Carlo) requires much longer run lengths
compared to the square or triangular case. This occurs 
due to the rugged free energy landscape associated with
the problem. Novel strategies, involving simultaneous
updation of multiple moments, ameliorate the situation
in specific cases.
\item The numerical implementation of the Langevin dynamics
scheme, using Suzuki-Trotter decomposition, breaks down when
the systematic torque on a site becomes identically zero.
This happens, for instance, for the Heisenberg model
on the 2d Kagome lattice. Hence, a more complicated 
discretization strategy is called for.
\end{itemize} 

\subsection{Adiabaticity and thermal noise}

\subsubsection{The adiabatic assumption}

Our approach has assumed that the characteristic timescale for
magnetic fluctuations is much greater than electronic timescales, 
in analogy with the electron-phonon problem \cite{sauri}.
In such a situation (i)~the electronic energy depends only on
the instantaneous magnetic configuration, and (ii)~the
leading contribution to electronic correlators can be computed 
without invoking retardation effects.
This argument holds good in the strong coupling regime, 
where the magnetic fluctuations operate on a scale of
$J_{eff} \sim t^{2}/U$ and the electrons are gapped at a scale $\sim U$.
However, as $U/t$ reduces, the former scale rises and the latter
diminishes due to closing of the gap. So, the argument isn't very good.
We also comment that the auxiliary field correlator, which we computed, 
reproduces the essential features of the real spin-spin correlator
$\langle {\bf \sigma}_{i}(t).{\bf \sigma}_{j}(t^{\prime}) \rangle$, 
measured in INS experiments as long as the adiabaticity assumption holds good.
This happens because the auxiliary field dynamics basically follows
the ${\bf \sigma}_{i}$ field, with the distinction that its magnitude
is not strictly bounded between 0 and 1. As a result, the respective 
intensities are different.

\subsubsection{The noise driving the dynamics}

The present method for accessing spin dynamics excludes the
effect of quantum fluctuations. This firstly results in the unphysical
\textit{freezing} of the moments at $T=0$ and makes the method unable to
access the ground state magnon spectrum. Furthermore, this feature 
limits the viability of the scheme at low temperatures for frustrated
geometries, where order by disorder phenomena are observed. 
To remedy this, the noise has to be consistently
generated with respect to the polarizability of the problem, which
itself will depend on the ${\bf m}_{i}(t)$ trajectories.

Using a Keldysh formulation of the original Hubbard model, and
decomposing the interaction term using an auxiliary vector field
${\bf m}_{i}$, we may subsequently assume this field to be slow
with respect to the electrons. This enables one to write an effective
equation of motion for ${\bf m}_{i,cl}$ of the following form-

\begin{eqnarray}
\Im\left[Tr\left(\hat{G}^{K}_{ii}(t,t)\vec{\sigma}\right)\right]
&=&{\bf m}_{i,cl}(t) + \vec{\xi}_{i}(t) \nonumber\\
\langle\xi^{a}_{i}(t)\xi^{b}_{j}(t^{\prime})\rangle &=& 
\left[\hat{\Pi}^{K}(t,t^{\prime})\right]^{ab}_{ij}
\end{eqnarray}

Here $G^{K}$ and $\Pi^{K}$ are the Keldysh Green's function and
(spin-dependent) polarizability of the electrons respectively.
In the adiabatic limit, each of these maybe expanded in a 
Kramers-Moyal series \cite{arijit}. On assuming that the coefficients
don't have any spatial dependence and the temperature is high 
enough compared to characteristic frequency scale of these, 
one arrives at a much simpler equation of the LLG form, 
which upon neglecting certain multiplicative noise 
terms reduces to Eq.1. 

To include the effect of quantum fluctuations, the high $T$ approximations
done on the coefficients of the Kramers-Moyal expansion need to be relaxed.
Basically, if the temperature approaches the energy scale of 
two-particle excitations, the memory-less assumption on the noise becomes
unjustified.

\section{Conclusions}

We've studied the dynamics of magnetic moments in the Mott insulating
phase of the half-filled Hubbard model on square and triangular
lattice geometries, using a Langevin dynamics
based real time technique. 
The method reproduces known results on 
the Heisenberg model in the strong coupling limit, and the RPA based
low-energy dispersion at low $T$ faithfully.  
We observe three broad regimes in the dynamics- 
(i)~weakly damped, where spin waves are dispersive and 
dampings are small, (ii)~strongly damped, where one can
see significant broadening due to mode coupling, but the
dispersive character survives, and (iii)~diffusive, where
the mode frequencies collapse to zero and the dampings span
the full bandwidth. The main results are twofold- (a)~ 
we obtain the deviation of low temperature dispersion from 
the Heisenberg results, and
(b)~we observe the onset of the thermal crossovers at significantly 
lower values of $T/J_{eff}$, compared to the Heisenberg case.
One also captures the effect of mild geometric frustration
on the mode damping, on going from the square to the triangle.
The method maybe applied to study equilibrium dynamics in 
fully frustrated lattices (e.g. pyrochlore) in near future.

We acknowledge use of the High Performance Computing Facility
at HRI.

\section*{Appendix A: Numerical details of the Langevin scheme}

All of our Langevin dynamics simulations are done by discretizing 
Eq.1 in real time and implemented in a Cartesian coordinate scheme.
The particular technique used to solve the equations is the Euler-Maruyama
method \cite{euler}. The time step is chosen to be $0.01 \tau_{mag}$.
At each step, the derivatives appearing in the RHS of Eq.1 are 
computed through exact diagonalization of the electronic problem.
The derivative $\frac{\partial \langle H_{SF} \rangle}{\partial{\bf m}_{i}}$ 
for our model is just $U({\bf m}_{i} - \langle {\bf \sigma}_{i} \rangle)$. 
Typically, the simulations are ran for $3\times 10^{6}$ steps.
We gave parallel runs for each temperature point, with the 
Hartree-Fock (HF) state as the initial condition for each value of 
the Hubbard coupling.  
The lattice size for the results shown for both the square and 
triangular cases is $18 \times 18$.

\begin{figure}[b]
\centerline{
\includegraphics[height=5cm,width=5.5cm]{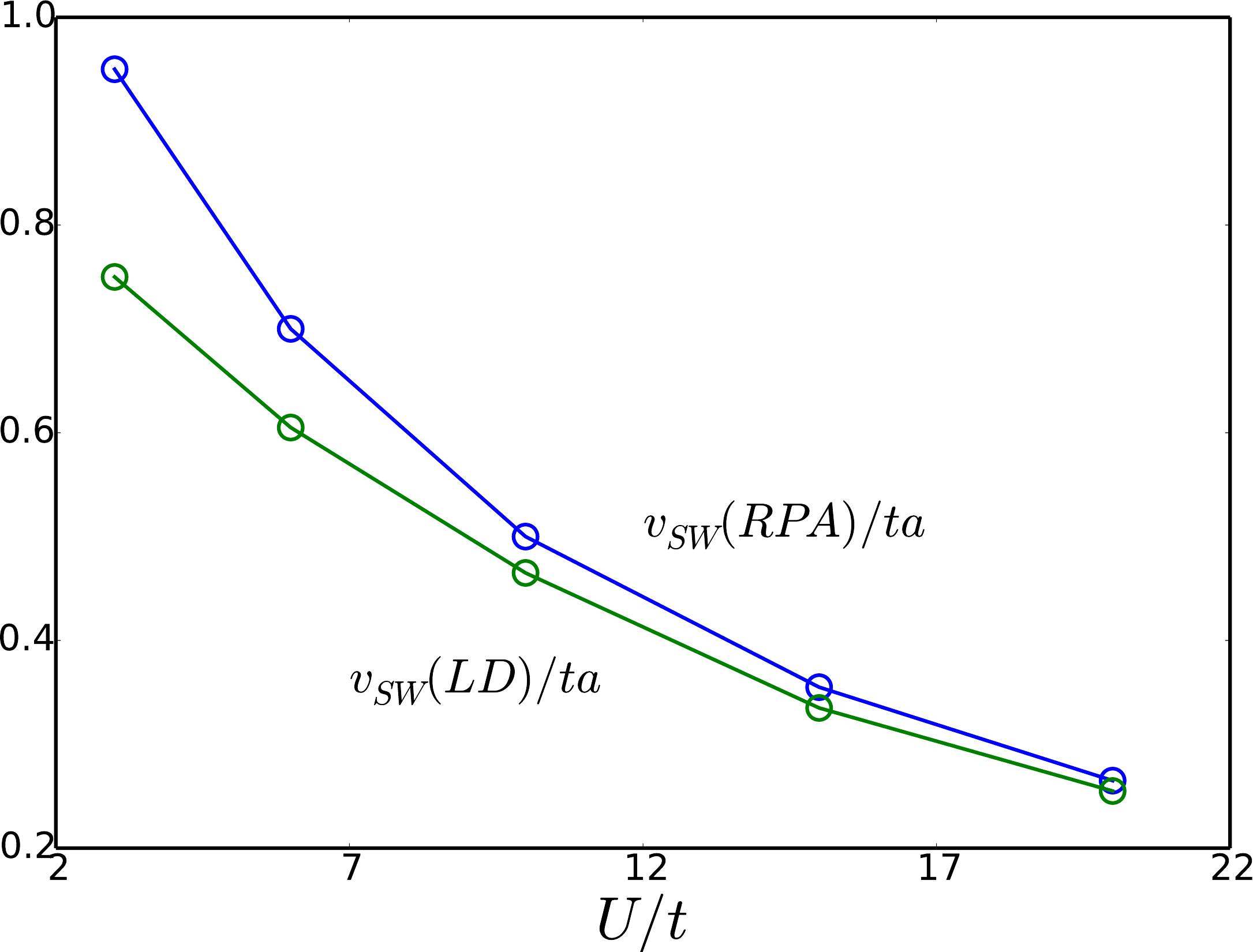}
}
\caption{Comparison of spin wave velocities ($v_{SW}$) computed using
our Langevin dynamics (LD) technique and the random phase approximation
(RPA) on the square lattice. We observe similar trends and quantitatively
lower values in LD compared to RPA. This is due to our assumption of 
classical spins.
}
\end{figure}

\begin{figure*}[t]
\centerline{
\includegraphics[height=3.6cm,width=15.7cm]{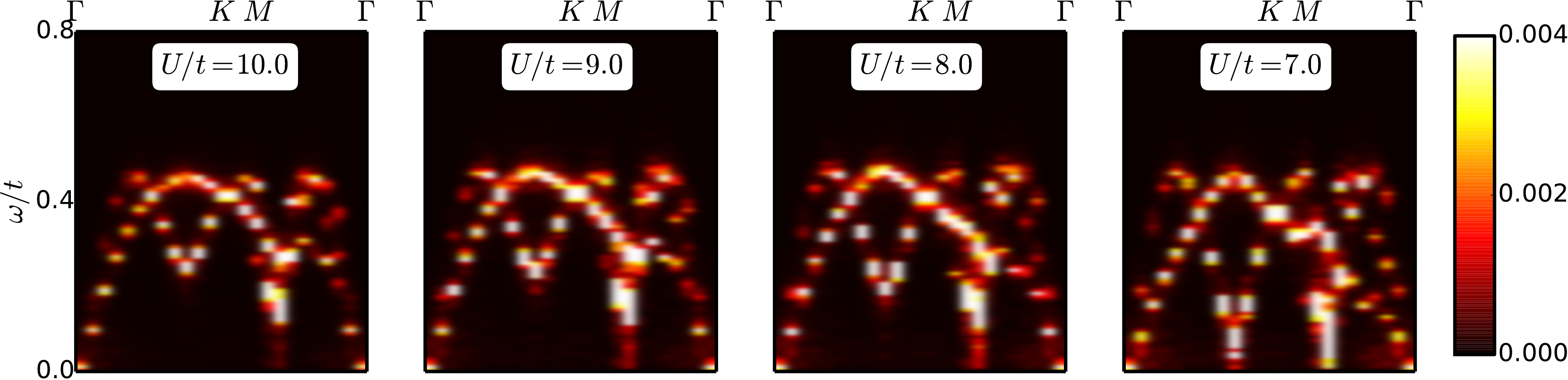}
}
\caption{Low temperature spectra on the triangular lattice on 
gradually lowering $U$, approaching the Mott transition. All
the couplings shown display order at zero temperature, with 
progressively smaller moment magnitudes. One observes a dramatic
softening of modes along the $\Gamma-K$ region in momentum space, 
albeit with a robust magnon bandwidth. 
}
\end{figure*}

\section*{Appendix B: Numerical details of the Monte Carlo scheme}

To benchmark the static properties obtained via the Langevin scheme, 
we used a competing Monte Carlo (MC) method. 
One first writes the Hubbard model in the Matsubara formalism 
and then decouples the quartic interaction in terms of the ${\bf m}_{i}$
field. Next, only the zero Matsubara mode of this field is retained, 
assuming $T \gtrsim J_{eff}$ and temporal fluctuations of the field
can be neglected. However, the thermal fluctuations and the associated
\textit{spatial correlations} are treated non-perturbatively.
This enables one to write an effective Hamiltonian for the auxiliary fields as-

\begin{eqnarray}
H_{eff}& = & -\frac{1}{\beta}log Tr e^{-\beta H_{el}} + 
U\sum_{i}|{\bf m}_{i}|^{2} \\ \nonumber 
H_{el}& = & -\sum_{<ij>\sigma}t_{ij}(c^{\dagger}_{i\sigma}c_{j\sigma} + h.c.) - U\sum_{i}{\bf m}_{i}.{\bf \sigma}_{i}
\end{eqnarray}

Finally, configurations of the ${\bf m}_{i}$ field are sampled using 
$P({\bf m}_{i})=Tr_{cc^{\dagger}}e^{-\beta H_{eff}}$ as the sampling
weight. These configurations are used for computing static structure 
factors and distribution of moment magnitudes, defined in Eq.5 and Eq.7
respectively and shown in Fig.1(a) and 1(b). 
We also mention that the correlation temperatures in 2(a) are
size-dependent, and will ultimately collapse logarithmically with
system size. However, we've still compared the MC and Langevin answers
for the \textit{same system size} to ensure that the latter method
faithfully reproduces the static properties.

\section*{Appendix C: Comparison of low temperature spectrum with RPA}

We compare the low temperature spectra obtained using our technique
with the standard spin wave theory (RPA) results for the square lattice 
in Fig.15. The spin wave velocities are quoted from the work of Singh et. al.
\cite{singh1}. One observes a fair agreement in terms of the trends. 
The RPA values are slightly higher. 
We ascribe this discrepancy to our assumption of classical
magnetic moments. However, since our main focus is on the finite
temperature dynamics, the quantitative mismatch isn't very important.
The agreement improves as one approaches the Heisenberg limit.

\section*{Appendix D: Approaching the Mott transition}

In the triangular lattice, there's a finite $U_{c} \sim 4.5t$ 
for the Mott transition. Close to the transition, one observes
complex large-period order \cite{hrk}. However, staying within
the 120\degree ordered state (restricting ourselves to large 
enough $U/t$ values where the ground state is the former), we 
observe signatures of proximity to $U_{c}$ in the spectrum. 
Fig.16 shows a marked softening of magnetic modes along the 
$\Gamma-K$ trajectory and a gradual \textit{linear} trend of the 
dispersion along $K-M$ as the coupling is lowered. We've already shown 
the spectra at $U/t=6$ in the main text, which is the lowest coupling
we've explored within the 120\degree ordered family. Ideally, 
the complex dynamics in the vicinity of the transition should 
also be capturable using our strategy, but requires considerably 
more numerical effort, as one needs to do a thermal annealing 
to even fix the initial state for the dynamics.

\begin{figure}[b]
\centerline{
\includegraphics[height=3.5cm,width=8.7cm]{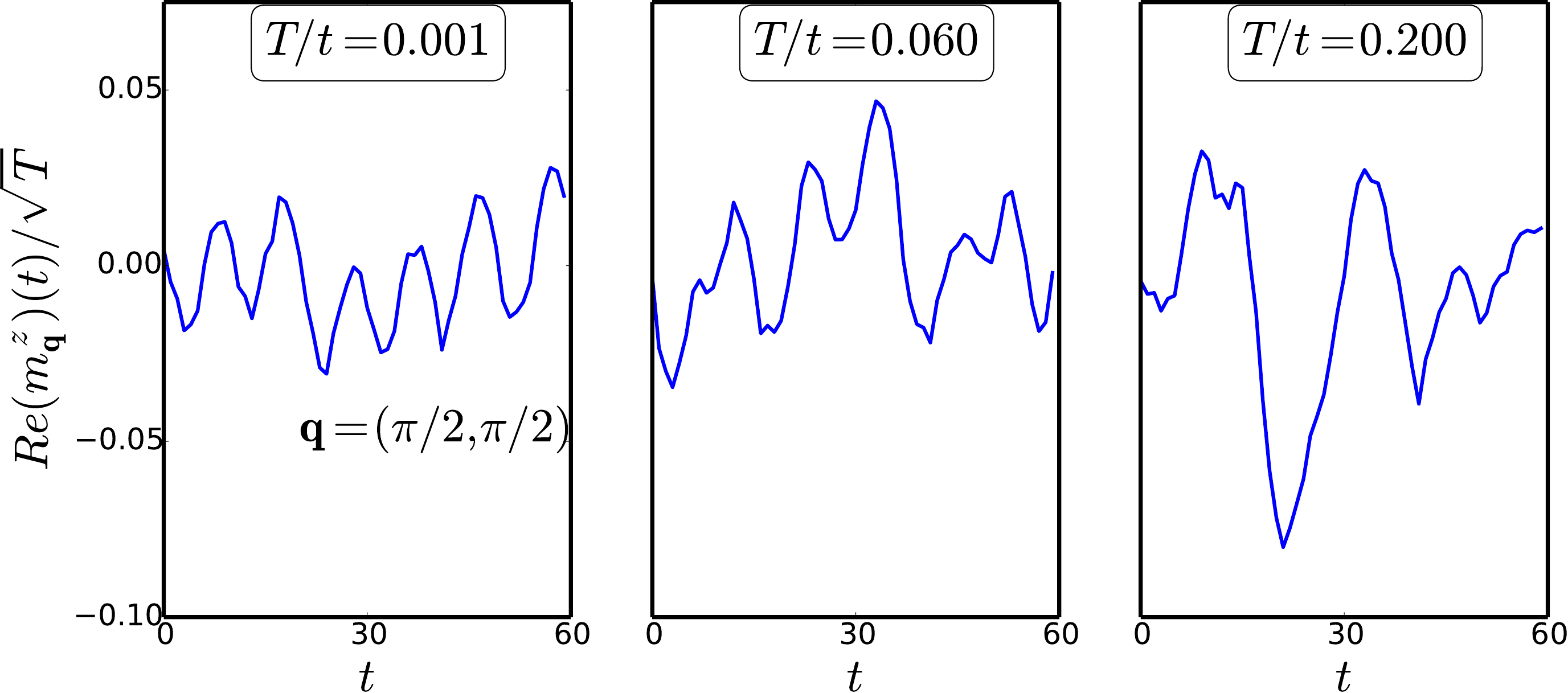}
}
\caption{Real time trajectories of $Re(m^{z}_{\bf q})(t)$
in three thermal regimes-
(i)~weakly damped ($T/t=0.001$),
(ii)~strongly damped ($T/t=0.06$) and (iii)~diffusive ($T/t=0.2$).
In (i), we see oscillations with timescale $\sim \tau_{mag}$
In (ii), some intermediate timescales emerge, but the earlier scale
is still visible. In (iii), the bare-oscillation scale is obliterated
and slow, large amplitude fluctuations dominate.
}
\end{figure}

\section*{Appendix E: Real time dynamics}

In Fig.17, we show the trajectory of the real part of 
$m^{z}_{\bf q}$ for a generic wavevector, ${\bf q}=(\pi/2,\pi/2)$, 
in real time for the three representative regimes- 
(i)~weakly damped, (ii)~strongly damped and (iii)~diffusive.
These are results for the square lattice Hubbard model at $U/t=10.0$.
We've also scaled the y-axis by $\sqrt{T}$, to gauge out the dominant
part of amplitude fluctuations.
At the lowest $T$, we see oscillatory behaviour, modified by weak noise. 
The characteristic timescale is
$\tau_{mag} \sim 1/J_{eff}$.
This corresponds to a well-defined lineshape in frequency.
In the second panel (regime (ii)), one observes the emergence
of some new timescales, but the earlier scale is still visible.
This translates in frequency space to broadened lineshapes 
centered around $\Omega_{\bf q}(T=0)$. 
On heating up further, thermal effects kill off the 
bare-oscillation timescale and slow oscillations dominate
the time series. The amplitude also increases
significantly, even after gauging the $\sqrt{T}$ factor.

\bibliographystyle{unsrt}

\begin{thebibliography}{99}

\bibitem{zhang}
X. Y. Zhang, M. J. Rozenborg, and G. Kotliar, 
Phys. Rev. Lett. {\bf 70}, 1666 (1993).

\bibitem{imada}
Masatoshi Imada, Atsushi Fujimori, and Yoshinori Tokura, 
Rev. Mod. Phys. {\bf 70}, 1039 (1998).

\bibitem{rozenberg}
Marcelo J. Rozenberg, R. Chitra, and Gabriel Kotliar, 
Phys. Rev. Lett. {\bf 83}, 3498 (1999).

\bibitem{capone}
Massimo Capone, Luca Capriotti Federico Becca, and Sergio Caprara,
Phys. Rev. B {\bf 63}, 085104 (2001).

\bibitem{park}
H. Park, K. Haule, and G. Kotliar, Phys. Rev. Lett. {\bf 101}, 186403 (2008).

\bibitem{ohashi2}
Takuma Ohashi, Tsutomu Momoi, Hirokazu Tsunetsugu, and Norio Kawakami, 
Phys. Rev. Lett. {\bf 100}, 076402 (2008).

\bibitem{sahebsara}
Peyman Sahebsara and David S\'{e}n\'{e}chal, 
Phys. Rev. Lett., {\bf 100}, 136402 (2008).

\bibitem{yamada1}
A. Yamada, Phys. Rev. B {\bf 89}, 195108 (2014).

\bibitem{ohashi1}
Takuma Ohashi, Norio Kawakami, and Hirokazu Tsunetsugu, 
Phys. Rev. Lett. {\bf 97}, 066401 (2006).

\bibitem{furukawa}
Yuta Furukawa, Takuma Ohashi, Yohta Koyama, and Norio Kawakami, 
Phys. Rev. B {\bf 82}, 161101(R) (2010).

\bibitem{bulut}
N. Bulut, W. Koshibae, and S. Maekawa, 
Phys. Rev. Lett. {\bf 95}, 037001 (2005).

\bibitem{yamada2}
A. Yamada, K. Seki, R. Eder, and Y. Ohta, 
Phys. Rev. B {\bf 83}, 195127 (2011).

\bibitem{kita}
Tomoko Kita, Takuma Ohashi, and Norio Kawakami, 
Phys. Rev. B {\bf 87}, 155119 (2013).

\bibitem{fujimoto}
Satoshi Fujimoto, Phys. Rev. B {\bf 64}, 085102 (2001).

\bibitem{normand}
B. Normand and Z. Nussinov, Phys. Rev. Lett. {\bf 112}, 207202 (2014).

\bibitem{swain}
Nyayabanta Swain, Rajarshi Tiwari, and Pinaki Majumdar, 
Phys. Rev. B {\bf 94}, 155119 (2016).

\bibitem{hirsch}
J. E. Hirsch, Phys. Rev. B {\bf 31}, 4403 (1985).

\bibitem{white}
S. R. White, D. J. Scalapino, R. L. Sugar, E. Y. Loh, J. E. Gubernatis, 
and R. T. Scalettar, Phys. Rev. B {\bf 40}, 506 (1989).

\bibitem{hrk}
H. R. Krishnamurthy, C. Jayaprakash, Sanjoy Sarker, and Wolfgang Wenzel, 
Phys. Rev. Lett. {\bf 64}, 950 (1990).

\bibitem{zimmermann}
Walter Zimmermann, Raymond Fresard, and Peter Wolfle, 
Phys. Rev. B {\bf 56}, 10097 (1997).

\bibitem{tasaki}
Hal Tasaki, J. Phys. Condens. Matter {\bf 10}, 4353 (1998).

\bibitem{fazekas}
Patrick Fazekas, \textit {Lecture notes on Electron Correlations and
Magnetism}, World Scientific (1999).

\bibitem{hochkeppel}
S. Hochkeppel, F. F. Assaad, and W. Hanke, 
Phys. Rev. B {\bf 77}, 205103 (2008).

\bibitem{watanabe}
T. Watanabe, H. Yokoyama, Y. Tanaka, and J. Inoue, 
Phys. Rev. B {\bf 77}, 214505 (2008).

\bibitem{yoshioka}
Takuya Yoshioka, Akihisa Koga, and Norio Kawakami, 
Phys. Rev. Lett. {\bf 103}, 036401 (2009).

\bibitem{tocchio1}
Luca F. Tocchio, Helene Feldner, Federico Becca, Roser Valenti, 
and Claudius Gros, Phys. Rev. B {\bf 87}, 035143 (2013).

\bibitem{tocchio2}
Luca F. Tocchio, Claudius gros, Roser Valenti, and Federico Becca, 
Phys. Rev. B {\bf 89}, 235107 (2014).

\bibitem{kokalj}
J. Kokalj and Ross H. McKenzie, Phys. Rev. Lett. {\bf 110}, 206402 (2013).

\bibitem{goto}
Shimpei Goto, Susumu Kurihara, and Daisuke Yamamoto, 
Phys. Rev. B {\bf 94}, 245145 (2016).

\bibitem{shirakawa}
Tomonori Shirakawa, Takami Tohyama, Jure Kokalj, Sigetoshi Sota, 
and Seiji Yunoki, Phys. Rev. B {\bf 96}, 205130 (2017).

\bibitem{li}
Shaozhi Li and Emanuel Gull, Phys. Rev. Research {\bf 2}, 013295 (2020).

\bibitem{DMFT}
A. Georges, G. Kotliar, W. Krauth, and M.J. Rozenberg, Rev. Mod. Phys.
{\bf 68}, 13 (1996).

\bibitem{bulla}
R. Bulla, Phys. Rev. Lett. {\bf 83}, 136 (1999).

\bibitem{zitko}
Rok Zitko, Janez Bonca, and Thomas Pruschke, 
Phys. Rev. B {\bf 80}, 245112 (2009).

\bibitem{eckstein}
Martin Eckstein, Marcus Kollar, and Philipp Werner, 
Phys. Rev. B {\bf 81}, 115131 (2010).

\bibitem{peters}
Robert Peters and Norio Kawakami, Phys. Rev. B {\bf 89}, 155134 (2014).

\bibitem{kitatani}
Motoharu Kitatani, Naoto Tsuji, and Hideo Aoki, 
Phys. Rev. B {\bf 92}, 085104 (2015).

\bibitem{kotliar1}
Gabriel Kotliar and Andrei E. Ruckenstein, 
Phys. Rev. Lett. {\bf 57}, 1362 (1986).

\bibitem{schulz}
H. J. Schulz, Phys. Rev. Lett. {\bf 65}, 2462 (1990).

\bibitem{ho}
Chang-Ming Ho, V. N. Muthukumar, Masao Ogata, and P. W. Anderson, 
Phys. Rev. Lett. {\bf 86}, 1626 (2001).

\bibitem{singh1}
Avinash Singh and Zlatko Tesanovic, Phys. Rev. B {\bf 41}, 614 (1990).

\bibitem{singh2}
Avinash Singh and Zlatko Tesanovic, Phys. Rev. B {\bf 41}, 11457 (1990).

\bibitem{singh3} 
Avinash Singh, Phys. Rev. B, {\bf 71}, 214406 (2005).

\bibitem{gunnarsson} 
O. Gunnarsson, T. Schafer, J.P.F. LeBlanc, E. Gull, 
J. Merino, G. Sangiovanni, G. Rohringer, and A. Toschi, 
Phys. Rev. Lett. {\bf 114}, 236402 (2015).

\bibitem{chern} 
Gia-Wei Chern, Kipton Barros, Zhentao Wang, Hidemaro Suwa and 
Cristian D. Batista, Phys. Rev. B {\bf 97}, 035120 (2018).

\bibitem{leblanc1}
J. P. F. LeBlanc, Shaozhi Li, Xi Chen, Ryan Levy, A. E. Antipov, 
Andrew J. Millis, and Emanuel Gull, Phys. Rev. B {\bf 100}, 075123 (2019).

\bibitem{cleveland}
Charles L. Cleveland and Rodrigo Medina A., Am. J. Phys., {\bf 44}, 44 (1976).

\bibitem{blume1}
R. E. Watson, M. Blume, and G. H. Vineyard, Phys. Rev. {\bf 181}, 180 (1969).

\bibitem{blume2}
M. Blume and J. Hubbard, Phys. Rev. B {\bf 1}, 3815 (1970).

\bibitem{takahashi}
Minoru Takahashi, J. Phys. Soc. Jpn. {\bf 52}, 3592 (1983).

\bibitem{gouvea}
M. E. Gouv\^{e}a, G. M. Wysin, A. R. Bishop, and F. G. Mertens, 
Phys. Rev. B {\bf 39}, 11840 (1989).

\bibitem{volkel}
A. R. V\"{o}lkel, G. M. Wysin, A. R. Bishop, and F. G. Mertens, 
Phys. Rev. B {\bf 44}, 10066 (1991).

\bibitem{peczak}
P. Peczak and D. P. Landau, Phys. Rev. B {\bf 47}, 14260 (1993).

\bibitem{costa}
J. E. R. Costa and B. V. Costa, Phys. Rev. B {\bf 54}, 994 (1996).

\bibitem{moessner}
R. Moessner and J. T. Chalker, Phys. Rev. Lett. {\bf 80}, 2929 (1998).

\bibitem{taillefumier} 
Mathieu Taillefumier, Julien Robert, Christopher L. Henley, Roderich Moessner, 
and Benjamin Canals, Phys. Rev. B {\bf 90}, 064419 (2014).

\bibitem{sherman}
Nicholas E. Sherman and Rajiv R. P. Singh, Phys. Rev. B {\bf 97}, 014423 (2018).

\bibitem{capriotti}
Luca Capriotti, Andreas Lauchli, and Arun Paramekanti, 
Phys. Rev. B {\bf 72}, 214433 (2005).

\bibitem{yang}
Hong-Yu Yang, Andreas M. Lauchli, Frederic Mila, and Kai Phillip Schmidt, 
Phys. Rev. Lett., {\bf 105}, 267204 (2010).

\bibitem{aeppli}
G. Aeppli, S. M. Hayden, H. A. Mook, Z. Fisk, S.-W. Cheong, D. Rytz, 
J. P. Remeika, G. P.Espinosa, and A. S. Cooper, 
Phys. Rev. Lett. {\bf 62}, 2052 (1989).

\bibitem{coldea}
R. Coldea, S. M. Hayden, G. Aeppli, T. G. Perring, C. D. Frost, T. E. Mason, 
S.-W. Cheong, and Z. Fisk, Phys. Rev. Lett. {\bf 86}, 5377 (2001).

\bibitem{stock}
C. Stock, R. A. Cowley, W. J. L. Buyers, C. D. Frost, J. W. Taylor, 
D. Peets, R. Liang, D. Bonn, and W. N. Hardy, 
Phys. Rev. B {\bf 82}, 174505 (2010).

\bibitem{kurosaki}
Y. Kurosaki, Y. Shimizu, K. Miyagawa, K. Kanoda, and G. Saito, 
Phys. Rev. Lett. {\bf 95}, 177001 (2005).

\bibitem{powell}
B. J. Powell and Ross H. McKenzie, Rep. Prog. Phys. {\bf 74}, 056501 (2011).

\bibitem{friedt}
O. Friedt, P. Steffens, M. Braden, Y. Sidis, S. Nakatsuji, and Y. Maeno, 
Phys. Rev. Lett. {\bf 93}, 147404 (2004).

\bibitem{steffens}
P. Steffens, O. Friedt, Y. Sidis, P. Link, J. Kulda, K. Schmalzl, S. Nakatsuji, and M. Braden, Phys. Rev. B {\bf 83}, 054429 (2011). 

\bibitem{shapiro}
M. C. Shapiro, Scott C. Riggs, M. B. Stone, C. R. de la Cruz, S. Chi, A. A. Podlesnyak, and I. R. Fisher, Phys. Rev. B {\bf 85}, 214434 (2012).

\bibitem{choi}
S. K. Choi, R. Coldea, A. N. Kolmogorov, T. Lancaster, I. I. Mazin, 
S. J. Blundell, P. G. Radaelli, Yogesh Singh, P. Gegenwart, K. R. Choi, 
S.-W. Cheong, P. J. Baker, C. Stock, and J. Taylor, 
Phys. Rev. Lett. {\bf 108}, 127204 (2012). 

\bibitem{tomiyasu}
Keisuke Tomiyasu, Kazuyuki Matsuhra, Kazuaki Iwasa, Masanori Watahiki, 
Seishi Takagi, Makoto Wakeshima, Yukio Hnatsu, Makoto Yokoyama, 
Kenji Ohoyama, and Kazuyoshi Yamada, 
J. Phys. Soc. Jpn. {\bf 81}, 034709 (2012).

\bibitem{bahr}
S. Bahr, A. Alfonsov, G. Jackeli, G. Khaliullin, A. Matsumoto, T. Takayama, 
H. Takagi, B. Buchner, and V. Kataev, Phys. Rev. B {\bf 89}, 180401(R) (2014).

\bibitem{bao}
Wei Bao, C. Broholm, M. Honig, P. Metcalf, and S. F. Trevino, 
Phys. Rev. B {\bf 54}, 3726(R) (1996).

\bibitem{kim}
Young-June Kim, A. P. Sorini, C. Stock, T. G. Perring, J. van den Brink, 
and T. P. Devereaux, Phys. Rev. B {\bf 84}, 085132 (2011).

\bibitem{chen}
Chih-Wei Chen, Weiyi Wang, Vaideesh Loganathan, Scott V. Carr, 
Leland W. Harriger, C. Georgen, Andriy H. Nevidomskyy, Pengcheng Dai, 
C. -L. Huang, and E. Morosan, Phys. Rev. B {\bf 99}, 144423 (2019).

\bibitem{QMC}
R. Blankenbecler, D. J. Scalapino, and R. L. Sugar, 
Phys. Rev. D {\bf 24}, 2278 (1981).

\bibitem{kotliar2}
G. Kotliar, S. Y. Savrasov, K. Haule, V. S. Oudovenko, 
O. Parcollet, and C. A. Marianetti, Rev. Mod. Phys. {\bf 78}, 865 (2006). 

\bibitem{fresard}
Vu Hung Dao and Raymond Fresard, Phys. Rev. B {\bf 95}, 165127 (2017).

\bibitem{leblanc2}
J.P.F. LeBlanc, Andrey E. Antipov, Federico Becca, Ireneusz W. Bulik, 
Garnet Kin-Lic Chan, Chia-Min Chung, Youjin Deng, Michel Ferrero, 
Thomas M. Henderson, Carlos A. Jiménez-Hoyos, E. Kozik, Xuan-Wen Liu, 
Andrew J. Millis, N.V. Prokof’ev, Mingpu Qin, Gustavo E. Scuseria, 
Hao Shi, B.V. Svistunov, Luca F. Tocchio, I.S. Tupitsyn, Steven R. White, 
Shiwei Zhang, Bo-Xiao Zheng, Zhenyue Zhu, and Emanuel Gull 
(Simons Collaboration on the Many-Electron Problem), 
Phys. Rev. X {\bf 5}, 041041 (2015).

\bibitem{ma}
Pui-Wai Ma and S. L. Dudarev, Phys. Rev. B {\bf 86}, 054416 (2012).

\bibitem{kirilyuk}
A. Kirilyuk, A. V. Kimel, and T. Rasing, Rev. Mod. Phys. {\bf 82}, 2731 (2010).

\bibitem{rebei1}
A. Rebei and G. J. Parker, Phys. Rev. B {\bf 67}, 104434 (2003).

\bibitem{brown}
William Fuller Brown, Jr., Phys. Rev. {\bf 130}, 1677 (1963).

\bibitem{rebei2}
A. Rebei, W. N. G. Hitchon, and G. J. Parker {\bf 72}, 064408 (2005).

\bibitem{mera}
B. Mera, V. R. Vieira, and V. K. Dugaev, Phys. Rev. B {\bf 88}, 184419 (2013).

\bibitem{tiwari}
Rajarshi Tiwari's thesis (2013).
http://www.hri.res.in/~libweb/theses/softcopy/rajarshi-tiwari.pdf

\bibitem{harris}
A. B. Harris, D. Kumar, B. I. Halperin, and P. C. Hohenberg, 
Phys. Rev. B {\bf 3}, 961 (1971).

\bibitem{tyc}
St{\'e}phane Ty and Bertrand I. Halperin, Phys. Rev. B {\bf 42}, 2096 (1990).

\bibitem{sauri}
Sauri Bhattacharyya, Sankha Subhra Bakshi, Samrat Kadge, and Pinaki Majumdar, 
Phys. Rev. B {\bf 99}, 165150 (2019).

\bibitem{arijit}
Arijit Dutta, Pinaki Majumdar, arXiv 2009.04533 (2020).

\bibitem{euler}
P. E. Kloeden and E. Platen, \textit{Numerical Solution of Stochastic
Differential Equations}, Springer, Berlin (1992).

\end{thebibliography}

\end{document}